\documentclass[12pt]{article}
\usepackage{amsmath,amsfonts,amssymb,amsthm,epsfig, graphicx, hyperref}
\usepackage[dvipsnames,table,dvipsnames*, svgnames*, hyperref]{xcolor}
\usepackage[linesnumbered,ruled]{algorithm2e}
\usepackage{natbib}
\usepackage{empheq}
\usepackage{caption}
\usepackage{subcaption}
\usepackage{babel,blindtext}
\usepackage{mwe}
\usepackage{float}
\usepackage{mathrsfs,enumitem,FCI_notations}
\usepackage{enumerate}
\usepackage{booktabs}
\usepackage{lscape}
\DeclareMathSizes{12}{12}{5}{3}
\usepackage{setspace}
\usepackage{caption}
\def\argmin{\mathop{\rm arg\,min}}%

\usepackage{pdfpages} 
\usepackage{pgffor} 

\makeatletter
\AtBeginDocument{\let\LS@rot\@undefined}
\makeatother

\def\supplementfilename{supplement.pdf}

\pdfximage{\supplementfilename}
\def\numbersupplementpages{\the\pdflastximagepages}

\newif\ifarXiv
\arXivtrue 

\newcommand{\blind}{1}

\begin{document}

\sloppy

\def\spacingset#1{\renewcommand{\baselinestretch}%
{#1}\small\normalsize} \spacingset{1}


\date{}

\if1\blind
{
  \title{Privacy-Preserving, Communication-Efficient, and Target-Flexible Hospital Quality Measurement\thanks{
  We thank Ambarish Chattopadhyay, Eric Cohn, Rui Duan, Zhu Shen, and Yi Zhang for their helpful comments and suggestions.}}

\author{
Larry Han\thanks{Department of Biostatistics, Harvard T.H. Chan School of Public Health; \url{larryhan@g.harvard.edu}.} \and 
Yige Li\thanks{Department of Biostatistics and CAUSALab, Harvard T.H. Chan School of Public Health.} 
\and  Bijan A. Niknam\thanks{Department of Health Care Policy and CAUSALab, Harvard University. }
\and Jos\'{e} R. Zubizarreta\thanks{Departments of Health Care Policy, Biostatistics, and Statistics, and CAUSALab, Harvard University. }}
  \maketitle
} \fi

\if0\blind
{
  \bigskip
  \bigskip
  \bigskip       
  \begin{center}
    {\LARGE \bf Privacy-Preserving, Communication-Efficient, and Target-Flexible Hospital Quality Measurement}
\end{center}
  \medskip
} \fi

\bigskip
\begin{abstract}
    Integrating information from multiple data sources can enable more precise, timely, and generalizable decisions. However, it is challenging to make valid causal inferences using observational data from multiple data sources. For example, in healthcare, learning from electronic health records contained in different hospitals is desirable but difficult due to heterogeneity in patient case mix, differences in treatment guidelines, and data privacy regulations that preclude individual patient data from being pooled. Motivated to overcome these issues, we develop a federated causal inference framework. We devise a doubly robust estimator of the mean potential outcome in a target population and show that it is consistent even when some models are misspecified. To enable real-world use, our proposed algorithm is privacy-preserving (requiring only summary statistics to be shared between hospitals) and communication-efficient (requiring only one round of communication between hospitals). We implement our causal estimation and inference procedure to investigate the quality of hospital care provided by a diverse set of $51$ candidate Cardiac Centers of Excellence, as measured by 30-day mortality and length of stay for acute myocardial infarction (AMI) patients.  We find that our proposed federated global estimator improves the precision of treatment effect estimates by $59\%$ to $91\%$ compared to using data from the target hospital alone. This precision gain results in qualitatively different conclusions about the estimated effect of percutaneous coronary intervention (PCI) compared to medical management (MM) in 63\% ($32$ of $51$) of hospitals.
We find that hospitals rarely excel in both PCI and MM, which highlights the importance of assessing performance on specific treatment regimens. These findings can help hospital managers identify areas of weakness relative to peer hospitals and can provide insights on potential resource allocation for improved quality of care.
\end{abstract}

\noindent%
\vfill

\newpage
\spacingset{1.9} 
\section{Introduction}
Accurate measurement of hospital performance matters to regulators, hospital managers, and patients. For regulators, the 2010 US Patient Protection and Affordable Care Act (Title III, Section 3001) mandated incentive payments (and penalties) to hospitals that ranked highly (or poorly) on certain quality performance benchmarks. Hospital managers must consider not only those financial incentives but also seek to understand their facility's performance to pinpoint and shore up areas of weakness relative to peer hospitals. Finally, patients can benefit from accurate information on hospital quality to help them to choose the best hospital for a particular treatment or procedure given their own health conditions and preferences. 

There are several practical and statistical challenges to hospital quality measurement \citep{normand2016league}. With typical methods, a set of hospitals must share individual-level patient data with each other in order to compare their performance relative to peers or improve the accuracy of their own measurements. However, such data sharing poses potential risks to individuals' privacy and is often incompatible with regulatory policies such as the Health Insurance Portability and Accountability Act (HIPAA) Privacy Rule in the United States and the General Data Protection Regulation (GDPR) in the European Union. Even if hospitals reached agreements to share individual-level patient data in a manner consistent with regulations, this effort may require investments in personnel and data security infrastructure beyond hospitals' scope of available resources. Thus, even before collecting data and collaborating with other hospitals to aggregate information that could improve clinical performance, managers must weigh the opportunity cost of doing so. An additional practical concern is that the iterative processes of sharing and updating these datasets and refining statistical models of hospital performance increase the number of necessary rounds of communication and therefore the chance of introducing human error in data protection or accuracy of results \citep{li2020federated}. 
Finally, it is desirable for the measurement approach to be flexible to accommodate the substantial heterogeneity in hospital patient populations so that decision-makers at each hospital are assured they are given information that is relevant to them. 

The causal inference framework can help overcome quality measurement challenges \citep{silber2014, keele2020hospital, longford2020performance}. It uses clear notation for the potential outcomes that patients may experience under alternative treatments and states precisely the estimands that investigators and stakeholders intend to measure. The framework also decouples the estimand from the estimation procedure and makes transparent the set of assumptions necessary for identification of treatment effects and valid inference. This clarity regarding assumptions allows decision-makers to evaluate their plausibility using their institutional knowledge and experience. Then, they can determine whether the evidence base is sufficient to translate the findings into action.

We propose a quality measurement approach that incorporates the causal inference framework while also overcoming the practical challenges of data sharing and communication. We describe an estimation method that leverages federated learning to overcome practical challenges to data sharing, asking each hospital to share only summary statistics describing its patient population and do so only once, thereby preserving patient privacy while also minimizing administrative burden and risk of error arising from iterative communication. To adjust for differences in patient case mix, we use these summary statistics to estimate hospital-specific density ratios relative to a target population of inference. Our proposed estimator of the target average treatment effect (TATE) ensembles estimators from each hospital by calculating data-adaptive weights. The resulting federated estimator makes an optimal bias-variance trade-off, up-weighting relevant peer hospitals that are helpful for improving estimator precision, while down-weighting or omitting hospitals that are dissimilar from the target hospital and may harm estimator accuracy. In so doing, our proposed methodology avoids negative transfer, i.e., incorporating data from other hospitals does not diminish the performance of the estimator for a target hospital. 
The proposed estimator can target specific populations, such as the hospital's own patient population for improved self-knowledge, an external target such as the national average for cross-hospital benchmarking, or specific populations of high importance for policy or patient stakeholder groups. This flexibility can make the estimator useful to a variety of stakeholders such as hospital managers, regulators, or at-risk patient populations. Finally, the framework can also help hospitals investigate their performance on specific treatments for a medical condition of interest. In summary, the proposed estimator is (a) privacy-preserving, (b) communication and resource-efficient, (c) doubly robust, and (d) population-targeted for flexibility and relevance to hospital managers, regulators, and patients.

We focus on the problem of hospital quality measurement in acute myocardial infarction (AMI) across a set of $51$ Cardiac Center of Excellence (CCE)-eligible hospitals across $29$ states in the U.S. \citep{centers}. Studying Medicare patients admitted to these 51 hospitals for AMI, we examine differences in outcomes between patients who received percutaneous coronary intervention (PCI) and those who received medical management (MM) after accounting for differences in patient case-mix. We show that estimators that only use a single hospital's own patient data lack power to distinguish treatment effects. In contrast, our federated estimator allows hospitals to estimate their treatment effects more precisely, with an $82\%$ median reduction in standard errors, ranging from $59\%$ to $91\%$. These precision gains are meaningful, altering the qualitative conclusion of the causal effect for PCI vs. MM in $63\%$ of hospitals. Importantly, we show that hospitals that perform well on PCI tend to not fare as well on MM, and vice versa. In fact, of the hospitals in the top quartile for PCI performance, not a single one ranked in the top quartile for MM performance.

The paper proceeds as follows. In Section 2 we provide a literature review. In Section 3, we detail the federated Medicare dataset for measuring the quality of care provided by CCE. Section 4 describes the problem set-up, identifying assumptions, estimation procedures, and results for inference. In Section 5, we demonstrate the performance of the federated global estimators in extensive simulation studies. In Section 6, we evaluate the quality of overall AMI care, PCI care, and MM care provided by  CCE. Section 7 concludes and discusses possibilities for future extensions of our work.

\section{Related Literature}
Dating back to the New York State registries in the late 1980s \citep{hannan2012}, data tracking of cardiac interventions has facilitated numerous insights not only into clinical performance but also the role of organizational processes in assigning treatments and improving patient outcomes. In particular, \citet{huckman2005} examined underlying staff factors contributing to variation in the use of PCI relative to the previous standard of coronary artery bypass graft (CABG) surgery. Their work showed that the relative seniority and experience of a hospital's interventional cardiologists vis-\`a-vis its cardiac surgeons strongly influenced the choice between the two treatments. However, less is known about heterogeneity in treatment assignment and treatment effects across hospitals between interventional treatments such as PCI and the alternative strategy of medical management (the predominant course of AMI treatment). 

A better understanding of treatment effects and clinical outcomes can yield insights into how organizations can improve their performance. Researchers have branched out to studying varied rates of learning across hospitals \citep{pisano2001organizational} and their underlying factors such as hospital-specific cardiac surgeon volume, termed firm specificity \citep{huckman2006firm, kc2012econometric}. Others have focused on the role of institutional teaching \citep{theokary2011empirical}, care team composition and familiarity among team members \citep{reagans2005individual, avgerinos2017team}, and principled learning from past successes and failures \citep{kc2013learning, ramdas2018variety}. These studies have furthered our understanding of operational practices that can improve the quality of heart attack care. Armed with better information on treatment effects and performance, hospital administrators can deploy strategies to improve on their weaknesses through targeted investments in clinical personnel, technology, or process improvements. 


Most statistical research into hospital quality measurement has concerned the forms and specifications of the models used to evaluate hospital performance; for example, the impact of hierarchical data structures \citep{austin2003comparing} or the properties of fixed-effects and random-effects models \citep{kalbfleisch2013monitoring}. Much consideration has also been given to strategies for producing stable and accurate estimates of performance in smaller hospitals with low case volumes, such as the shrinkage models used by the Centers for Medicare and Medicaid Services (CMS) in their Hospital Compare models \citep{normand2016league}. However, these models all rely on individual-level patient data to provide hospital-specific performance and are agnostic to the appropriateness of the peer hospitals they borrow information from. While recent advances in causal inference can address the former problem by leveraging summary-level data from federated data sources \citep{han2021federated, vo2021federated, xiong2021federated}, they can be biased if appropriate peer hospitals are not correctly identified. Until these challenges are addressed, hospital quality measurement cannot fully benefit from the privacy and communication efficiency advantages of federated methods. In addition, while various stakeholders could benefit from being able to specify flexible target populations in their models, current approaches cannot do so in the federated setting. Finally, models for hospital performance on medical conditions such as AMI do not consider both the hospital selection process and the treatment assignment process upon admission. 

\section{Measuring Quality Rendered by Cardiac Centers of Excellence}
\vspace{0.1in}
\subsection{Care Alternatives} 
Acute myocardial infarction (AMI) is one of the ten leading causes of hospitalization and death in the United States \citep{hcup2017,benjamin2017}. Consequently, hospital quality measurement in AMI has been closely studied, with risk-adjusted mortality rates reported by CMS since 2007 \citep{krumholz2006}. Numerous accreditation organizations release reports on AMI performance and designate high-performing hospitals as Cardiac Centers of Excellence (CCE) \citep{centers}. These reports typically describe a hospital's overall performance for all patients admitted for AMI. 

AMI patients can receive different types of treatments depending on the severity of disease, their age or comorbid conditions, and the admitting hospital's technical capabilities and institutional norms. For example, a cornerstone treatment for AMI is percutaneous coronary intervention (PCI), which is a cardiac procedure that restores blood flow to sections of the heart affected by AMI. PCI has been called one of the ten defining advances of modern cardiology \citep{braunwald2014} and is considered an important part of the AMI treatment arsenal. In fact, many accreditation organizations require hospitals to perform a minimum annual volume to be considered for accreditation as a CCE \citep{centers}. 

While PCI is generally considered the standard of care for ST-elevation MI (STEMI), the guidelines are less clear for non-ST-elevation MI (NSTEMI), with a recommendation rating of 5 on a scale of 1 to 9 \citep{acc2017}. As NSTEMIs comprise the majority of heart attacks in older patients, this can lead to heterogeneity in practice styles between cardiologists and across hospitals. For NSTEMIs and specific cases with contraindications for PCI, medical management (MM) is typically used. MM can include medications such as thrombolytic agents to break up blood clots, vasodilators to expand blood vessels, and anticoagulants to thin blood and enhance circulation, with the patient observed until deemed stable for discharge. Since these alternative treatment regimens draw on different clinical skills, there can be variation within a hospital in skill at rendering alternative treatments, and a hospital may compare favorably to its peers on one, but not another. 

\subsection{Building Medicare Records}
Our federated dataset consists of a $20\%$ sample of fee-for-service Medicare beneficiaries who were admitted to short-term acute-care hospitals for AMI from January 1, 2014 through November 30, 2017. A strength of this dataset is that it is representative of the entire fee-for-service Medicare population in the United States. Our Medicare claims data include complete administrative records from inpatient, outpatient, and physician providers. For each patient, we used all of their claims from the year leading up to their hospitalization to characterize the patient's degree of disease severity upon admission for AMI. ICD-9/10 codes were used to identify AMI admissions and PCI treatment status. Mortality status was validated for the purpose of accurate measurement of our outcome.

To ensure data consistency, we excluded patients under age 66 at admission or who lacked 12 months of complete fee-for-service coverage in the year prior to admission. We also excluded admissions with missing or invalid demographic or date information and patients who were transferred from hospice or another hospital. After exclusions, we randomly sampled one admission per patient.

\subsection{Identifying CCE-Eligible Hospitals}  
We examine treatment rates and outcomes for AMI across hospitals that have a sufficient annual volume of PCI procedures to be eligible to be certified as CCE. Insurers require that a hospital perform at least 100 PCIs per year \citep{centers}, which translates to a minimum of 80 PCI procedures in our four-year 20\% sample of Medicare patients. 51 hospitals met the minimum volume threshold for certification as a CCE. Collectively, these 51  CCE treated 11,103 patients, were distributed across 29 U.S. states, and included both urban and rural hospitals. The set of hospitals also displayed diverse structural characteristics as defined by data from the Medicare Provider of Services file \citep{posfile}, and included academic medical centers, non-teaching hospitals, not-for-profit, for-profit, and government-administered hospitals, as well as hospitals with varying levels of available cardiac technology services \citep{silber2018}. Thus, although CCE share a common designation, they can be heterogeneous in terms of their characteristics, capacity, and capabilities. 

\subsection{Patient Outcomes, Treatment Assignment, and Baseline Covariates}
We studied two outcomes, a binary indicator for all-cause, all-location mortality within 30 days of admission, and length of hospital stay measured in days. For the latter, rather than simply counting the number of days the patient was hospitalized (i.e., ``process" length of stay), we used the ``outcome" length of stay definition, which assigns a greater value to patients who die in the hospital. This substitution ensures that hospitals whose patients more frequently die early in the hospitalization are not spuriously considered to be more efficient than other hospitals. Accordingly, we assigned patients who died in the hospital either the $99^{th}$ percentile value of length of stay in the dataset (26 days) or, if longer, their actual length of stay. 

To define treatment with PCI versus MM, we used ICD-9/10 procedure codes from the index admission claim to determine whether each patient underwent PCI \citep{ahrqpci}. We adjusted for 10 baseline covariates that are considered predictive of both probability of receiving PCI as well as risk of mortality. Because older patients are less likely to receive PCI, we adjusted for age at admission. To account for treatment trends over time, we adjusted for admission year. As noted above, the STEMI and NSTEMI subtypes are strong predictors of treatment assignment, so we included them in the adjustment. We also adjusted for gender and prior history of PCI or CABG procedures, all of which can influence the likelihood of receiving PCI. To determine the presence of key clinical risk factors, we used diagnoses indicated as present on admission as well as comprehensive Inpatient, Outpatient, and Part B claims from the entire year prior to admission to document patient history of dementia, heart failure, unstable angina, or renal failure \citep{krumholz2006, cmsreport2021}.

\section{Methods}
\vspace{0.1in}
\subsection{Setting and Notation} 

For each patient $i$, we observe an outcome $Y_{i}$, which can be continuous or discrete, a $p$-dimensional baseline covariate vector $\bX_{i} = (X_{i1},...,X_{ip})^\top$, and a binary treatment indicator $A_{i}$, with $A_{i}=1$ denoting treatment and $A_{i}=0$ denoting control. 
Under the potential outcomes framework \citep{neyman1923application, rubin1974estimating}, we have $Y_{i} = A_{i} Y_{i}^{(1)} + (1-A_{i})Y_{i}^{(0)}$, where $Y_{i}^{(a)}$ is the potential outcome for patient $i$ under treatment $A_{i}=a$, $a=0,1$.

Suppose data for a total of $N$ patients are stored at $K$ independent study hospitals, where the $k$-th hospital has sample size $n_k$ so $N = \sum_{k=1}^K n_k$. Let $R_{i}$ be a hospital indicator such that $R_{i} = k$ indicates that patient $i$ is in the hospital $k$. We summarize the observed data at each hospital $k$ as $D_k = \{(Y_{i}, \bX_{i}^{^\top} , A_{i})^\top; R_i = k\},$ where each hospital has access to its own patient-level data but cannot share this data with other hospitals. Let $T \subseteq \{1, \dots, K\}$ indicate hospitals that comprise the target population and $S = \{1, \dots, K\} \setminus T$ indicate hospitals that comprise the source population.

\subsection{Causal Estimands and Identifying Assumptions}
Our goal is to estimate the mean potential outcome for each treatment option,
\begin{equation}
    \mu^{(a)}_T = \E[Y^{(a)} \mid R \in T], \quad a=0,1,
\end{equation}
where the expectation is taken over the distribution of potential outcomes in the target population. The target average treatment effect (TATE) is then $$\Delta_T = \mu_T^{(1)} - \mu_T^{(0)}.$$ Depending on how one specifies the target population T, the TATE can correspond to different goals for different decision-makers. For example, a hospital manager may specify the target population T to be the covariate profiles of the patients admitted to their hospital, whereas a regulator may specify T to include all covariate profiles of patients admitted to hospitals within a geographic region.

To identify the mean potential outcome under each treatment in the target population, we require the following causal assumptions:

\begin{assumption}[Consistency]\label{ass::1} 
If $A = a$, then $Y = Y^{(a)}$.
\end{assumption}

\begin{assumption}[Positivity of treatment assignment]\label{ass::2} 
$P(A=a \mid \bX=\bx, R=k) \in (0,1),$ for all $a$ and for all $\bx$ with positive density, i.e., $f(\bx \mid R=k) > 0$.
\end{assumption}

\begin{assumption}[Positivity of hospital selection]\label{ass::3} 
$P(R=k \mid \bX=\bx) > 0,$ for all $\bx$ with positive density.
\end{assumption}

\begin{assumption}[Mean exchangeability over treatment assignment]\label{ass::4}
$\E[Y^{(a)} \mid \bX = \bx, R = k, A=a] = \E[Y^{(a)} \mid \bX = \bx, R=k]$.
\end{assumption}

\begin{assumption}[Mean exchangeability over hospital selection]\label{ass::5} 
$\E[Y^{(a)} \mid \bX = \bx, R=k]= \E[Y^{(a)} \mid \bX = \bx]$. 
\end{assumption}

\begin{remark}
Assumption \ref{ass::1} states that the observed outcome for patient $i$ under treatment $a$ is the patient's potential outcome under the same treatment. Assumption \ref{ass::2} is the standard treatment overlap assumption \citep{rosenbaum1983central}, which states that the probability of being assigned to each treatment, conditional on baseline covariates, is positive in each hospital. This assumption is plausible in our case study because every hospital performs PCI and also renders MM, and no baseline covariate is an absolute contraindication for PCI. Assumption \ref{ass::3} states that the probability of being observed in a hospital, conditional on baseline covariates, is positive. This too is plausible because all patients in the study have the same insurance and none of the 51 hospitals deny admission to AMI patients on the basis of any of the baseline covariates. Assumption \ref{ass::4} states that in each hospital, the potential mean outcome under treatment $a$ is independent of treatment assignment, conditional on baseline covariates. Assumption \ref{ass::5} states that the potential mean outcome is independent of hospital selection, conditional on baseline covariates. We show that Assumption 5 is sufficient but not necessary, since our data-adaptive procedure for weighting hospitals can screen out hospitals that severely violate the assumption.
\end{remark}


\subsection{Targeting the Target of the Decision-maker}
When the primary interest lies in transporting causal inferences from source hospitals to a target hospital, we can define the target population to be a particular hospital, e.g., $T = k$. This would be particularly relevant to hospital managers focused on understanding treatment effects on their particular patient population, such as an underrepresented population for which limited samples result in a lack of power to make meaningful inferences for these marginalized groups. When primary interest lies in generalizing causal inferences across multiple hospitals, we can define the target population to comprise all hospitals, i.e., $T = \{1,...,K\}$ or a subset of hospitals, i.e., $T \subset \{1,...,K\}$. This can be particularly relevant for regulators who may care less about the causal effect of a treatment in a particular hospital and more about the causal effect across the hospitals in a geographic region. More generally, one can define the target population to be any population for which summary data is available.

An initial estimation strategy is to use only the target hospital data to estimate the mean potential outcome. For example, one could use a doubly robust estimator given by 
\begin{align}
    \hat{\mu}^{(a)}_{T} 
    &= \frac{1}{n_T} \sum_{i=1}^N I(R_i \in T) \left[m_{a,T}(\bX_i;\hat{\bgb}_{a,T}) + \frac{I(A_i=a)}{\pi_{a,T}(\bX_i;\hat{\bga}_T)}\{Y_i-m_{a,T}(\bX_i;\hat{\bgb}_{a,T})\} \right],
\end{align}
where $\pi_{a,T}(\bX_i; {\bga}_T)$ is a propensity score model for $P(A_i=a \mid \bX_i, R_i \in T)$ based on a model with finite-dimensional parameter $\bga_T$, which can be estimated with a model, denoted $\pi_{a,T}(\bX; \hat{\bga}_T)$, where $\hat{\bga}_T$ is a finite-dimensional parameter estimate; $m_{a,T}(\bX_i; {\bgb}_{a,T})$ is an outcome regression model for $\E[Y \mid \bX,A=a, R \in T]$, where $\bgb_{a,T}$ is a finite-dimensional parameter that can also be estimated by fitting a model, denoted $m_{a,T}(\bX; \hat{\bgb}_{a,T})$, where $\hat{\bgb}_{a,T}$ is a finite-dimensional parameter estimate. This augmented inverse probability weighted (AIPW) estimator \citep{robins1994estimation} is doubly robust in the sense that it is consistent when either the outcome regression model $m_{a,T}(\bX_i; \widehat{\bgb}_{a,T})$ or the propensity score model $\pi_{a,T}(\bX_i; \widehat{\bga}_T)$ is correctly specified, but not necessarily both. 

While this strategy will give an unbiased estimator when at least one model is correctly specified, it can often be imprecise, i.e., giving uninformative confidence intervals, since it leverages only the samples in the target population. It is desirable to leverage data from other source populations to 
provide more precise uncertainty quantification. However, when incorporating source data, one must be careful to 1) account for differences in patient case-mix (to avoid introducing bias through negative transfer) and 2) protect patient privacy. We consider strategies to overcome both challenges in turn.

Specifically, to estimate the mean potential outcome $\mu^{(a)}_T$ using source data, the covariate shifts between the target and source hospitals need to be accounted for in order to avoid bias in the estimator. In other words, the patient case-mix from source hospitals should be adjusted to match that of the target hospital. We correct for this potential covariate shift by estimating the density ratio, $\omega_k(\bX) = f_T(\bX) / f_k(\bX)$, where $f_T(\bX)$ denotes the density function of $\bX$ in the target hospital $T$ and $f_k(\bX)$ is the density function of $\bX$ in source hospital $k \in S$. We propose a  semiparametric model for the density ratio $\omega_k(\bX)$, such that
\begin{equation} \label{eq:expo}
    f_T(\bx) = f_k(\bx) \omega_k(\bx; \bgg_k)
\end{equation}
where the function $\omega_k$ satisfies $\omega_k(\bx;0) = 1$ and $\int f_k(\bx) \omega_k(\bx; \bgg_k)d\bx = 1$. We choose $\omega_k(\bx;\bgg_k) = \exp \left(-\bgg_k^\top \psi(\bx) \right),$
where $\psi(\cdot)$ is some $d$-dimensional basis with $1$ as its first element. This is known as the exponential tilt density ratio model \citep{qin1998inferences}. Choosing $\psi(\bx)= \bx$ recovers the entire class of natural exponential family distributions. By including higher order terms, the exponential tilt model can also account for differences in dispersion and correlations, which has great flexibility in characterizing the heterogeneity between two populations \citep{duan20201fast}.

Since patient-level data cannot be shared across hospitals, from (\ref{eq:expo}) we observe that
\begin{equation} \label{eq:4}
     \underbrace{\int \psi(\bx) f_T(\bx) d \bx}_{\E[\psi(\bX) \mid R \in T]} = \underbrace{\int \psi(\bx) f_k(\bx) \omega_k(\bx; \bgg_k) d \bx}_{\E[\psi(\bX) \omega_k(\bX; \bgg_k)|R = k]}.
\end{equation}
Thus, we propose to estimate $\bgg_k$ by solving the following estimating equations:
\begin{equation} \label{eq:5}
    n_T^{-1} \sum_{i=1}^{N} I(R_i \in T)\psi(\bX_i)  = n_k^{-1} \sum_{i=1}^{N} I(R_i = k) \psi(\bX_i) \omega_k(\bx_k;\bgg_k).
\end{equation}
Choosing $\psi(\bx)= \bx$ , the target hospital broadcasts its $p$-vector of covariate means to the source hospitals. Each source hospital $k \in S$ then solves the above estimating equations using only its own patient data.
Finally, the density ratio weight $\omega_k(\bx) = f_T(\bx) / f_k(\bx)$ is estimated as $\omega_k(\bx; \hat{\bgg}_k)$.  With the estimated density ratio weights $\omega_k(\bx; \hat{\bgg}_k)$, a designated processing center can calculate a doubly robust estimate of the mean potential outcome as
 \begin{align}
\hat{\mu}^{(a)}_{k}&= n_T^{-1}\sum_{i=1}^{N}I(R_i \in T)m_{a,T}(\bX_i;\hat{\bgb}_{a,T}) + n_k^{-1}\sum_{i=1}^{N}\frac{I(R_i=k,A_i=a)\omega_k(\bX_i;\hat{\bgg}_k)}{\pi_{a,k}(\bX_i;\hat{\bga}_k)}\{Y_i-m_{a,k}(\bX_i;\hat{\bgb}_{a,k})\}.
\end{align}

The processing site can be any one of the $K$ hospitals or another entity entirely, such as a central agency or an organization to which the hospitals belong.

\subsection{Federated Global Estimator}
We now turn toward the question of how to optimally combine the estimators from each hospital, i.e., learn ensemble weights. Without loss of generality, we take the target population to be a single hospital, but our methodology can be easily extended to the setting where the target consists of multiple hospitals.

To adaptively combine estimators from each hospital, we propose the following global estimator for the mean potential outcome, 
\begin{equation} \label{eq:2}
      \hat{\mu}_{T,Fed}^{(a)} = \hat{\mu}^{(a)}_{T} + \sum_{k \in S} \eta_k \left(\hat{\mu}^{(a)}_{k} - \hat{\mu}^{(a)}_{T} \right),
\end{equation}
where $\hat{\mu}^{(a)}_{T}$ is the estimated mean potential outcome in treatment group $A=a$ using data from the target hospital and $\hat{\mu}^{(a)}_{k}$ is the estimated mean potential outcome in $A=a$ incorporating data in source hospital $k \in S$. In (\ref{eq:2}), $\eta_k$ are adaptive ensemble weights that satisfy $\sum_{k=1}^K \eta_k = 1$ with $\eta_k \geq 0$. 

\begin{remark}
Our proposed global estimator $\hat{\mu}^{(a)}_{T,Fed}$ in (\ref{eq:2}) leverages information from both the target and source hospitals. It can be interpreted as a linear combination of the estimators in each of the $K$ hospitals, where the relative weight assigned to each hospital is $\eta_k$. For example, in the case of a single target hospital $(R=1)$ and source hospital $(R=2)$, the global estimator can be written equivalently as $$\hat{\mu}^{(a)}_{T,Fed} = \underbrace{\hat{\mu}^{(a)}_{1} + \eta(\hat{\mu}^{(a)}_{2} - \hat{\mu}^{(a)}_{1})}_{\textrm{Anchor and augment}} =  \underbrace{(1-\eta)\hat{\mu}^{(a)}_{1} + \eta \hat{\mu}^{(a)}_{2}}_{\textrm{Linear combination}},$$
where $\eta \in [0,1]$ determines the relative weight assigned to the target hospital and source hospital estimate. The left-hand equation makes it clear that we ``anchor" on the estimator from the target hospital, $\hat{\mu}^{(a)}_{1}$, and augment it with a weighted difference between the target hospital estimator and the source hospital estimator. The right-hand equation shows the estimator re-written from the perspective of a linear combination of two estimators.

Since it is likely that some source hospitals may present large discrepancies in estimating the mean potential outcome in the target hospital, $\eta_k$ should be estimated in a data-adaptive fashion, i.e., to downweight source hospitals that are markedly different. In Section \ref{sec:opt.combo}, we describe a data-adaptive method to optimally combine the $K$ hospital estimators. 
\end{remark}

\subsection{Optimal Combination and Inference}
\label{sec:opt.combo}
We now describe how the processing site can estimate the adaptive ensemble weights $\hat{\eta}_k$ such that it optimally combines estimates of the mean potential outcome in the target hospital $\hat{\mu}^{(a)}_{T}$ and source hospitals $\hat{\mu}^{(a)}_{k}$ for efficiency gain when the source hospital estimates are sufficiently similar to the target estimate, and shrinks the weight of unacceptably different source hospital estimates toward $0$. In order to safely leverage information from source hospitals, we anchor on the estimator from the target hospital, $\hat{\mu}^{(a)}_{T}$. 
When $\hat{\mu}^{(a)}_{k}$ is similar to ${\mu}^{(a)}_T$, we would seek to estimate $\eta_k$ to minimize their variance. But if $\hat{\mu}^{(a)}_{k}$ for any $k$ is too different from $\mu^{(a)}_T$, a precision-weighted estimator would inherit this bias. By examining the mean squared error (MSE) of the data-adaptive global estimator to the limiting estimand of the target-hospital estimator, the MSE can be decomposed into a variance term that can be minimized by a least squares regression of influence functions from an asymptotic linear expansion of $\hat{\mu}^{(a)}_{T}$ and $\hat{\mu}^{(a)}_{k}$, and an asymptotic bias term of $\hat{\mu}^{(a)}_{k}$ for estimating the limiting estimand $\bar{\mu}^{(a)}_{T,}$. More formally, define
\begin{align}
    \sqrt{N}(\hat{\mu}^{(a)}_{T} - \mu^{(a)}_{T}) &= \frac{1}{\sqrt{N}}\sum_{i = 1}^N \xi^{(a)}_{i,T} + o_p(1), \\ \sqrt{N}(\hat{\mu}^{(a)}_{k} - \mu^{(a)}_{k}) &= \frac{1}{\sqrt{N}}\sum_{i=1}^N \xi^{(a)}_{i,k} + o_p(1),
\end{align}
where $\xi^{(a)}_{i,T}$ is the influence function for the target hospital and $\xi^{(a)}_{i,k}$ is the influence function for source hospital $k$. The form of these influence functions is derived in the Appendix.

To estimate $\eta_k$, we minimize a weighted $\ell_1$ penalty function,
\begin{equation} \label{eq:l1}
    \hat{Q}_a(\boldsymbol{\eta}) = \sum_{i=1}^N \left[ \hat{\xi}^{(a)}_{i,T} - \sum_{k \in S} \eta_k( \hat{\xi}^{(a)}_{i,T} -  \hat{\xi}_{i,k}^{(a)} - \hat{\delta}_k)\right]^2 + \lambda \sum_{k \in S} |\eta_k|\hat{\delta}_k^2,
\end{equation}
where $\hat{\delta}_k = \hat{\mu}^{(a)}_{k} - \hat{\mu}^{(a)}_{T}$ is the estimated bias from source hospital $k$, $\lambda$ is a tuning parameter that determines the level of penalty for a source hospital, and  $\sum_{k=1}^K \eta_k = 1$ with $\eta_k \geq 0$. We call this estimator GLOBAL-$\ell_1$. 

\begin{remark}
    We show that given a suitable choice for $\lambda$, then $\hat{\eta}_k = \argmin_{\eta_k} \hat{Q}_a(\boldsymbol{\eta})$ are adaptive weights such that  $\hat{\eta}_k - \eta_k^* = O_p(n_k^{-1/2})$ when $\bar{\delta}_k = 0$ and $P(\hat{\eta}_k = 0) \to 1$ when $\bar{\delta}_k \neq 0$ (Appendix). In words, this means that (i) biased source site augmentation terms have zero weights (i.e., they are completely removed) with high probability; and (ii) regularization on the weights for unbiased source site augmentation terms is asymptotically negligible. We also show that we can solve for the $\eta_k$ that minimizes the $Q_a(\boldsymbol{\eta})$ function without sharing patient-level information from the influence functions (Appendix).
\end{remark}

\begin{remark}
The GLOBAL-$\ell_1$ estimator may be preferable in `sparse' settings where few source hospitals are similar to the target hospital TATE. Furthermore, the GLOBAL-$\ell_1$ estimator has the practical advantage of `selecting' peer hospitals for comparison. For example, in multi-year studies, all hospitals could be used in year one, and only those hospitals with non-zero $\eta_k$ weights can be used in the ensuing years. This could result in substantial resource savings in cost and time. In settings where the TATE estimates from source hospitals have relatively low biases compared to the target hospital, we would not wish to shrink the weight of any hospital to zero.  In this case, hospital-level weights $\tilde{\boldsymbol{\eta}}$ can minimize a penalty function where $\eta_k^2$ replaces $|\eta_k|$ in the penalty term of (\ref{eq:l1}). We call this estimator GLOBAL-$\ell_2$. 
\end{remark}

\subsubsection{Tuning parameter}
We propose sample splitting for the optimal selection of $\lambda$. Specifically, we split the data into training and validation datasets across all hospitals. In the training dataset, we estimate our nuisance parameters $\bga_k$, $\bgb_a$, and $\bgg_k$ and influence functions, and solve $Q_a(\boldsymbol{\eta})$ distributively for a grid of $\lambda$ values. Using the associated $\eta_k$ weights from each value of $\lambda$, we estimate the MSE in the validation data. We set the value of the optimal tuning parameter, $\lambda_{\mathrm{opt}}$, to be the value that minimizes the MSE in the validation data.

\subsubsection{Inference}
We propose estimating SEs for $\hat{\mu}^{(a)}_{T,Fed}$ using the influence functions for $\hat{\mu}^{(a)}_{T}$ and $\hat{\mu}^{(a)}_{k}$ for $k \in S$. By the central limit theorem, $\sqrt{N}(\hat{\mu}^{(a)}_{T,Fed}-\bar{\mu}^{(a)}_{T}) \overset{d}{\rightarrow} \mathcal{N}(0,\Sigma),$
where $\Sigma = E\left[\sum_{k=1}^K \bar{\eta}_k \xi^{(a)}_{i,k}\right]^2$ and $\bar{\eta}_k$ is the limiting value of $\hat{\eta}_k$. We estimate the SE of $\hat{\mu}^{(a)}_{T,Fed}$ as $\sqrt{\hat{\Sigma} / N}$, where 
\begin{equation} \label{eq:se}
   \hat{\Sigma} = N^{-1} \sum_{k=1}^K \sum_{i=1}^{n_k} \left(\hat{\eta}_k \hat{\xi}^{(a)}_{i,k} \right)^2. 
\end{equation}
Given the SE estimate for the global estimator, pointwise confidence intervals (CIs) can be constructed based on the normal approximation.

\subsection{Summary of the Workflow}

\begin{figure}[ht]
    \centering
    \vspace{-0.35in}
    \includegraphics[scale=0.72]{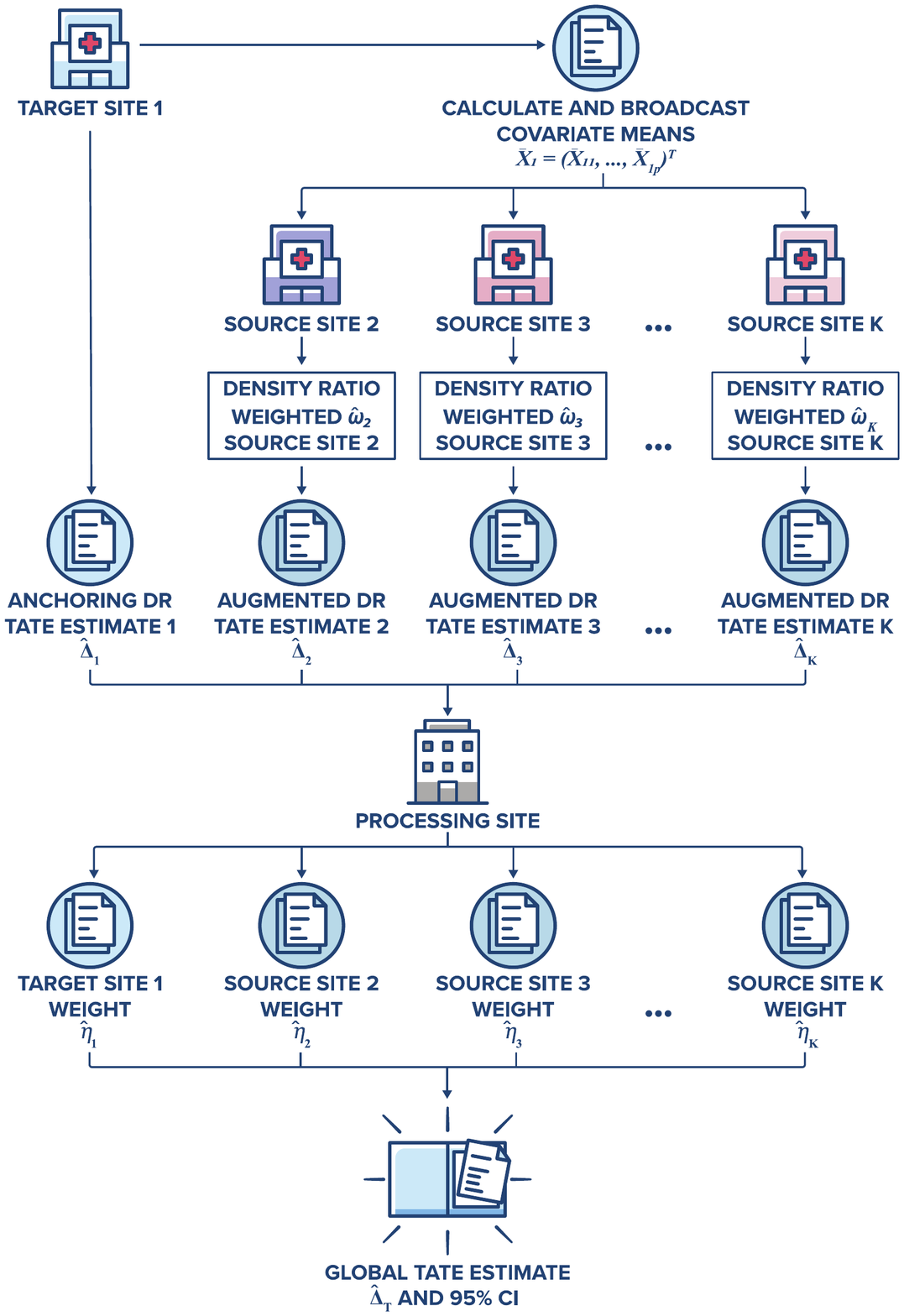}
    \vspace{-0.45in}
    \caption{Flowchart of the estimation procedure. The target site estimates its TATE with its own data, and shares its covariate means with source sites to enable them to calculate their own TATEs. A processing site then collects the estimates to determine the adaptive ensemble weights and produce the global estimate for the target site.}
    \label{fig:flow}
\end{figure}

The workflow is outlined in Algorithm 1. Figure \ref{fig:flow} provides a flowchart of the procedure. For ease of presentation, the target population is depicted as a single hospital.  First, the target hospital calculates its covariate mean vector, $\bar{\bX}_T$, and transfers it to the source hospitals. In parallel, the target hospital estimates its outcome regression model $m_{a,T}(\bx;\hat{\bgb}_{a,T})$ and its propensity score model $\pi_{a,T}(\bx;\hat{\bga}_T)$ to calculate the TATE $\hat{\Delta}_{T}$ and likewise transfers it to the processing site. The source hospitals use $\bar{\bX}_T$ obtained from the target hospital to estimate the density ratio parameter $\bgg_k$ by fitting an exponential tilt model and obtain their hospital-specific density ratio, $\omega_k(\bx) = {f_T(\bx)}/{f_k(\bx)}$, as $\exp(\hat{\bgg}_k^\top \bX)$. In parallel, the source hospitals estimate their outcome regression models as $m_{a,k}(\bX_i; \hat{\bgb}_{a,k})$ and propensity score models as $\pi_{a,k}(\bX_i;\hat{\bga}_k)$ to compute $\hat{\Delta}_{k}$. These model estimates are then shared with the processing site. Finally, the processing site computes a tuning parameter $\lambda$, adaptive weights $\eta_k$, the global TATE $\Delta_{T,Fed} = \mu_{T,Fed}^{(1)} - \mu_{T,Fed}^{(0)}$, and $95\%$ CI.

 \begin{algorithm}[ht]
 \KwData{For $k=1,...,K$ hospitals, $(Y_{ik},\bX_{ik}^\top,A_i)^\top, i=1,...,n_k$}
 
 \For{Target hospital $T$}{
  Calculate $\bar{\bX}_T = (\bar{\bX}_{T,1},...,\bar{\bX}_{T,p})$ and transfer to source hospitals. Estimate $\bga_T$, $\pi_{a,T}(\bx;\hat{\bga}_T)$, $\bgb_{a,T}$, and $m_{a,T}(\bx;\hat{\bgb}_{a,T})$. Calculate TATE as $\widehat{\Delta}_{T} = \hat{\mu}_{T}^{(1)} - \hat{\mu}_{T}^{(0)}$ and transfer to processing site.

 }
 \For{Source hospitals $k \in S$}{
     Solve for $\bgg_{k}$, calculate $\omega_k(\bx;\hat{\bgg}_k) = \exp(\hat{\bgg}_k^\top \bx)$ and transfer to processing site. Estimate $\bga_k$, $\pi_{a,k}(\bx,\hat{\bga}_k)$, $\bgb_{a,k}$, $m_{a,k}(\bx;\hat{\bgb}_{a,k})$ and transfer to processing site. 

 }
  \For{processing site}{
        Calculate the TATE estimator from each source hospital as $\widehat{\Delta}_{k} = \hat{\mu}_{k}^{(1)} - \hat{\mu}_{k}^{(0)}$. Estimate $\eta_k$ by solving the penalized regression in (\ref{eq:l1}). Construct the final global estimator as $\widehat{\Delta}_{T,Fed} = \hat{\mu}^{(1)}_{T,Fed} - \hat{\mu}^{(0)}_{T,Fed}$ by (\ref{eq:2}) and variance by (\ref{eq:se}) and construct 95\% CI.

 }
 
 \KwResult{Global TATE estimate, $\hat{\Delta}_T$ and $95\%$ CI }
 \caption{Pseudocode to obtain global estimator leveraging all hospitals}
\end{algorithm}

\clearpage


\section{Simulation Study}
We evaluate the finite-sample performance of our proposed federated global estimators and compare them to an estimator that leverages target hospital data only and two sample size-adjusted estimators that use all hospitals, but do not adaptively weight them. We examine the empirical absolute bias, root mean square error (RMSE), coverage of the $95\%$ CI, and length of the $95\%$ CI for alternative data generating mechanisms and various numbers of source hospitals, running $1000$ simulations for each setting.

\subsection{Data Generating Process}

We examine two generating mechanisms: the dense $\mathcal{D}_{\text{dense}}$ and sparse $\mathcal{D}_{\text{sparse}}$ data settings, where $\mathcal{D}_{\text{sparse}}$ means that fewer source hospitals have the same covariate distribution as the target hospital, and the proportion of such source hospitals declines as the number of source hospitals increases. 

To simulate heterogeneity in the covariate distributions across hospitals, we consider skewed normal distributions with varying levels of skewness for each hospital. Specifically, the covariates $\bX_{kp}$ are generated from a skewed normal distribution $\mathcal{SN}(x; \Xi_{kp}, \Omega^2_{kp}, \mathrm{A}_{kp})$, where $k = 1,...,K$ indexes the hospitals and $p=1,...,P$ indexes the covariates. $\Xi_{kp}$ is the location parameter, $\Omega_{kp}$ is the scale parameter, and $\mathrm{A}_{kp}$ is the skewness parameter. The distribution follows the density function $f(x) = 2\phi\left(\frac{x-\Xi_{kp}}{\Omega_{kp}}\right)\Phi\left(\mathrm{A}_{kp}\frac{x-\Xi_{kp}}{\Omega_{kp}}\right)$, where $\phi(\cdot)$ is the standard normal probability density function and $\Phi(\cdot)$ is the standard normal cumulative distribution function. Using these distributions, we examine and compare the dense $\mathcal{D}_{\text{dense}}$ and sparse $\mathcal{D}_{\text{sparse}}$ settings. In this section, we examine in detail the $\mathcal{D}_{\text{sparse}}$ setting with $P = 2$. To study the likely impact of increasing  $P$ and to show that the algorithm accommodates both continuous and binary covariates, we consider $P = 10$ continuous covariates for $\mathcal{D}_{\text{dense}}$, and for $\mathcal{D}_{\text{sparse}}$, we consider $P = 10$ with two continuous and eight binary covariates. The covariate distribution of the target hospital is the same in each setting.

For the target hospital $k=1$, we specify its sample size as $n_1 = 100$ patients. We assign sample sizes to each source hospital using the distribution $$\{n_k\}_{k = 2}^K \sim \min\{\text{Gamma}(16, 0.08), 50\},$$ specifying that the gamma distribution have a mean of $200$, a standard deviation of $50$, and a minimum volume threshold of $50$ patients. 

\subsection{Simulation Settings}
The true potential outcomes are generated as 
$$Y_k(1) = (X_k - \mu_1)\beta_{11} + X_k^{\circ 2}\beta_{21} + 3 + \varepsilon_k,$$
where $X_k^{\circ 2}$ denotes $X_k$ squared element-wise, $\beta_{11} = 3\times(0.4,.., 1.2)/P$ is a $P$-vector of equal increments, $\beta_{21} = \boldsymbol{0}$, $\varepsilon_k \sim \mathcal{N}(0, \frac{9}{4}P^2)$, and $$Y_k(0) = (X_k - \mu_1)\beta_{10} + X_k^{\circ 2}\beta_{20} + \varepsilon_k,$$ where $\beta_{10} = (0.4,.., 1.2)/P$ is a $P$-vector of equal increments, and $\beta_{20} = \boldsymbol{0}$.

The true propensity score model is generated as $$A_k \mid X=x \sim \text{Bernoulli}(\pi_k), \quad \pi_k = \text{expit} (X_k\alpha_{1} + X_k^{\circ 2}\alpha_{2}),$$ with $\alpha_{1} = (0.5, ..., -0.5)$, $\alpha_{2} = \boldsymbol{0}$. 

We consider five different model specification settings. In Setting I with  $\mathcal{D}_{\text{dense}}$, we study the scenario where both the outcome model and propensity score model are correctly specified. Setting II differs in that we have a $P$-vector of equal increments $\beta_{21} = \beta_{20} = (0.2,.., 0.4)$, so that the true $Y_k(1)$ and $Y_k(0)$ include quadratic terms, which we misspecify by fitting a linear outcome model. Setting III differs from Setting I in that we have a $P$-vector of equal decrements $\alpha_{2} = (0.15, ..., -0.15)$, which we misspecify by fitting a logistic linear propensity score model. Setting IV includes both $P$-vectors of equal increments $\beta_{21} = \beta_{20} = (0.2,.., 0.4)$ and $P$-vectors of equal decrements $\alpha_{2} = (0.15, ..., -0.15)$, which we misspecify by fitting a linear outcome regression model and a logistic linear propensity score model, respectively. Finally, Setting V has $\beta_{21} = \beta_{20} = (0.2,.., 0.4)$, $\alpha_{2} = \boldsymbol{0}$ for the target hospital and half the source hospitals $n_k \in (\text{Q1}(\{n_k\}_{k = 1}^K), \text{Q3}(\{n_k\}_{k = 1}^K))$, but $\alpha_{2} = (0.15, ..., -0.15)$ for the remaining source hospitals, thereby misspecifying the outcome model in all hospitals and the propensity score model in half the source hospitals. For $\mathcal{D}_{\text{sparse}}$, the five settings are generated similarly. Details on the generating mechanisms are provided in the Appendix. 

In the simulations, we consider all ten combinations of the model specifications and covariate density setting with $K \in \{10, 20, 50\}$ total hospitals, and five estimators: 1) an estimator using data from the target hospital only (Target-Only), 2) an estimator using all hospitals that weights each hospital proportionally to its sample size and assumes homogeneous covariate distributions across hospitals by fixing the density ratio to be $1$ for all hospitals (SS naive), 3) an estimator using all hospitals that weights each hospital proportionally to its sample size but correctly specifies the density ratio weights (SS), 4) the GLOBAL-$\ell_1$ federated estimator, and 5) the GLOBAL-$\ell_2$ federated estimator.

We choose the tuning parameter $\lambda$ from among $\{0, 10^{-4}, 10^{-3}, 10^{-2}, 0.1, 0.25, 0.5, 1, 2, 5, 10\}$ as follows. In ten folds, we split the simulated datasets into two equal-sized samples, with each containing all hospitals, using one sample for training and the other for validation. The $Q_a(\boldsymbol{\eta})$ function is evaluated as the average across those ten splits.

\subsection{Diagnostics for Assessing Patient Case-Mix Balance}
\begin{figure}[ht]
    \centering
    \includegraphics[scale=0.35]{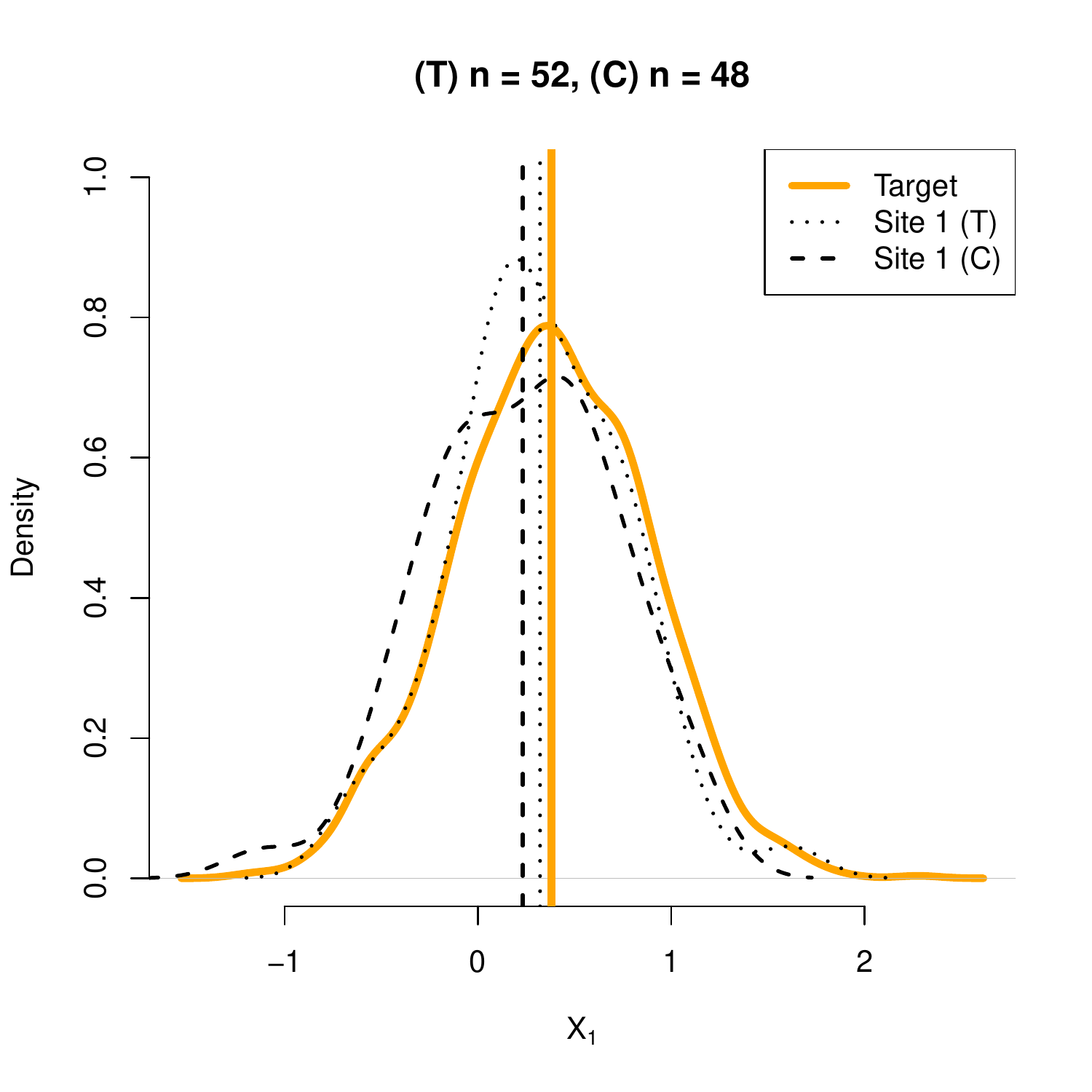}
     \includegraphics[scale=0.35]{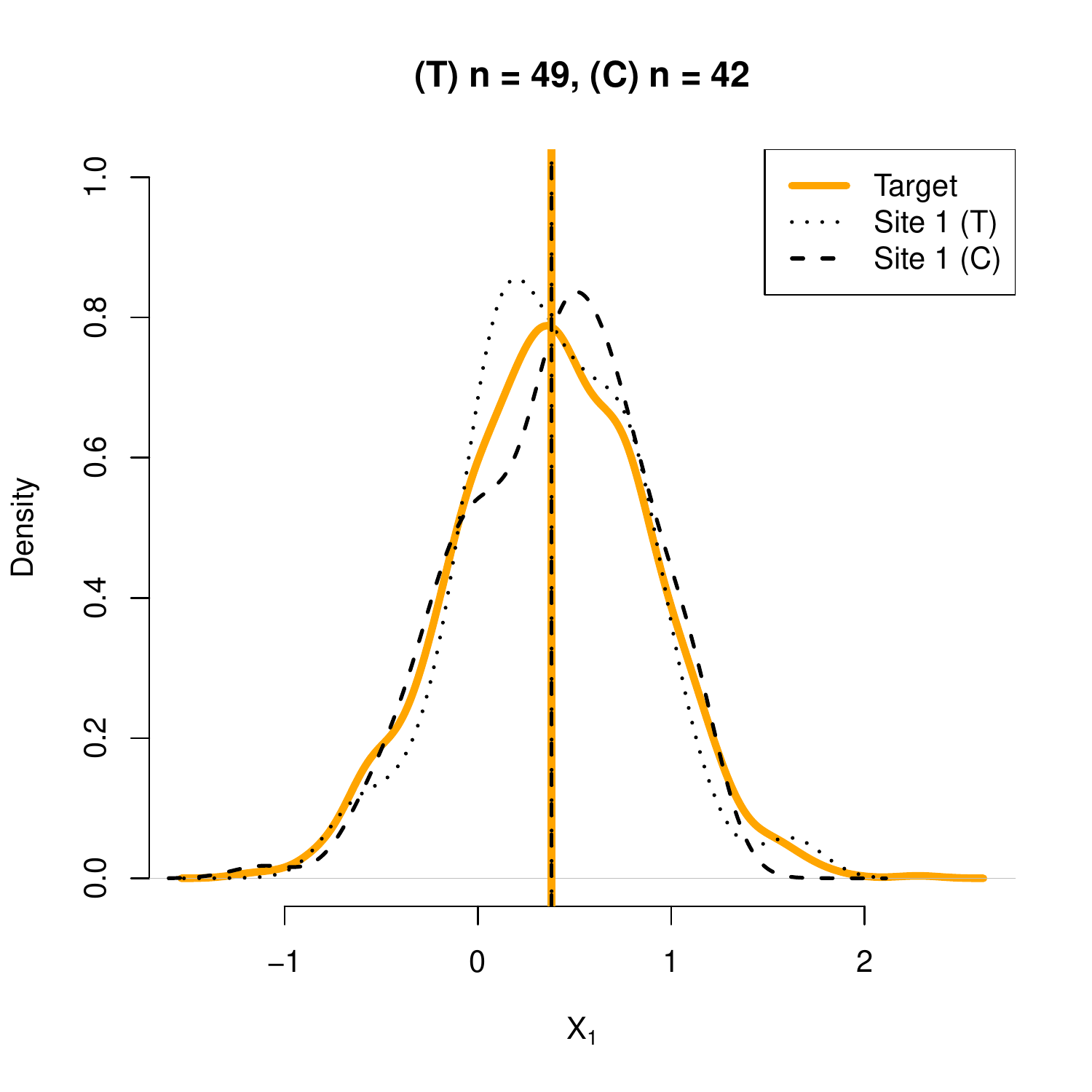}
     \includegraphics[scale=0.35]{graphics_temp/X1_density_weighted_target_only_seed8.pdf} \\
     \includegraphics[scale=0.35]{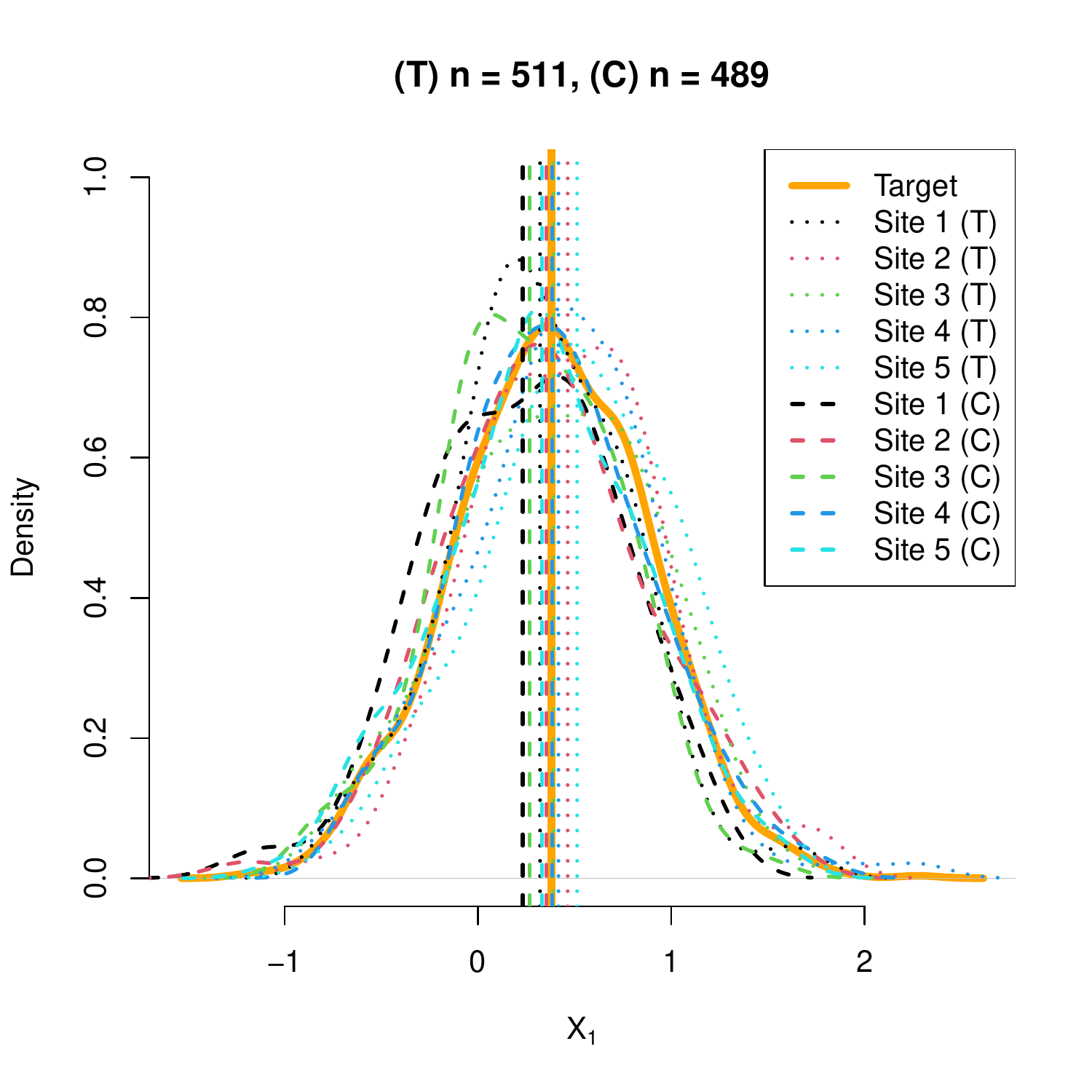}
     \includegraphics[scale=0.35]{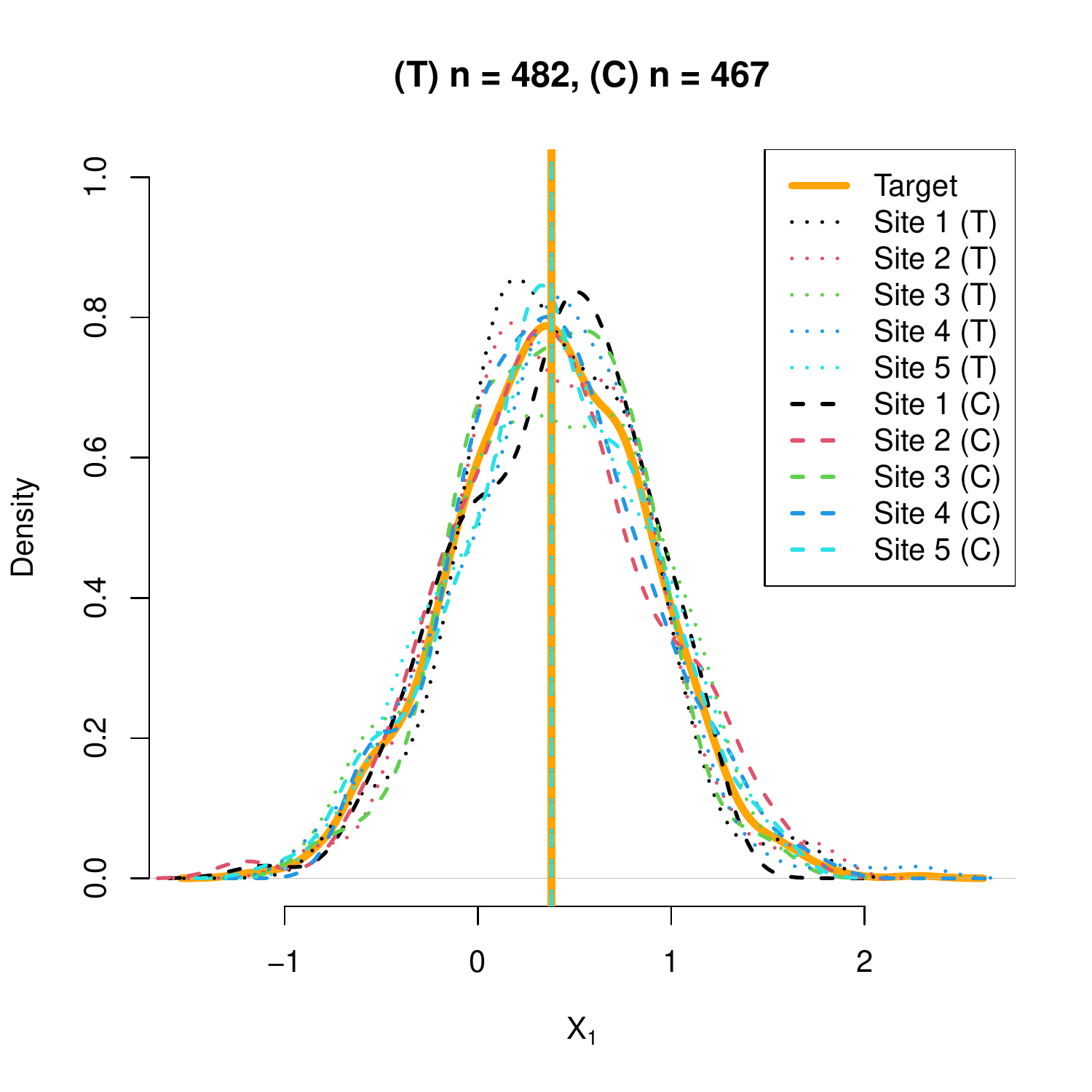}
    \includegraphics[scale=0.35]{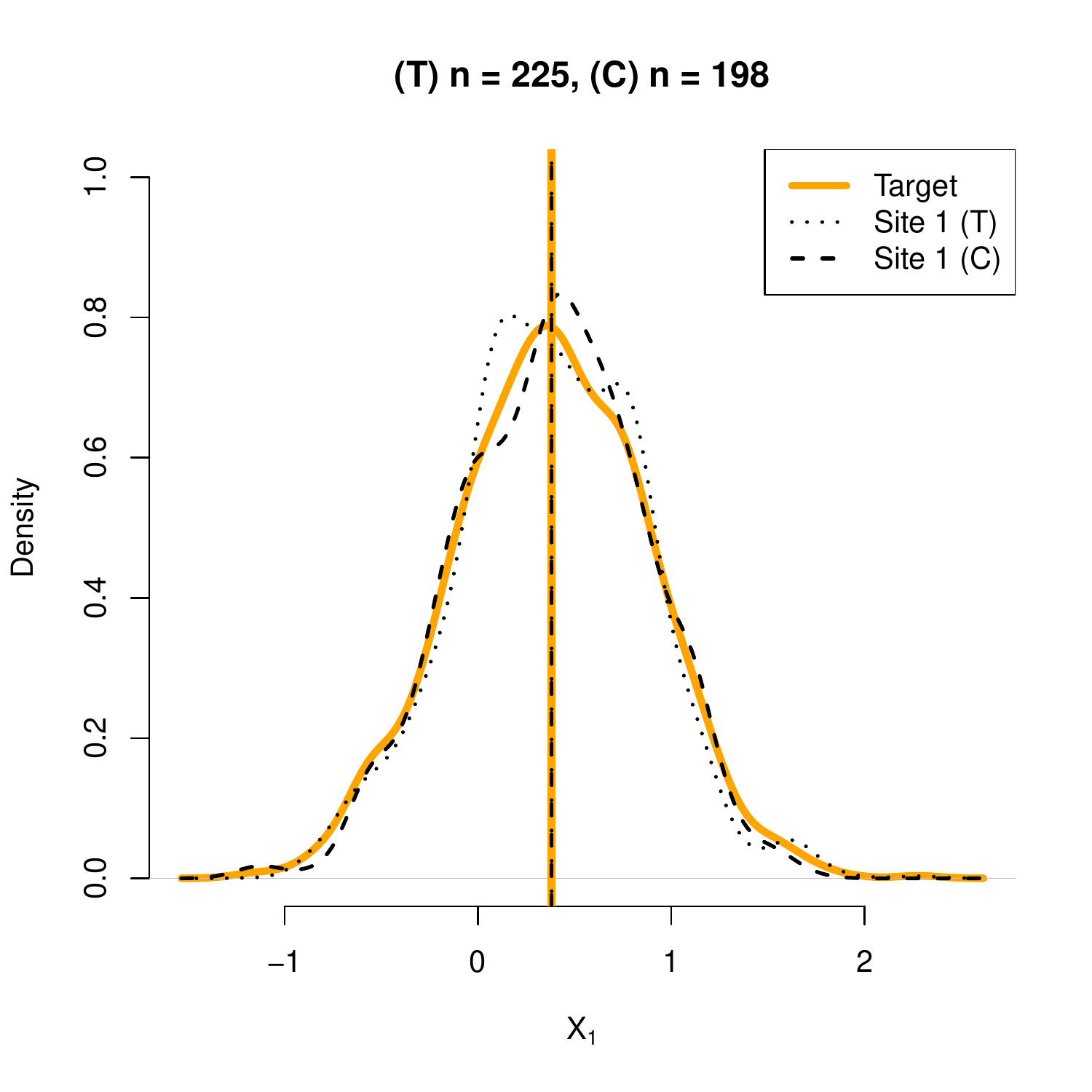} \\
    \begin{subfigure}{0.325\textwidth}
     \includegraphics[scale=0.35]{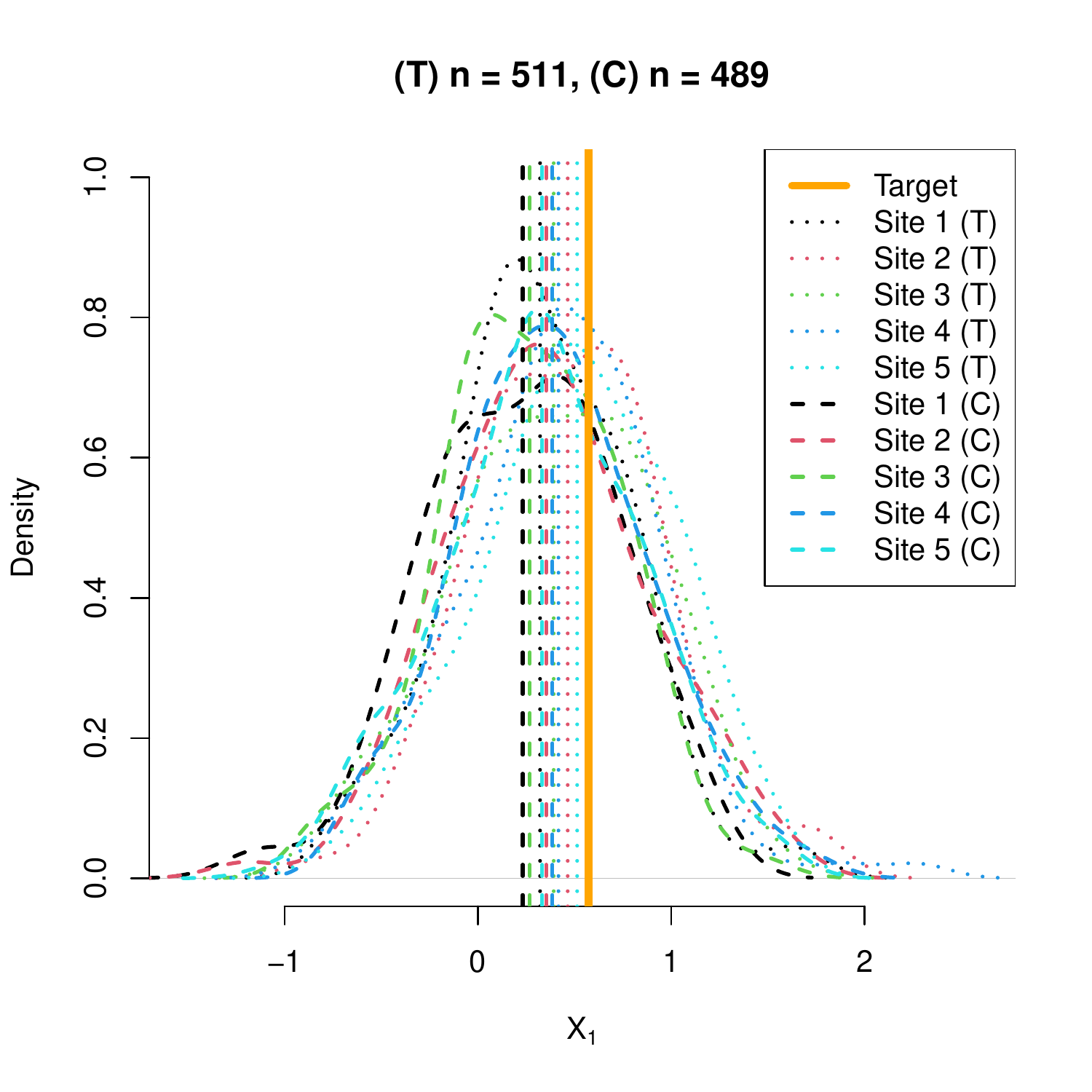}
     \caption{Before adjustment}
     \end{subfigure}
     \begin{subfigure}{0.325\textwidth}
     \includegraphics[scale=0.35]{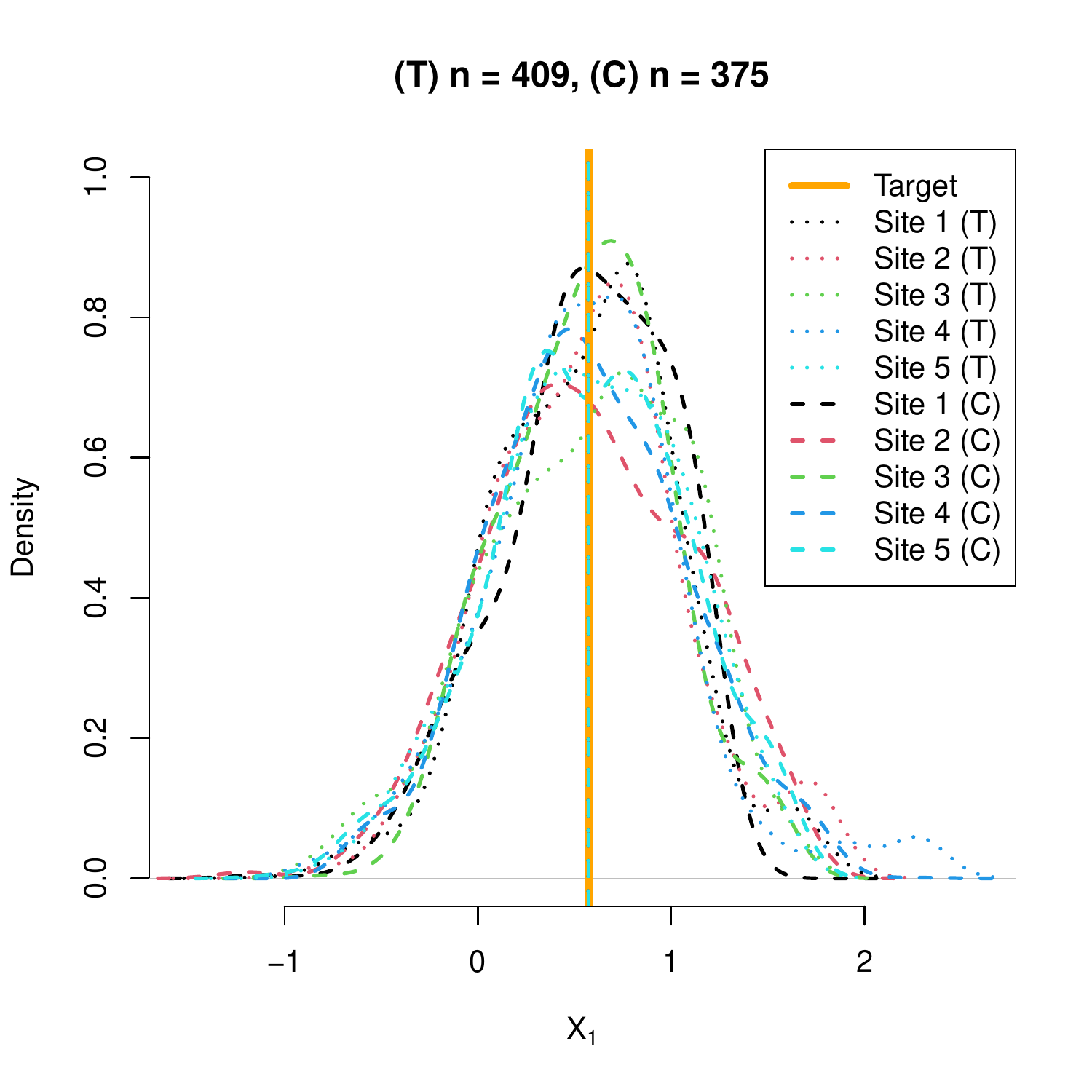}
     \caption{After adjustment}
     \end{subfigure}
     \begin{subfigure}{0.325\textwidth}
    \includegraphics[scale=0.35]{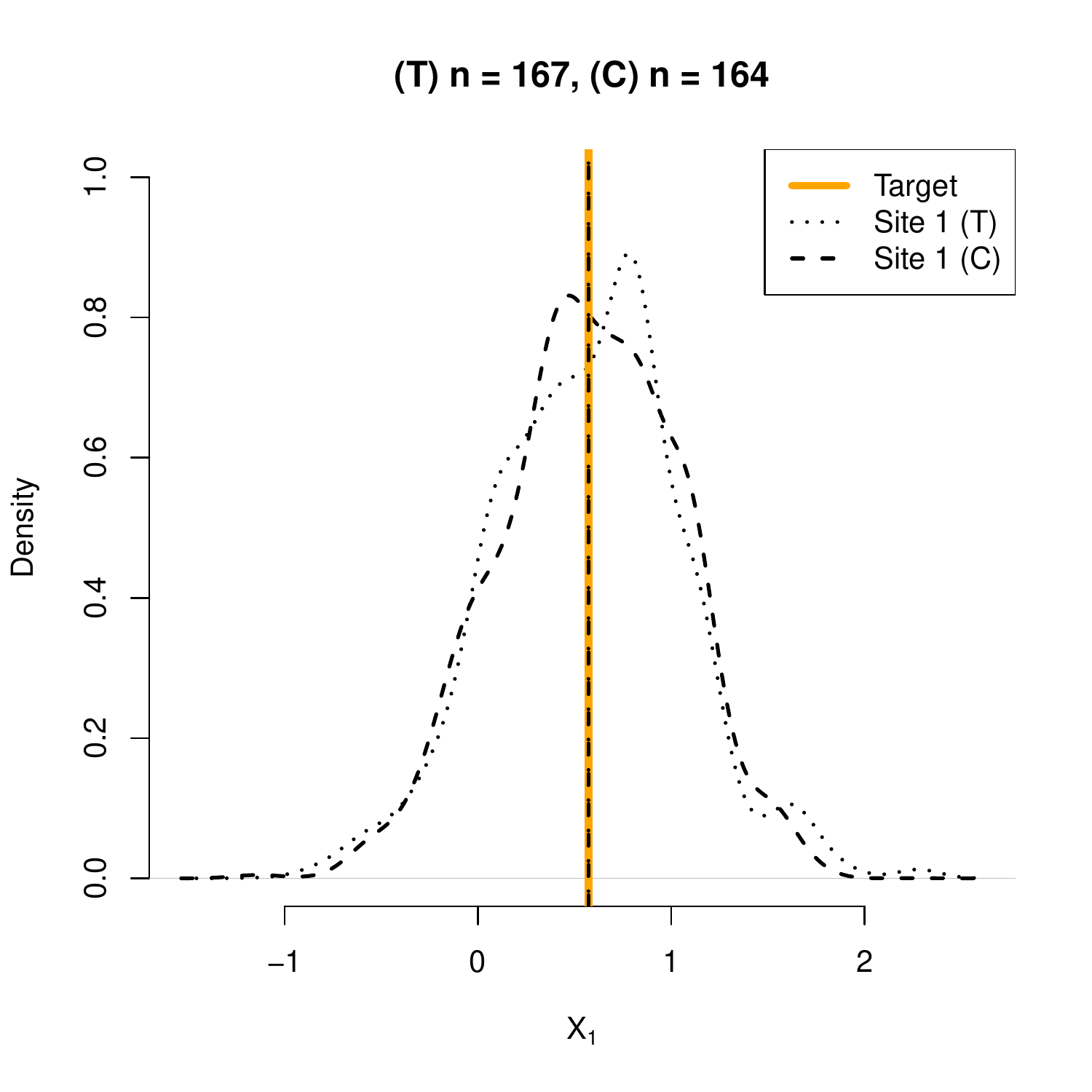}
    \caption{Combining sites}
    \end{subfigure}
    \caption{
    Shifts of covariate distribution before and after adjustment of two estimation procedures. The first row illustrates the shifts of covariate distribution in the target-only estimator. The target site estimates its target mean potential outcomes by weighting the treated group (T) and the control group (C) respectively toward the target population. The second row illustrates a combined estimator (Global-$\ell_1$), where the target site and the source sites are combined for estimation by weighting their treated group and control group toward the target population respectively. The combined estimator is shown to be more precise than the target-only estimator and has a larger underlining effective sample size (225 versus 49 for the treated group, 198 versus 42 for the control group). The third row shows that the Global-$\ell_1$ estimator can adjust to a specified target population.}
    \label{fig:curve_example}
\end{figure}

Figure \ref{fig:curve_example} shows how the implied weights from two estimators (target-only and Global-$\ell_1$) adjust the covariate distributions so that the weighted covariate distributions of the treated group and the control group approach their target population in a simulated example. 

\subsection{Simulation Results}


Throughout this section, we describe in detail the simulation results for model specifications I--V with $P = 2$ in the $\mathcal{D}_{\text{sparse}}$ setting evaluating indirect standardization. We address the $\mathcal{D}_{\text{dense}}$ setting and extensions to $P = 10$ covariates and report detailed numerical results for these alternative settings in the Appendix. A detailed simulation study about direct standardization is also presented in the Appendix, comparing the proposed approach with the traditional fixed-effects regression.

Table \ref{tab_sparse_2} reports results for $\mathcal{D}_{\text{sparse}}$ with $P = 2$ across $1000$ simulations. 

\begin{center}
\renewcommand{\arraystretch}{0.7}
\begin{table}[ht]
\caption{Results from 1000 simulated datasets for covariate distribution $\mathcal{D}_{\text{sparse}}$ when $P = 2$ with varying simulation specifications and numbers of source sites.}
\label{tab_sparse_2}
\centering
\footnotesize
\setlength{\tabcolsep}{4pt}
\begin{tabular}{lrrrrrrrrrrrr}
  \hline
   & \multicolumn{12}{c}{Simulation scenarios} \\
     \cmidrule(lr){2-13} 
  &  \multicolumn{4}{c}{$K = 10$} &  \multicolumn{4}{c}{$K = 20$} &  \multicolumn{4}{c}{$K = 50$} \\
  \cmidrule(lr){2-5}  \cmidrule(lr){6-9} \cmidrule(lr){10-13} 
     & Bias & RMSE & Cov. & Len. & Bias & RMSE & Cov. & Len. & Bias & RMSE & Cov. & Len. \\
   \cmidrule(lr){1-1}  \cmidrule(lr){2-5}  \cmidrule(lr){6-9} \cmidrule(lr){10-13} 
Specification I &  &  &\\ 
\hspace{.15cm} Target-Only & 0.00 & 0.69 & 98.20 & 3.10 & 0.00 & 0.69 & 98.10 & 3.10 & 0.00 & 0.69 & 98.20 & 3.10\\ 
\hspace{.15cm} SS (naive) & 0.87 & 0.88 & 24.90 & 1.41 & 0.90 & 0.91 & 1.20 & 0.84 & 0.91 & 0.92 & 0.00 & 0.45\\ 
\hspace{.15cm} SS & 0.01 & 0.40 & 99.60 & 2.40 & 0.01 & 0.29 & 99.20 & 1.57 & 0.00 & 0.20 & 96.50 & 0.91 \\ 
\hspace{.15cm} GLOBAL-$\ell_2$ & 0.16 & 0.34 & 98.80 & 1.78 & 0.17 & 0.29 & 96.30 & 1.23 & 0.17 & 0.24 & 84.40 & 0.72\\ 
\hspace{.15cm} GLOBAL-$\ell_1$ & 0.05 & 0.48 & 97.30 & 2.01 & 0.07 & 0.41 & 97.20 & 1.60 & 0.09 & 0.35 & 94.60 & 1.20\\ 
   \cmidrule(lr){1-1}  \cmidrule(lr){2-5}  \cmidrule(lr){6-9} \cmidrule(lr){10-13} 
Specification II &  &  &\\ 
\hspace{.15cm} Target-Only & 0.00 & 0.72 & 97.60 & 3.18 & 0.00 & 0.72 & 97.50 & 3.18 & 0.00 & 0.72 & 97.60 & 3.18 \\ 
\hspace{.15cm} SS (naive) & 0.87 & 0.89 & 26.40 & 1.44 & 0.90 & 0.91 & 2.20 & 0.86 & 0.91 & 0.92 & 0.00 & 0.46\\
\hspace{.15cm} SS & 0.01 & 0.40 & 99.50 & 2.44 & 0.01 & 0.29 & 99.20 & 1.59 & 0.00 & 0.21 & 96.30 & 0.92\\ 
\hspace{.15cm} GLOBAL-$\ell_2$ & 0.18 & 0.35 & 98.60 & 1.86 & 0.18 & 0.30 & 96.00 & 1.27 & 0.19 & 0.25 & 84.30 & 0.74 \\ 
\hspace{.15cm} GLOBAL-$\ell_1$ & 0.06 & 0.49 & 98.00 & 2.05 & 0.08 & 0.42 & 97.00 & 1.63 & 0.10 & 0.35 & 94.60 & 1.20\\ 
   \cmidrule(lr){1-1}  \cmidrule(lr){2-5}  \cmidrule(lr){6-9} \cmidrule(lr){10-13} 
Specification III &  &  &\\ 
\hspace{.15cm} Target-Only & 0.04 & 0.70 & 96.60 & 3.09 & 0.04 & 0.70 & 96.50 & 3.09 & 0.04 & 0.70 & 96.60 & 3.09 \\ 
\hspace{.15cm} SS (naive) & 0.84 & 0.86 & 28.50 & 1.46 & 0.87 & 0.88 & 2.10 & 0.88 & 0.89 & 0.89 & 0.00 & 0.46 \\ 
\hspace{.15cm} SS  & 0.02 & 0.42 & 99.90 & 2.56 & 0.01 & 0.31 & 99.20 & 1.69 & 0.03 & 0.22 & 97.40 & 0.96 \\ 
\hspace{.15cm} GLOBAL-$\ell_2$ & 0.12 & 0.33 & 99.00 & 1.83 & 0.13 & 0.28 & 97.50 & 1.27 & 0.14 & 0.22 & 88.30 & 0.73\\ 
\hspace{.15cm} GLOBAL-$\ell_1$ & 0.01 & 0.50 & 97.00 & 2.04 & 0.03 & 0.42 & 96.00 & 1.62 & 0.05 & 0.35 & 93.40 & 1.20 \\ 
   \cmidrule(lr){1-1}  \cmidrule(lr){2-5}  \cmidrule(lr){6-9} \cmidrule(lr){10-13} 
Specification IV &  &  &\\ 
\hspace{.15cm} Target-Only & 0.10 & 0.74 & 96.80 & 3.18 & 0.10 & 0.74 & 96.90 & 3.18 & 0.10 & 0.74 & 96.80 & 3.18\\ 
\hspace{.15cm} SS (naive) & 0.82 & 0.83 & 34.70 & 1.53 & 0.85 & 0.86 & 3.30 & 0.91 & 0.86 & 0.87 & 0.00 & 0.48\\ 
\hspace{.15cm} SS  & 0.05 & 0.43 & 99.80 & 2.60 & 0.04 & 0.31 & 99.20 & 1.71 & 0.06 & 0.23 & 97.20 & 0.98\\ 
\hspace{.15cm} GLOBAL-$\ell_2$ & 0.10 & 0.33 & 99.00 & 1.92 & 0.11 & 0.27 & 97.70 & 1.32 & 0.12 & 0.21 & 91.30 & 0.76\\ 
\hspace{.15cm} GLOBAL-$\ell_1$ & 0.03 & 0.52 & 96.70 & 2.09 & 0.01 & 0.44 & 95.50 & 1.65 & 0.02 & 0.36 & 92.90 & 1.20\\ 
   \cmidrule(lr){1-1}  \cmidrule(lr){2-5}  \cmidrule(lr){6-9} \cmidrule(lr){10-13} 
Specification V &  &  &\\ 
\hspace{.15cm} Target-Only & 0.00 & 0.72 & 97.60 & 3.18 & 0.00 & 0.72 & 97.50 & 3.18 & 0.00 & 0.72 & 97.60 & 3.18\\ 
\hspace{.15cm} SS (naive) & 0.85 & 0.86 & 27.90 & 1.44 & 0.87 & 0.88 & 2.50 & 0.87 & 0.89 & 0.89 & 0.00 & 0.46 \\ 
\hspace{.15cm} SS & 0.02 & 0.43 & 99.70 & 2.59 & 0.02 & 0.31 & 99.30 & 1.71 & 0.03 & 0.22 & 96.90 & 0.98\\ 
\hspace{.15cm} GLOBAL-$\ell_2$ & 0.17 & 0.35 & 98.90 & 1.87 & 0.17 & 0.29 & 96.30 & 1.29 & 0.18 & 0.24 & 86.30 & 0.74\\ 
\hspace{.15cm} GLOBAL-$\ell_1$ & 0.06 & 0.50 & 97.80 & 2.07 & 0.07 & 0.43 & 96.80 & 1.63 & 0.09 & 0.35 & 94.50 & 1.20 \\ 
\hline
\end{tabular}
Abbreviations: RMSE = Root mean square error; Cov. = Coverage, Len. = Length of $95\%$ CI; \\ SS = Sample Size.
\vspace{.25cm}
\footnotesize{
\begin{flushleft}
\end{flushleft}
}
\end{table}
\end{center}

In this setup, the GLOBAL-$\ell_1$ estimator produces sparser weights and has substantially lower RMSE than the Target-Only estimator in every setting where at least one hospital has a correctly specified model (Settings I, II, III, and V). The GLOBAL-$\ell_2$ estimator produces estimates with larger biases, but also with the lowest RMSE, with the RMSE advantage increasing with $K$. Relative to the global estimators, the SS (naive) estimator demonstrates very large biases, while the SS estimator that incorporates the density ratio weights has less bias and lower RMSE, but has longer CIs. A notable difference in the alternative $\mathcal{D}_{\text{dense}}$ setting is that the GLOBAL-$\ell_2$ estimator produces more uniform weights and has better performance (see Appendix Table 3). 
In the Appendix, the distribution of patient-level observations is visualized for the $K=10$, $P = 2$ case.

\begin{figure}[ht]
    \centering
    \includegraphics[scale=0.7]{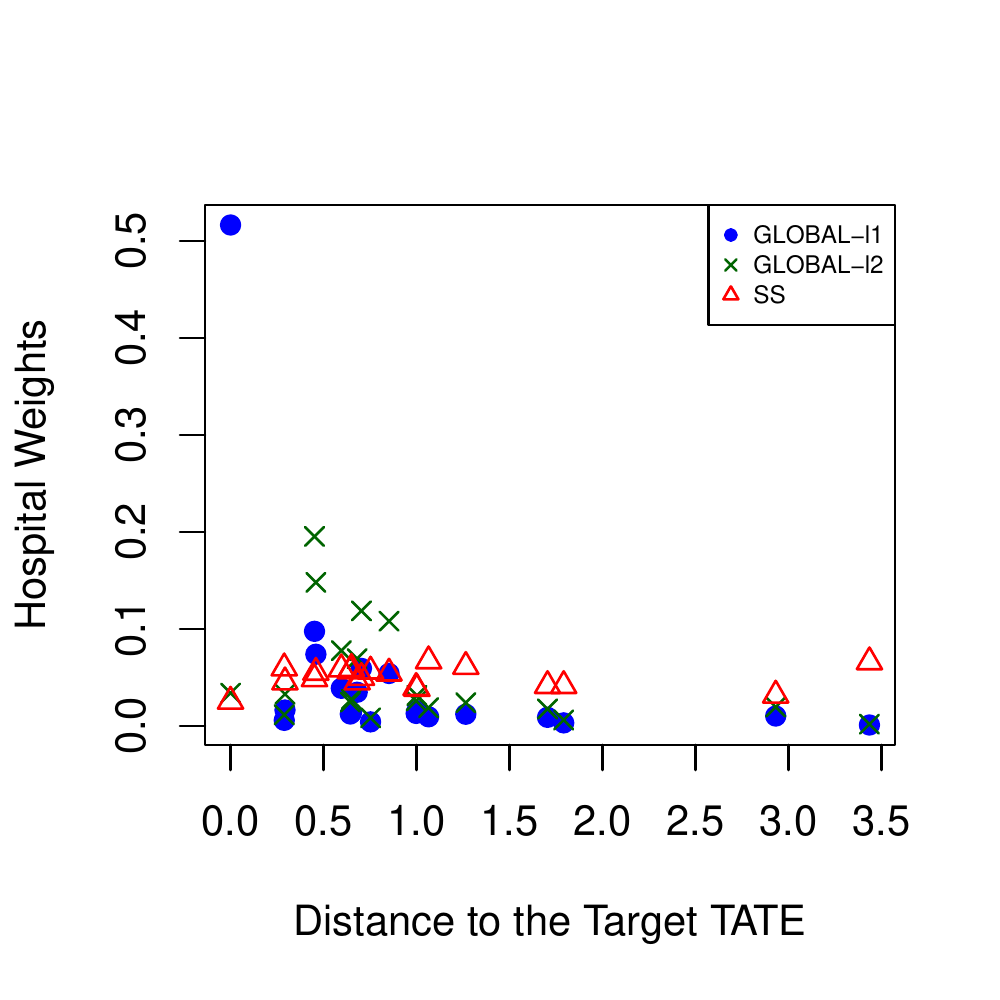}
    \includegraphics[scale=0.7]{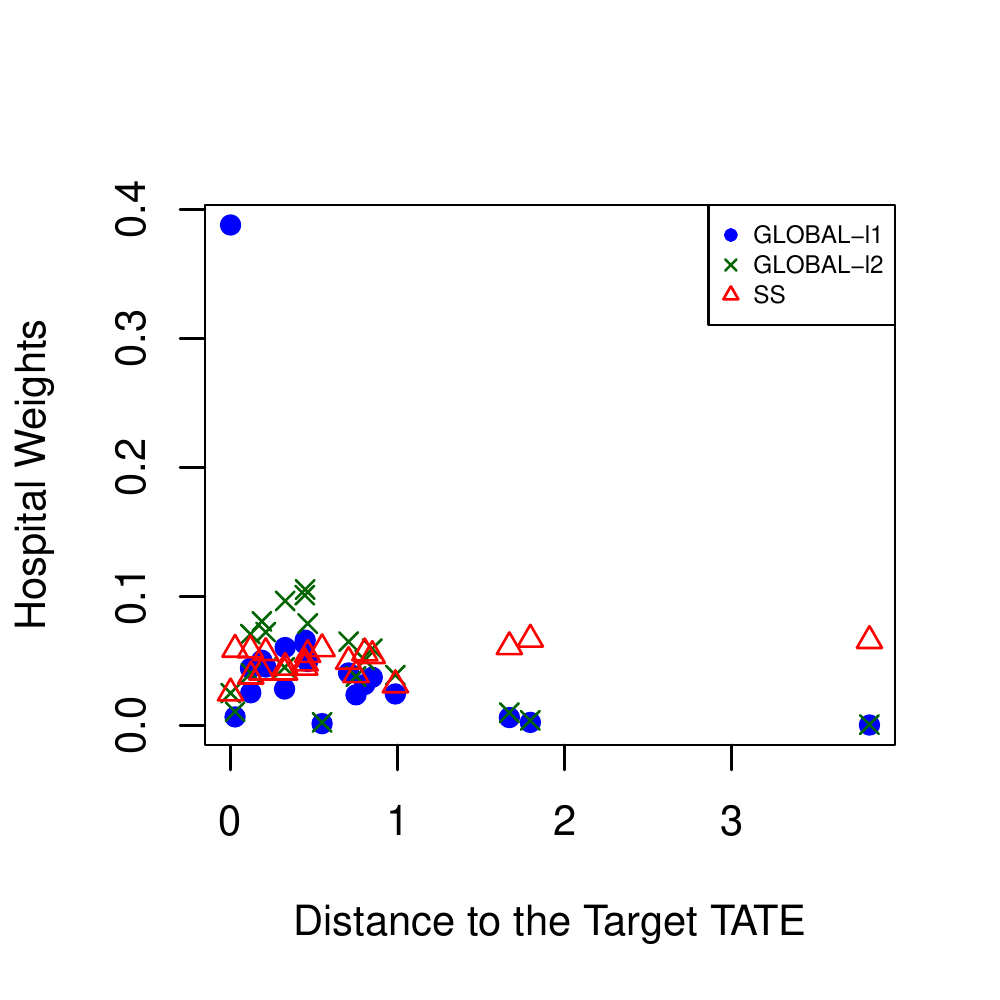}
    \caption{At left, we plot the distance from each source hospital to the target hospital TATE vs. hospital weights in the $\mathcal{D}_{\text{sparse}}$ setting where $p = 2$, showing how the global estimator upweights hospitals with similar TATE estimates. At right, we plot for the $\mathcal{D}_{\text{dense}}$ setting, showing the same phenomenon.}
    \label{fig:glmnet}
\end{figure}

To highlight the difference in $\eta_k$ weights obtained from the different methods, we plot the weights of the GLOBAL-$\ell_1$, GLOBAL-$\ell_2$, and SS estimators as a function of the distance to the target hospital TATE. Figure \ref{fig:glmnet} illustrates the $\eta_k$ weights for $k=1,...,K$ when $K = 20$ hospitals in the $\mathcal{D}_\text{sparse}$ and $\mathcal{D}_\text{dense}$ settings where $P = 2$. The GLOBAL-$\ell_1$ estimator places about half the weight on the target hospital and drops three hospitals entirely that have large bias compared to the target hospital TATE. The SS estimator has close to uniform weights. The GLOBAL-$\ell_2$ estimator produces weights between the GLOBAL-$\ell_1$ estimator and the SS estimator. 

In both $\mathcal{D}_\text{sparse}$ and $\mathcal{D}_\text{dense}$ settings, the GLOBAL-$\ell_1$ and GLOBAL-$\ell_2$ estimators are preferred because of smaller RMSEs.

\section{Performance of Cardiac Centers of Excellence}
We used our federated causal inference framework to evaluate the performance of $51$  CCE. In the following section, we showcase the ability of our methodology to (i) balance covariate distributions between hospitals and treatment groups, (ii) provide transparent hospital-level ensemble weights that can be used to construct comparator groups, (iii) increase the precision of the estimated treatment effect for target hospitals by a median of $82\%$, (iv) highlight that few hospitals performed well on both PCI and MM, and (v) understand the hospital-level attributes that helped explain the heterogeneity in causal estimates, namely not-for-profit status, teaching status, and availability of cardiac technology services.

\subsection{Balance Diagnostics}
\begin{figure}[ht]
    \centering
    \vspace{-0.1in}
    \includegraphics[scale=0.65]{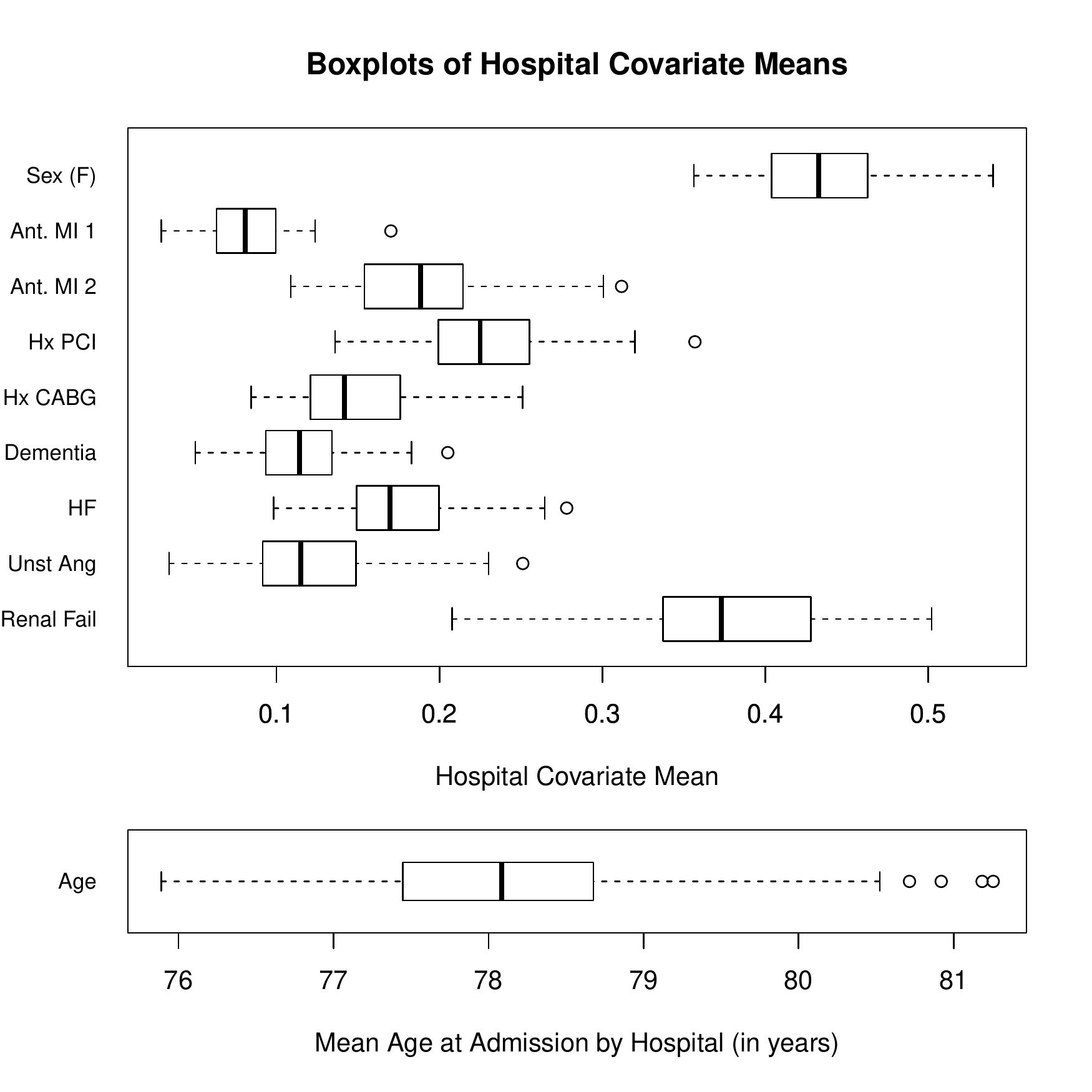}
    \vspace{-0.1in}
    \caption{Variation in hospital case-mix across $51$  CCE.}
    \label{fig:base}
\end{figure}

Despite the common designation as CCE, there was substantial variation in the distribution of baseline covariates across hospitals (Figure \ref{fig:base}). For example, the proportion of patients with renal failure varied from one-fifth to one-half of patients in a hospital. To hospitals, this signifies that despite their shared candidate CCE designation, fellow hospitals may not be appropriate comparators. To show that our methodology is able to properly adjust for these differences when making comparisons, we run covariate balance diagnostics.

\begin{figure}[ht]
    \centering
    \begin{subfigure}{0.475\textwidth}
    \includegraphics[scale=0.5]{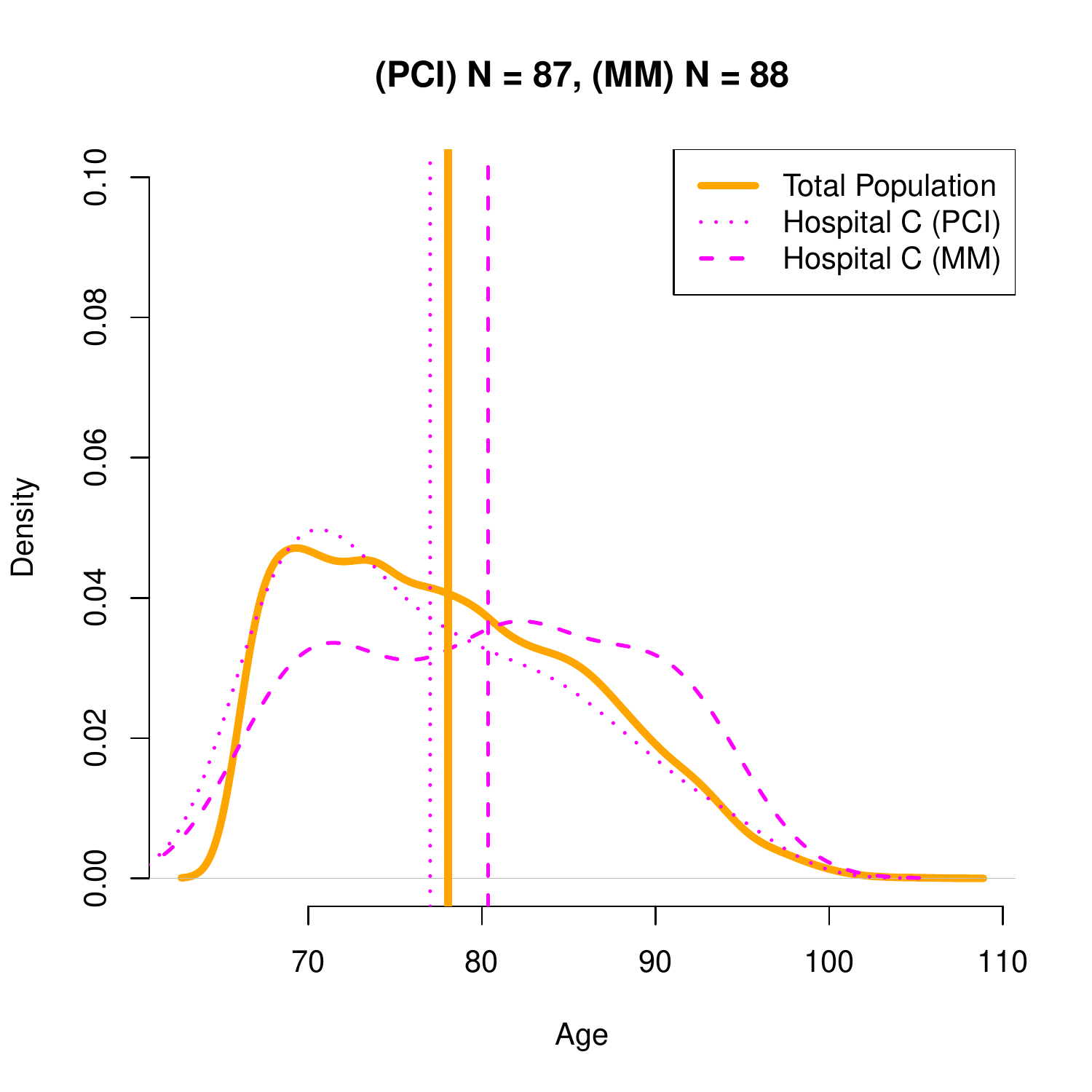}
    \caption{Age distribution of PCI/MM patients in Hospital C}
    \end{subfigure}
    \begin{subfigure}{0.475\textwidth}
    \includegraphics[scale=0.5]{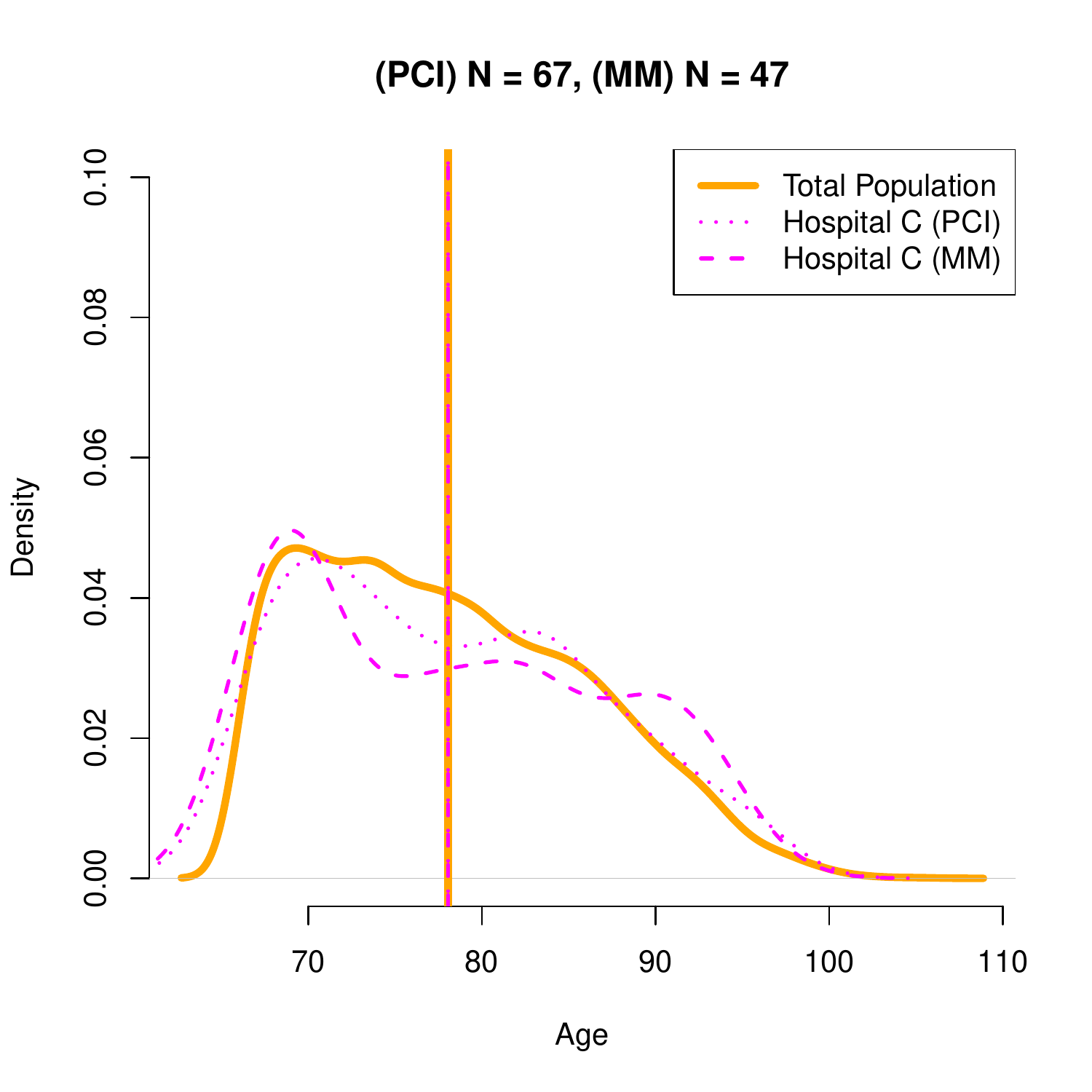} 
    \caption{Weighted age distribution of PCI/MM patients in Hospital C}
    \end{subfigure}
    
    \begin{subfigure}{0.475\textwidth}
    \includegraphics[scale=0.5]{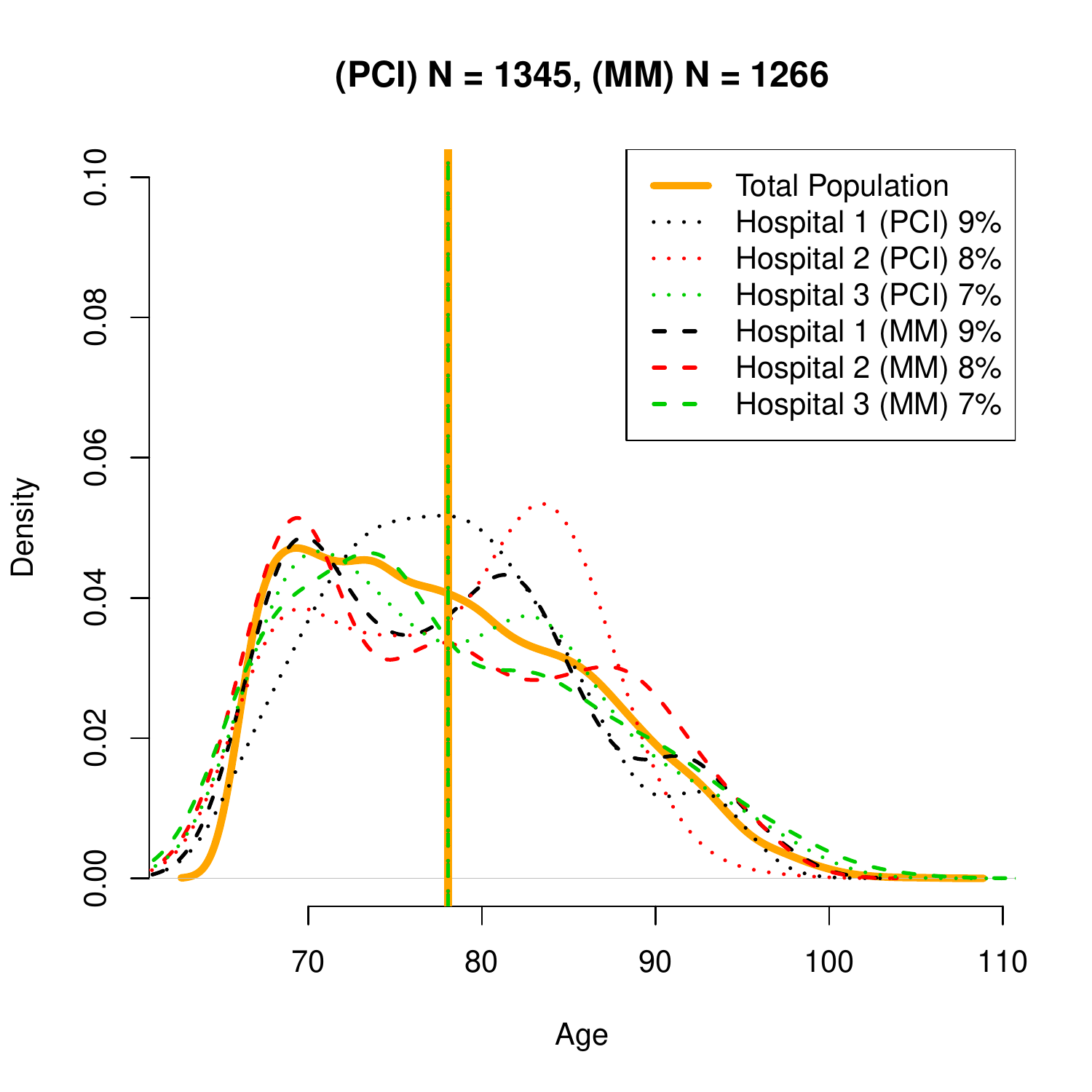}
    \caption{Weighted age distribution of PCI/MM patients in Hospital 1-3 among 50 source hospitals}
    \end{subfigure}
    \begin{subfigure}{0.475\textwidth}
    \includegraphics[scale=0.5]{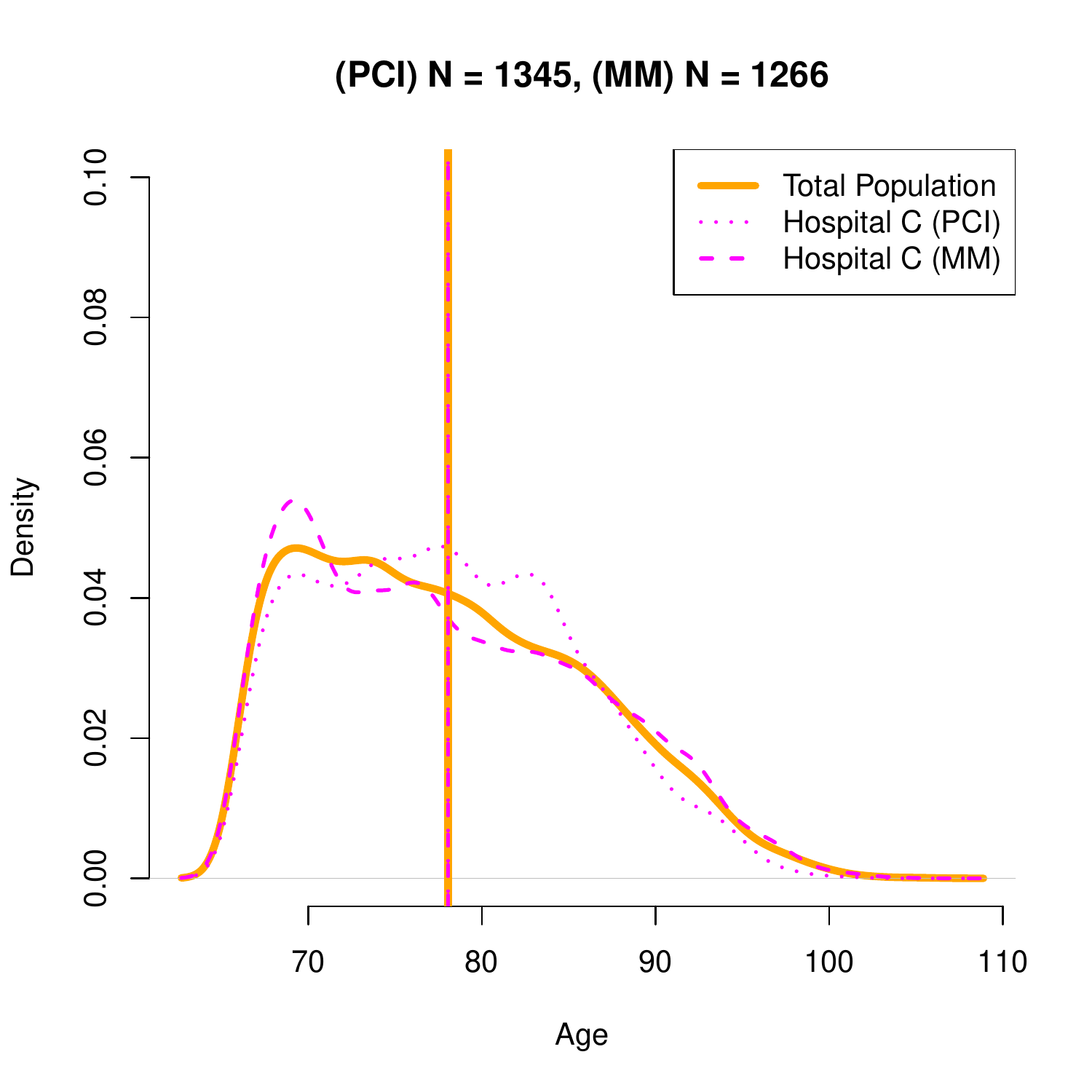}
    \caption{Combined weighted age distribution of PCI/MM patients from 50 source hospitals \\}
    \end{subfigure}
    \caption{Shifts of age distribution before and after adjustment. The first row shows the shifts in age distribution in the target-only estimator for Hospital C. The second row illustrates the Global-$\ell_1$ estimator. The latter is more precise and has a larger underlining effective sample size (1345+1266 vs 67+47).}
    \label{fig:curve_case}
\end{figure}

Figure \ref{fig:curve_case} shows how the implied weights from two estimators (target-only and Global-$\ell_1$) adjust the age distributions so that the weighted age distributions of the treated group and the control group approach their target population in the real data. Detailed covariate balance diagnostics for all variables are provided in the Appendix.

\subsection{Data-Adaptive Hospital Weights}
\begin{figure}[ht]
    \centering
    \includegraphics[width=0.95\textwidth]{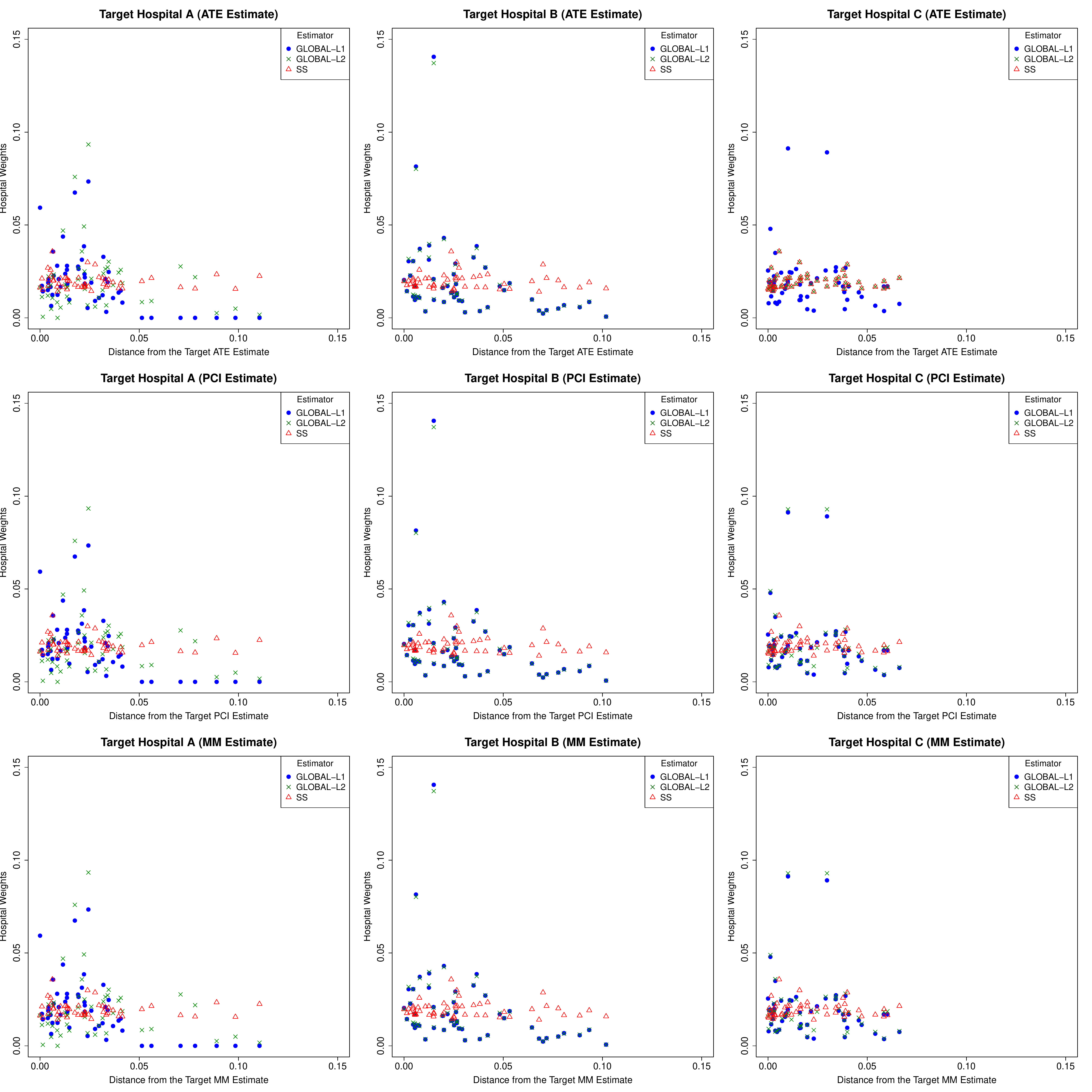}
  \caption{30-day mortality: The global estimators place more weight on source hospitals with more similar TATE or $\hat{\mu}^{(a)}$ estimates, whether the target hospital is an urban major teaching hospital (A), an urban for-profit hospital (B), or a rural non-teaching hospital (C).}
    \label{fig:weights}
\end{figure}

We illustrate how our proposed global estimators can data-adaptively select relevant peer hospitals and how this improves upon existing sample-size weighted estimators, which are unable to adapt to the selected target hospital. First, we show that there is a marked difference in the hospital ensemble weights $\boldsymbol{\eta}$ obtained from GLOBAL-$\ell_1$, GLOBAL-$\ell_2$, and SS. We plot the absolute bias obtained from each source hospital's TATE or $\hat{\mu}^{(a)}$ estimate against the corresponding $\eta_k$ weights for that estimate (Figure \ref{fig:weights}). Intuitively, source hospitals with smaller absolute bias should receive larger weights relative to source hospitals with larger absolute bias. Relevant peer hospitals can then be selected on the basis of source hospitals that receive larger weights. As examples, we showcase our method when the target hospital is selected to be one of three diverse hospitals. Hospital A is an urban major academic medical center with extensive cardiac technology, Hospital B is urban and for-profit, and Hospital C is rural and non-teaching.

Figure \ref{fig:weights} shows that the SS weights are close to uniform for all $51$ hospitals. Indeed, regardless of which hospital serves as the target hospital, the SS weights are the same in each case, showing an inability to adapt to the specified target hospital. Thus, despite potential systemic differences in the types of patients served by these three very different hospitals, SS-based estimators are indifferent to this variation. 

In contrast, the GLOBAL-$\ell_1$ estimator places more weight on hospitals that are closer to the target hospital TATE or $\hat{\mu}^{(a)}$ and `drops' hospitals once a threshold bias is crossed. Practically, the GLOBAL-$\ell_1$ estimator makes a data-adaptive bias-variance trade-off, reducing variance and increasing the effective sample size at the cost of introducing slight bias in estimates. Therefore each of these three different hospitals not only benefits from a gain in estimation precision of its own performance, but is also reassured that the source hospitals providing that precision gain were more relevant bases for comparison.

The GLOBAL-$\ell_2$ estimator produces weights in between the GLOBAL-$\ell_1$ weights and the SS weights. This follows the expected relationship outlined in Section 3, as the method for obtaining GLOBAL-$\ell_2$ emphasizes retaining all hospitals in the analysis while still placing additional weight on source hospitals that are more similar to the target hospital. 

\subsection{Precision Gain of Federated Global Estimators}
\begin{figure}[ht]
    \centering
    \includegraphics[scale=0.32]{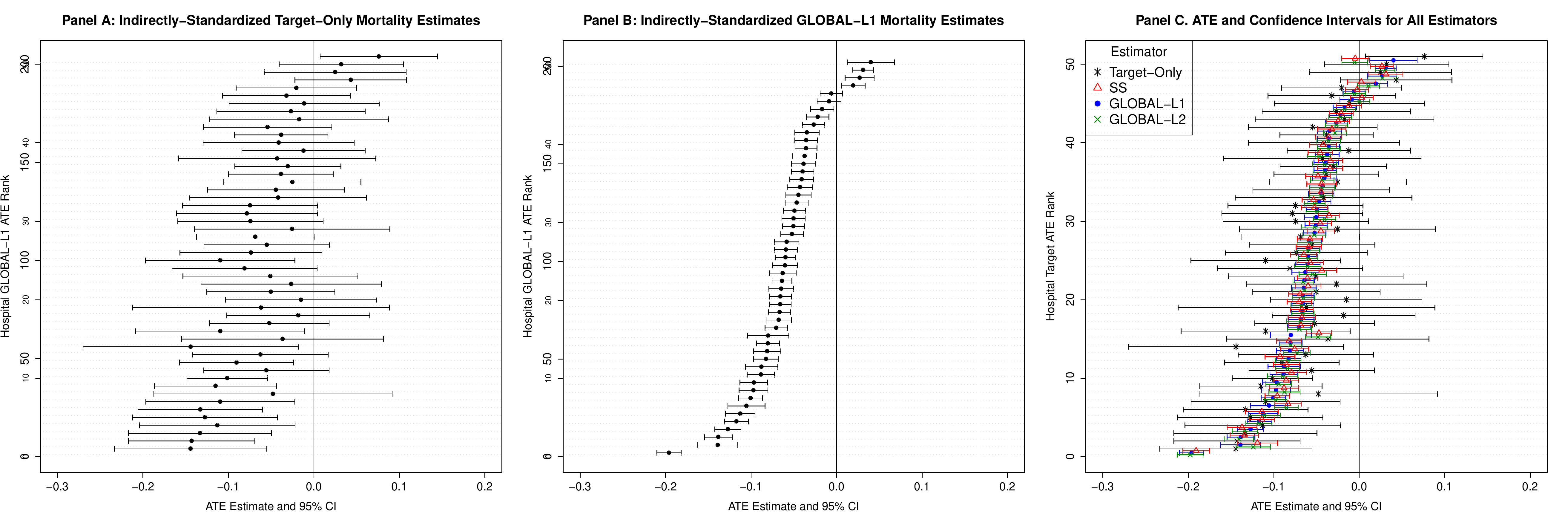}
    \caption{30-day mortality: TATE estimates for all $51$ hospitals show substantial precision gain of GLOBAL-$\ell_1$ compared to the Target-Only estimator and better accuracy relative to the GLOBAL-$\ell_2$ and SS estimators.}
    \label{fig:spline}
\end{figure}

Estimators that only use the target hospital's own data often lack power to distinguish treatment effects when the effect size is relatively small, potentially leading hospitals to misinterpret their performance. Thus, the appeal of the federated causal inference framework is that it helps the target hospital estimate its treatment effects more precisely. 
To demonstrate the efficiency gain from using a global estimator to estimate the TATE, we plot the TATE estimate for each hospital using the target-only estimator (left panel), the GLOBAL-$\ell_1$ estimator (middle panel), and overlaying the target-only estimator, GLOBAL-$\ell_1$ estimator, GLOBAl-$\ell_2$ estimator, and SS estimator (right panel) (Figure \ref{fig:spline}). Each row represents a different target hospital, i.e., each hospital takes its turn as the target hospital, with the other 50 hospitals serving as the source hospitals. The GLOBAL-$\ell_1$ estimator yields substantial variance reduction for each target hospital compared to the target-only estimator while introducing a smaller bias compared to the GLOBAL-$\ell_2$ and SS estimators. Due to this efficiency gain, the qualitative conclusion regarding the mortality effect of PCI treatment relative to MM changes from not statistically significant to statistically significant in about 63\% ($32/51$) of the hospitals. In $88\%$ ($45/51$) of the hospitals, the causal effect of PCI versus MM is statistically significant when using the GLOBAL-$\ell_1$ estimator which, unlike the target-only estimator, supports conventional wisdom that PCI reduces 30-day mortality.

\begin{figure}[ht]
    \centering
    \includegraphics[scale=0.135]{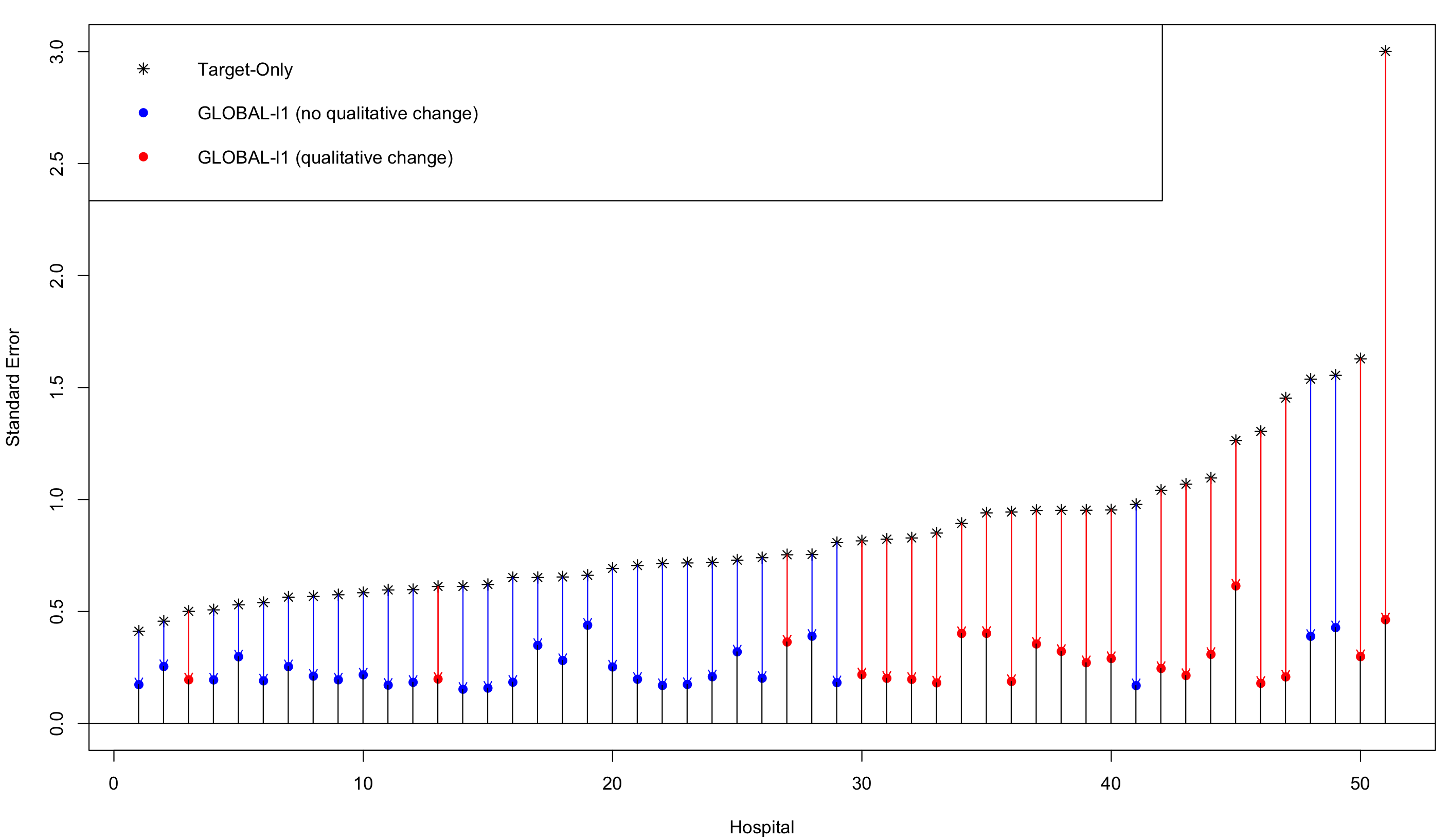}
    \caption{30-day mortality: The proportion of each vertical line that is in color represents the percent reduction in SE using the GLOBAL-$\ell_1$ estimator for each hospital's TATE estimate. Red signifies a change in interpretation from no treatment effect to a significant treatment effect.} 
    \label{fig:std}
\end{figure}

The precision gain for each hospital using GLOBAL-$\ell_1$ compared to the Target-Only estimator is substantial, with a $82\%$ median reduction in the TATE SE, ranging from $59\%$ to $91\%$ (Figure \ref{fig:std}). Moreover, this precision gain was not accompanied by a noticeable loss of accuracy. In Figure \ref{fig:spline}, GLOBAL-$\ell_1$ has a smaller bias to the Target-Only estimates relative to the GLOBAL-$\ell_2$ and SS estimators. Taken together, these findings show that the GLOBAL-$\ell_1$ estimator can provide hospitals with more precise yet still accurate guidance on their performance. The lower accuracy of the GLOBAL-$\ell_2$ estimates provide evidence that not all hospitals should be used as peer hospitals, especially in sparse settings where the heterogeneity across hospitals is large. Nevertheless, the GLOBAL-$\ell_2$ estimator still demonstrates substantial precision gains over the Target-Only estimator. The GLOBAL-$\ell_2$ estimator can be a useful approach in denser settings, i.e., when the heterogeneity across hospitals is small.

\subsection{Assessing Hospital Performance on Different Treatments}
\begin{figure}[ht]
    \centering
    \includegraphics[scale=0.5]{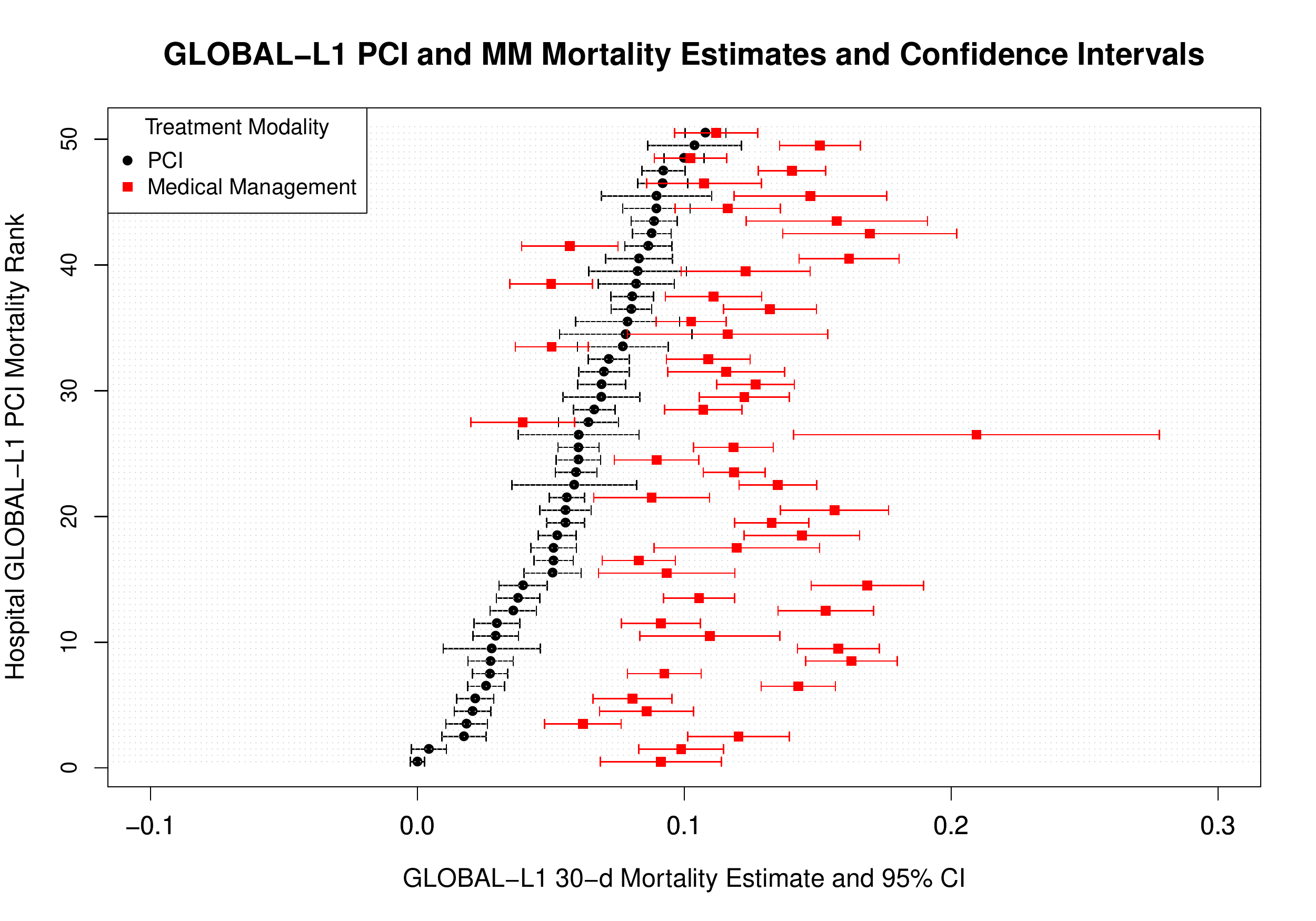}
    \caption{30-day mortality: GLOBAL-$\ell_1$ hospital mortality estimates and $95\%$ CIs show that while most hospitals achieved lower mortality rates with PCI, MM mortality estimates varied considerably even among hospitals with similar PCI performance.}
    \label{fig:ranks}
\end{figure}

In addition to examining hospitals based on their TATE performance, we also used the GLOBAL-$\ell_1$ estimator to compare hospitals on their performance had all patients received PCI $\mu_{T,Fed}^{(1)}$, or had all patients received MM $\mu_{T,Fed}^{(0)}$. This guidance on specific AMI treatments can be useful both to hospitals and prospective patients. Figure \ref{fig:ranks} shows hospital mortality estimates for PCI and MM, with hospitals sequenced from the lowest (best) to highest (worst) PCI mortality. Hospital performance on MM appeared to be more variable than on PCI. In some cases, pairs of hospitals with adjacent PCI mortality estimates displayed up to a twofold difference in MM mortality. These findings suggest that the staffing, skill, and resource inputs that translate to better performance in interventional cardiology differ from those that drive excellent MM practices. An alternative explanation is that the hospitals with the best PCI performance could have assigned sicker patients (e.g., more severe AMI subtype, comorbidities, older age, etc.) to MM compared to the hospitals with the worst PCI performance; however, we have controlled for several plausible confounders in our outcome regression models, including patient age, gender, admission year, AMI diagnosis subtype, history of PCI or CABG procedures, and history of a number of comorbidities.

\subsection{Hospital-level Attributes and Outcome  Performance}
To understand which hospital-level attributes help explain the heterogeneity in causal estimates, we regressed the mean potential outcome and TATE estimates on various hospital-level factors in a series of linear regression models. In Table \ref{tab:mortality_tree}, we show the estimated coefficients of various hospital-level characteristics. These coefficients correspond to indirectly standardized estimates of the mean potential outcome for PCI and MM according to the characteristics of each hospital's own patient population. Not-for profit (NFP) teaching hospitals on average had lower estimated mortality for PCI treatment compared to other hospitals. 

To directly compare hospital performance, we also considered directly standardized estimates of the mean potential outcome for PCI and MM, with each hospital benchmarked on a fixed target population. In these analyses, NFP teaching hospitals on average had lower estimated mortality for PCI treatment compared to other hospitals (-1.83, 95\% CI [-3.44,-0.21]). 

\newpage
\begin{figure}[ht]
    \centering
    \begin{subfigure}{0.32\textwidth}
    \includegraphics[scale=0.5]{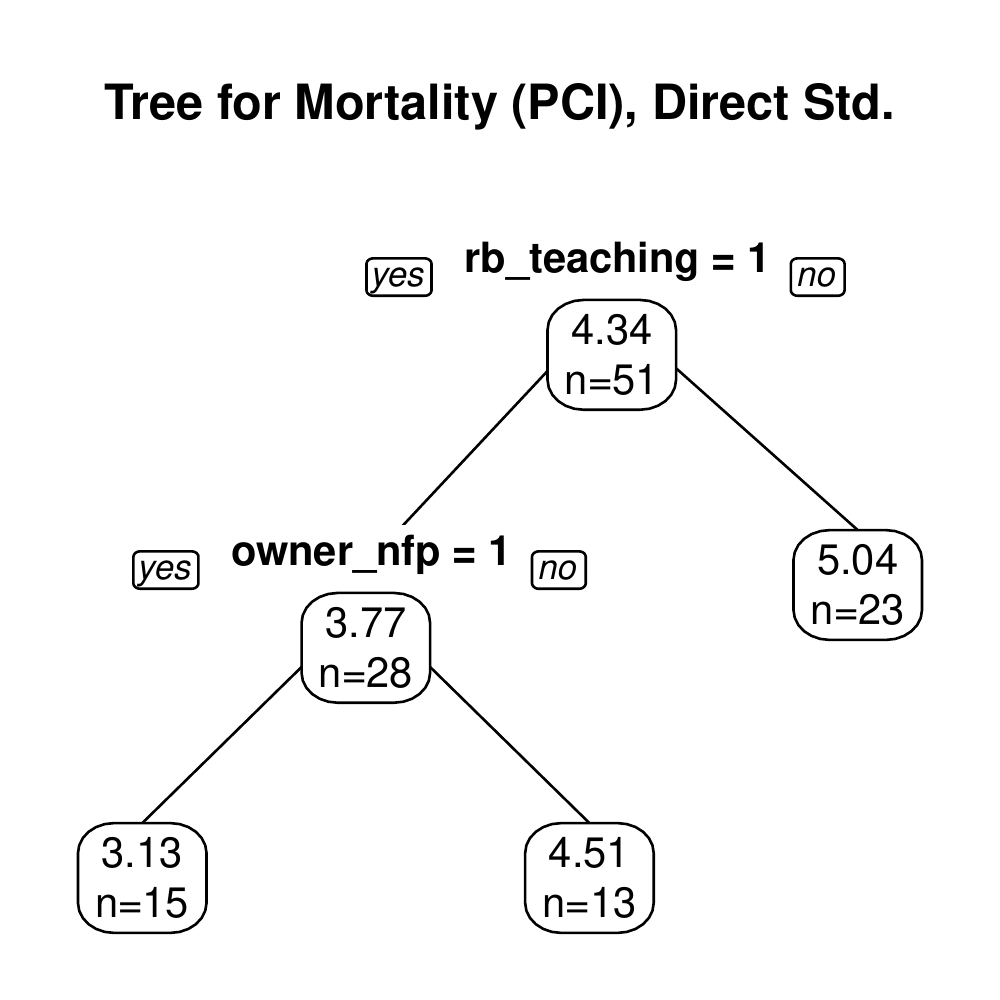}
    \end{subfigure}
    \begin{subfigure}{0.32\textwidth}
    \includegraphics[scale=0.5]{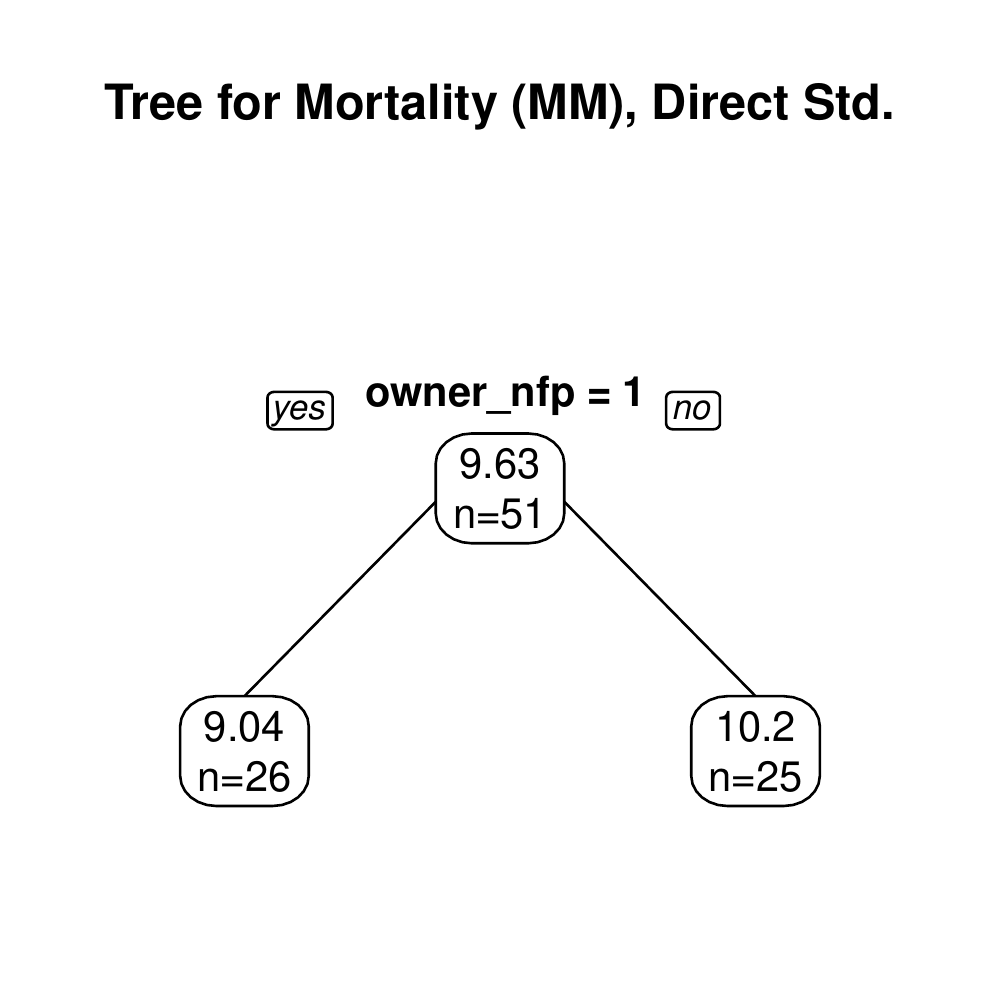} 
    \end{subfigure}
    \begin{subfigure}{0.32\textwidth}
    \includegraphics[scale=0.5]{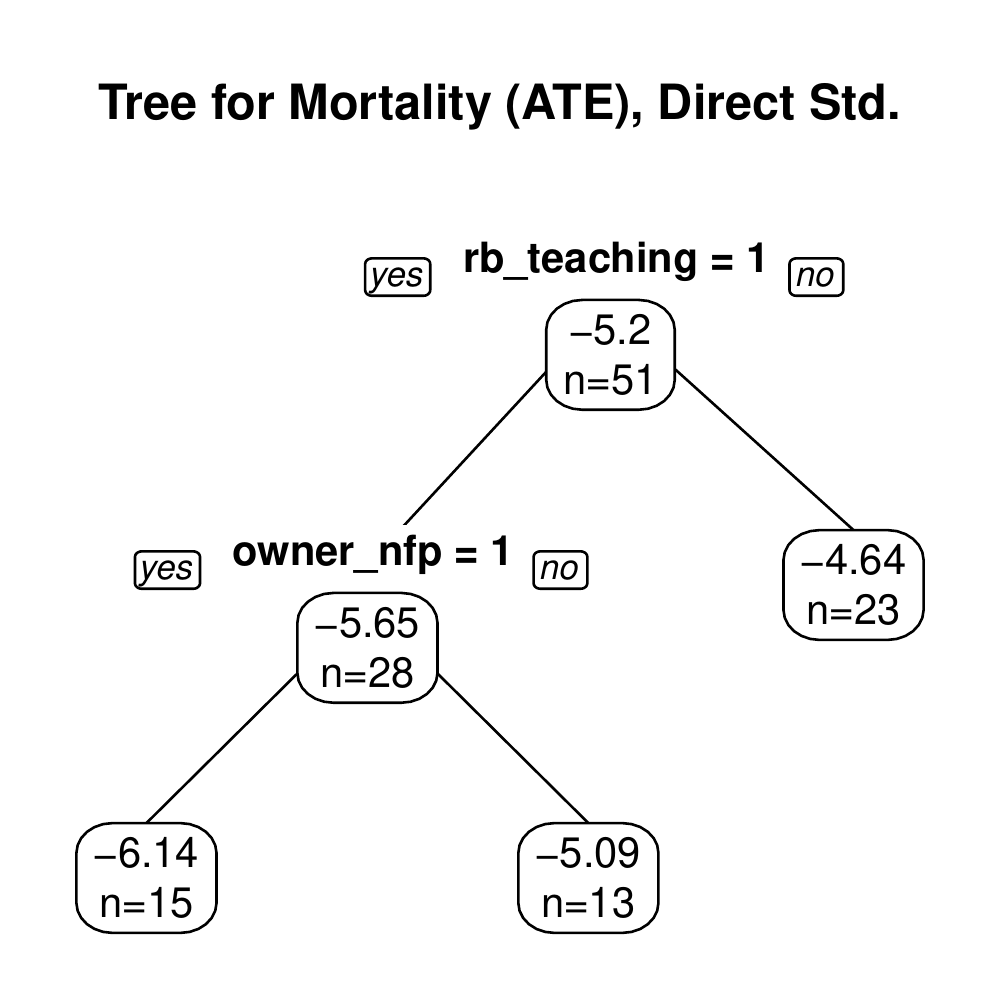} 
    \end{subfigure}
    
    \begin{subfigure}{0.32\textwidth}
    \includegraphics[scale=0.5]{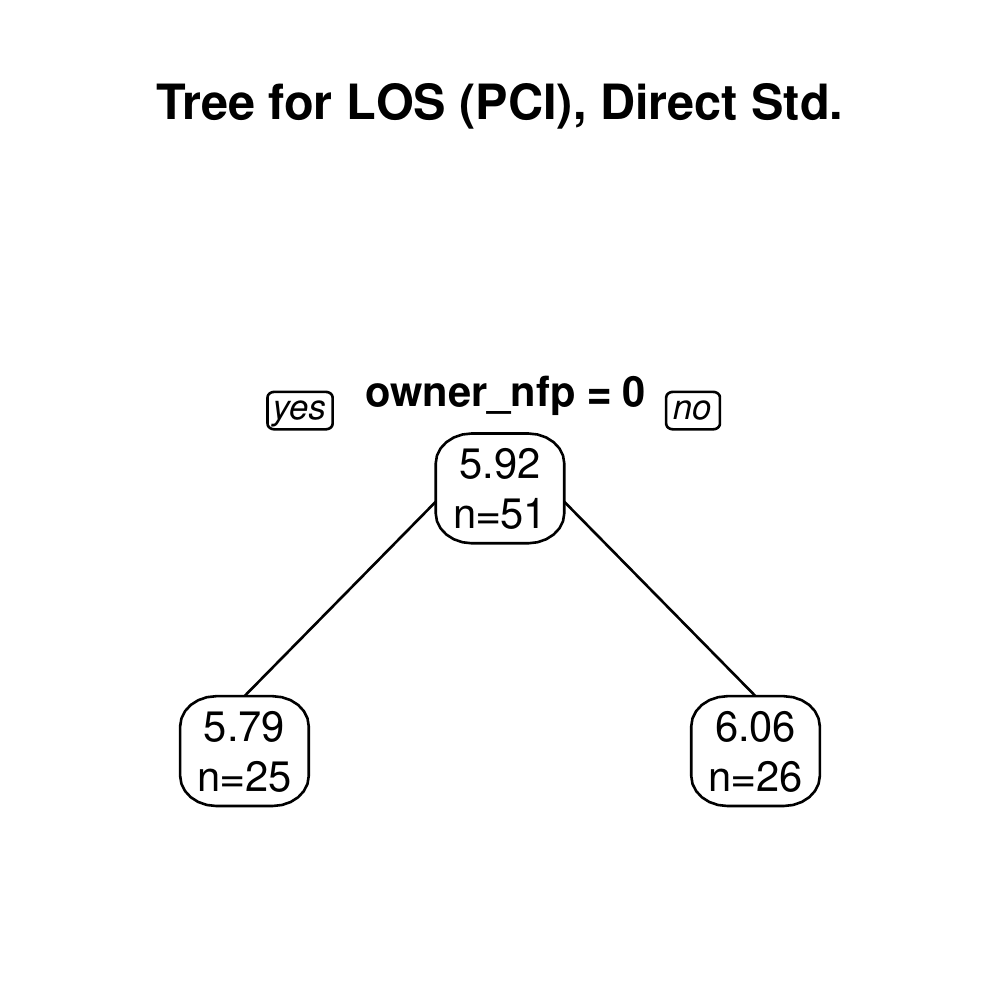}
    \end{subfigure}
    \begin{subfigure}{0.32\textwidth}
    \includegraphics[scale=0.5]{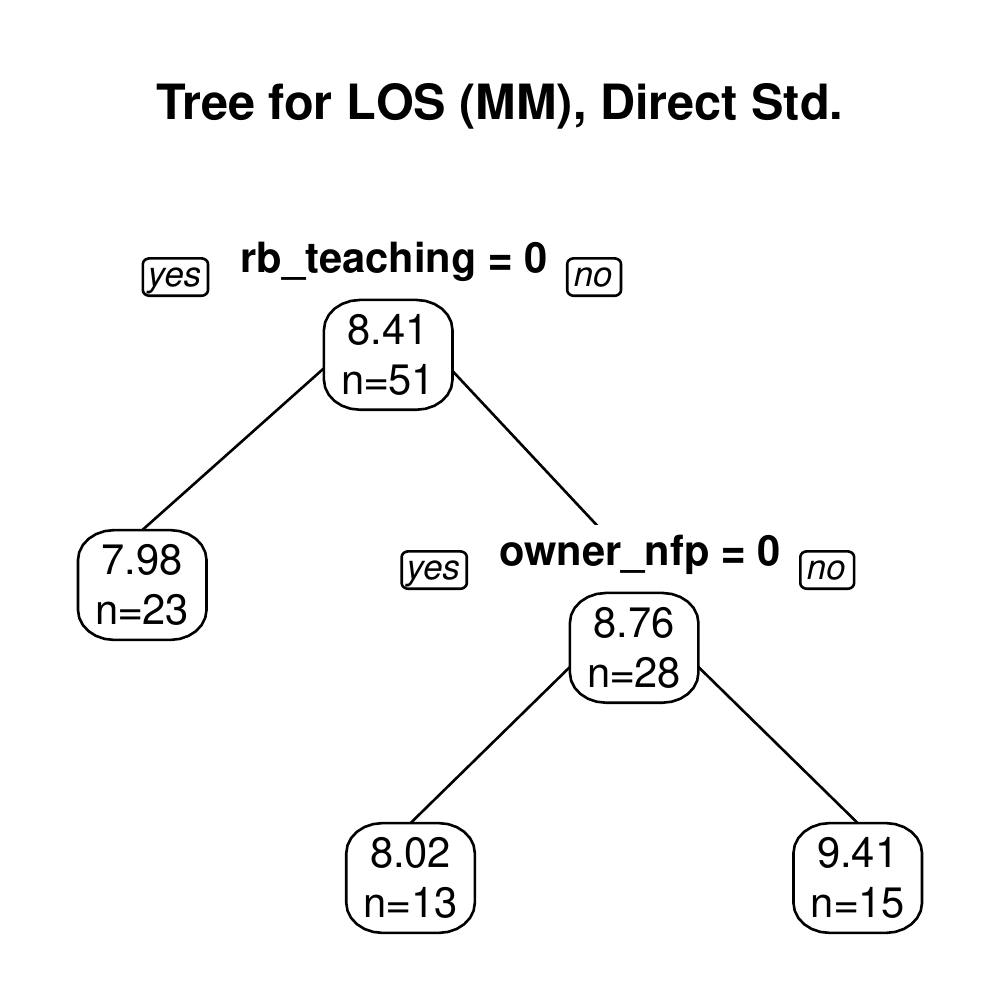} 
    \end{subfigure}
    \begin{subfigure}{0.32\textwidth}
    \includegraphics[scale=0.5]{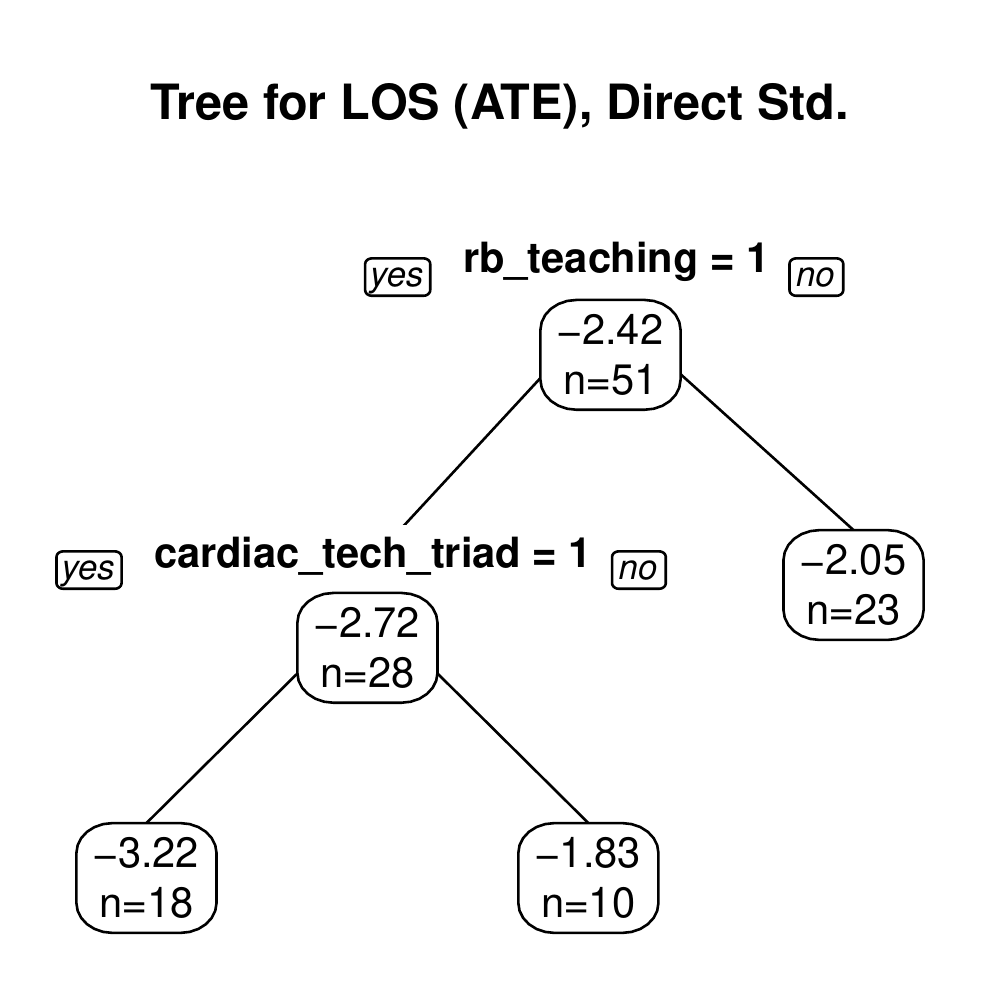} 
    \end{subfigure}
    
    \caption{Estimated regression trees of causal effects on hospital-level attributes for mortality (row 1) and length-of-stay (row 2) show that not-for-profit (NFP) status, teaching status (rb teaching), and the availability of cardiac technology services (cardiac tech triad) are important factors.}
    \label{fig:tree}
\end{figure}

\newpage

\begin{center}
\renewcommand{\arraystretch}{1}
\begin{table}[ht]
\caption{Summary statistics, coefficients of linear regression models on hospital-level attributes, and estimated causal effects for different types of hospitals.}\label{tab:mortality_tree}
\centering
\setlength{\tabcolsep}{2pt}
\begin{tabular}{lrrrr}
  \hline
  & \multicolumn{4}{c}{Estimates} \\
     \cmidrule(lr){2-5} 
  &  \multicolumn{2}{c}{Not-for profit} &  \multicolumn{2}{c}{For-profit \& others} \\
  \cmidrule(lr){2-3}  \cmidrule(lr){4-5} 
  & Teaching &  Non-Teaching & Teaching &  Non-Teaching\\
\cmidrule(lr){1-1} \cmidrule(lr){2-3}  \cmidrule(lr){4-5}
Summary Statistics &  &  &\\
\hspace{.15cm}  No. of Hospitals & 15 & 11 & 13 & 12  \\
\hspace{.15cm}  No. of  Patients & 3293 & 2425 & 2636 & 2749\\
\hspace{.15cm}  No. of PCI Patients & 1463 & 1071 & 1247 & 1223 \\
  \cmidrule(lr){1-1} \cmidrule(lr){2-3}  \cmidrule(lr){4-5}
Patient Covariates &  &  &\\ 
\hspace{.15cm}  Age & *78.46 & 78.08 & *77.64 & 77.93 \\ 
\hspace{.15cm}  Age $|$ PCI & 76.72 & 76.24 & 76.58 & 76.65 \\ 
\hspace{.15cm}  Age $|$ MM & *79.86 & 79.55 & *78.60 & *78.96 \\ 
\cmidrule(lr){1-1} \cmidrule(lr){2-3}  \cmidrule(lr){4-5}
Patient Outcomes &  &  &\\ 
\hspace{.15cm}  Mortality & 9.84 & 9.53 & 9.41 & 10.51\\
\hspace{.15cm}  Mortality $|$ PCI  & *4.65 & 5.79 & 5.93 & 6.62\\
\hspace{.15cm}  Mortality $|$ MM  & 13.99 & 12.48 & 12.53 & 13.63 \\
 \cmidrule(lr){1-1} \cmidrule(lr){2-3}  \cmidrule(lr){4-5}
Hospital Characteristics &  &  &\\ 
\hspace{.15cm}  PCI Rate & 0.45 & 0.45 & 0.47 & 0.45  \\
\hspace{.15cm}  Nurse-to-beds & 1.99 & 1.98 & 1.79 & 1.83\\
\cmidrule(lr){1-1} \cmidrule(lr){2-3}  \cmidrule(lr){4-5}
Raw Estimates &  &  &\\
\hspace{.15cm}  Mortality (PCI) & 4.60 & 5.93 & 5.97 & 6.59 \\ 
\hspace{.15cm}  Mortality (MM) & 14.13 & 12.51 & 12.48 & 13.79  \\ 
\cmidrule(lr){1-1} \cmidrule(lr){2-3}  \cmidrule(lr){4-5}
Target-only (Indirect) &  &  &\\
\hspace{.15cm}  Mortality (PCI) & *4.16 & 5.91 & 6.38 & *8.15\\ 
\hspace{.15cm}  Mortality (MM) & 12.13 & 10.98 & 11.03 & 12.54  \\ 
\cmidrule(lr){1-1} \cmidrule(lr){2-3}  \cmidrule(lr){4-5}
Global-$\ell_1$ (Indirect) &  &  &\\ 
\hspace{.15cm}  Mortality (PCI) & 4.59 & 5.92 & 6.08 & *7.45  \\ 
\hspace{.15cm}  Mortality (MM) & 12.26 & 10.78 & 11.26 & 12.07  \\
\cmidrule(lr){1-1} \cmidrule(lr){2-3}  \cmidrule(lr){4-5}
Target-only (Direct) &  &  &\\
\hspace{.15cm}  Mortality (PCI) & *3.18 & 5.39 & 4.63 & 6.19 \\ 
\hspace{.15cm}  Mortality (MM) & 9.11 & 8.42 & 10.40 & 10.45 \\ 
\cmidrule(lr){1-1} \cmidrule(lr){2-3}  \cmidrule(lr){4-5}
Global-$\ell_1$ (Direct) &  &  &\\ 
\hspace{.15cm}  Mortality (PCI) & 3.13 & 4.56 & 4.51 & 5.48 \\ 
\hspace{.15cm}  Mortality (MM) & 9.36 & 8.59 & 9.85 & 10.67 \\
\hline
\end{tabular}
\vspace{.25cm}
\footnotesize{
\begin{flushleft}
$*$: The group is significantly different from the others combined (two sample two-sided T-test, level = 0.05).

\end{flushleft}
}
\end{table}
\end{center}

\section{Conclusions}
\label{sec:disc}

We developed a federated and adaptive method that leverages summary data from multiple hospitals to safely and efficiently estimate target hospital treatment effects for hospital quality measurement. 
Our global estimation procedure preserves patient privacy and requires only one round of communication between target and source hospitals. 
We used our federated causal inference framework to investigate quality among $51$  CCE in the U.S. We obtained accurate TATE estimates accompanied by substantial precision gains, ranging between $36\%$ and $88\%$ relative to the estimator using only target hospital data. 

Our global estimator can be used in other federated data settings, including transportability studies in which some hospitals have access to randomized trial data and other hospitals have only observational data (e.g., EHR, insurance claims, etc.). In this setting, one could anchor the estimates on a hospital with randomized trial data and enhance the data with observational data from other hospitals. Alternatively, if one were primarily interested in transporting causal estimates to an observational study, one could treat the observational study as the target study and use the randomized trial as a source study within our framework.

Importantly, our method enables a quality measurement framework that benchmarks hospitals on their performance on specific alternative treatments for a given diagnosis as opposed to only aggregate performance or an isolated treatment.  
This comprehensive analysis enabled us to show that among patients admitted for AMI, superiority in one treatment domain does not imply success in another treatment.  
Equipped with these methods and results, hospitals can make informed strategic decisions on whether to invest in shoring up performance on specific medical treatments where they are less successful than their peers, or to allocate even more space and resources to treatment paradigms where they have a comparative advantage. Our treatment-specific assessments reflect the nuance that different hospitals sometimes have comparative advantages and disadvantages, and in so doing, these separate rankings can inform hospitals on which forms of treatment may require quality improvement efforts.

Moreover, this information may also be useful to patients. Specifically, if a patient has a strong preference for one treatment over another, they can opt for a hospital that excels at their preferred course of treatment. This feature of the framework may be even more useful in elective clinical domains where patient agency over facility choice is not diminished by the need for emergent treatment, such as oncology, where treatment options vary substantially and are carefully and collaboratively considered by physicians and their patients. For example, if a patient requires cancer resection followed by adjuvant therapy but does not want chemotherapy, they can instead opt for treatment at a cancer center that showcases strong performance when rendering surgery with adjuvant radiation.

The limitations of our approach present opportunities for future advancements.
In the multi-hospital setting, even if a common set of covariates are universally known and acknowledged, it may be difficult to agree upon a single functional form for the models. 
Therefore, researchers in different hospitals might propose different candidate models. 
For example, treatment guidelines can differ across CCE based on variation in patient populations or the resources at each hospital's disposal. For increased robustness, researchers at different hospitals can propose different propensity score and outcome regression models using multiply robust estimators \citep{han2013estimation}. While the risk factors we included are based on a comprehensive longitudinal record of Medicare claims, they are reliant on the integrity and completeness of those claims. There are also additional risk factors that could be included in future work that extends this methodology to higher-dimensional settings with greater \emph{p}. Additionally, while we studied a number of hospital structural characteristics that have been frequently shown to be associated with quality such as teaching \citep{silber2020} and nurse staffing \citep{lasater2021}, other factors such as physician experience and credentials may also be important drivers of treatment and outcomes. Future work could incorporate and study the effect of these covariates to provide further managerial insights into the role of physician characteristics on PCI selection and quality.

\bibliographystyle{agsm}

\bibliography{_federated_arxiv.bib}

\clearpage

\ifarXiv
    \foreach \x in {1,...,28}
    {
        \includepdf[pages={\x}]{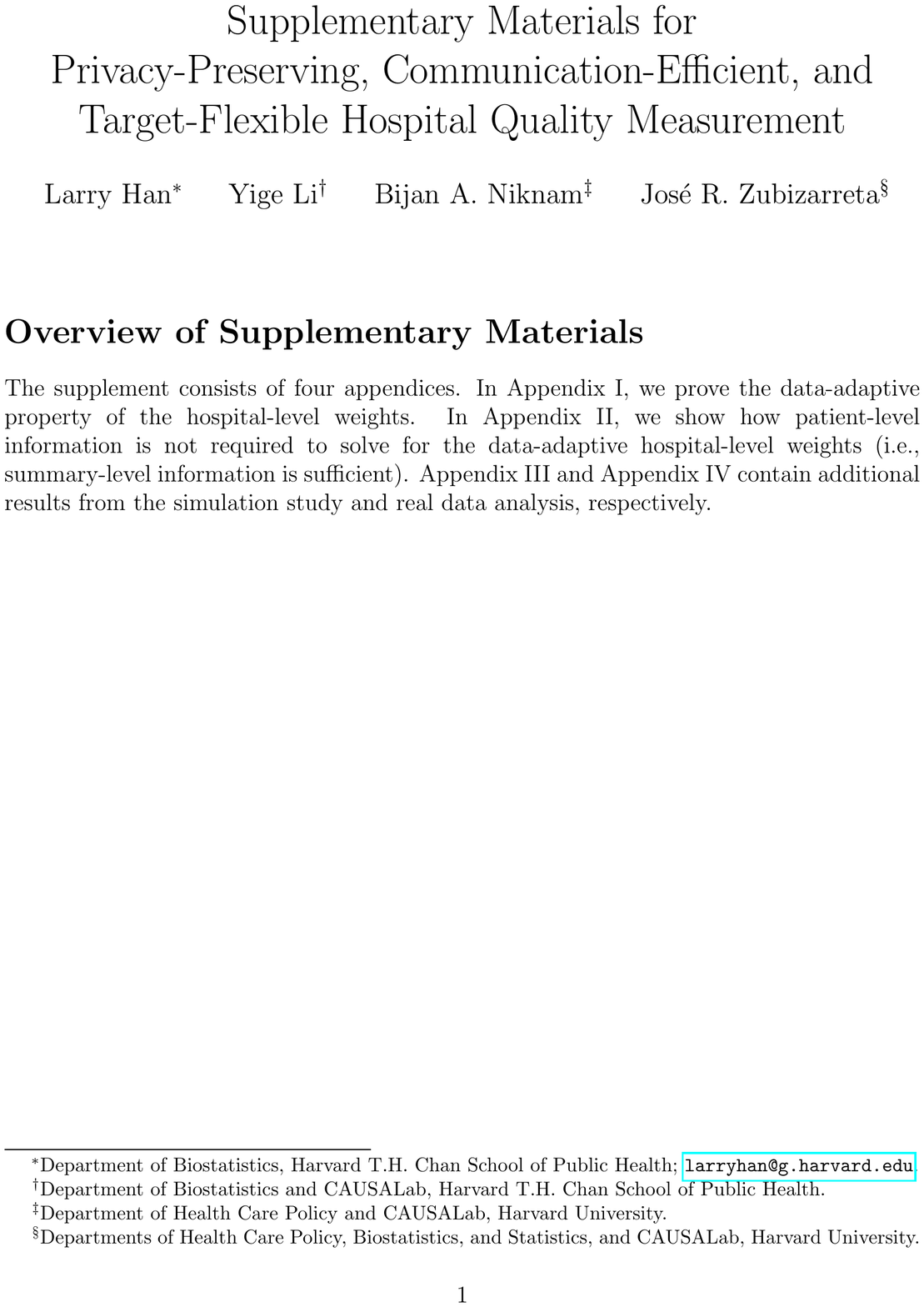}
    }
\fi

\end{document}


\def\spacingset#1{\renewcommand{\baselinestretch}%
{#1}\small\normalsize} \spacingset{1}


\if1\blind
{
  \title{\bf Privacy-Preserving and Communication-Efficient Causal Inference for Hospital Quality Measurement Supplementary Materials}
  \author{Larry Han$^{1}$, Yige Li$^{1}$, Bijan Niknam$^{2}$,  Jos\'{e} Zubizarreta$^{1,2,3}$ \hspace{.2cm}\\
  %
    1 Department of Biostatistics, Harvard T.H. Chan School of Public Health\\
    2 Department of Health Policy, Harvard Medical School \\
    3 Department of Statistics, Harvard University }
  \maketitle
} \fi

\if0\blind
{
  \bigskip
  \bigskip
  \bigskip       
  \begin{center}
    {\LARGE\bf Federated Causal Inference for Hospital Quality Measurement}
\end{center}
  \medskip
} \fi

\bigskip

\newpage
\spacingset{1.45} 
\section*{Overview of Supplementary Materials}
The supplement consists of five appendices. In Appendix I, the exact form of the influence functions are derived in the case where GLM is used for outcome regression models, logistic regression is used for propensity score models, and exponential tilting models are used for density ratio models. In Appendix II, we prove the data-adaptive property of the site-level weights. In Appendix III, we show how patientn-level information is not required (i.e., summary level information is sufficient) to solve for the data-adaptive site-level weights. Appendix IV and Appendix V contains additional results from the simulation study and real data analysis, respectively.

\section*{Appendix I. Derivation of Influence Functions}
\subsection*{General Form of Influence Functions}
In this section, we summarize the general form of the influence functions for the target and source site estimators. The influence function in the target site $\xi_{i,T}^{(a)}$ is
\begin{align*}
    \sqrt{N_\tgt}(\hat{\mu}^{(a)}_{T,T} - {\mu}^{(a)}_\tgt) &= \frac{1}{\sqrt{N}} \sum_{i=1}^{N} \frac{I(R_i=T)}{\rho_\tgt} \xi_{i,T}^{(a)} + o_p(1),
\end{align*}
where
\begin{align*}
    \xi_{i,T}^{(a)} &= \frac{I(A_i=a)Y_i}{\pi_a(X_i,\bga^*)} - \left(\frac{I(A_i=a)}{\pi_a(\bX_i,\bga^*)}-1\right)m_a(\bX_i,\bgb_a^*) - \mu_\tgt^{(a)} \\
    &\quad+ D_1(\bga^*,\bgb_a^*) \phi_\tgt(\bX_i,\bga^*) + D_2(\bga^*,\bgb_a^*) I_\tgt \triangledown f(\bX_i,Y_i,\bgb_a^*),
\end{align*}
where $D_1(\bga^*,\bgb_a^*) = \E\left[(Y_i-m_a(\bX_i,\bgb_a^*)) (-\bX_i^\top \exp(\bga^{*\top} \bX_i) I(A_i=a))\right],$ \\
$D_2(\bga^*,\bgb_a^*) = \E\left[\bX_i^T \{\frac{I(A_i=a)}{\pi_a(\bX_i,\bga^*)} - 1\}\right]$, \\
$I_\tgt = \E\left[\triangledown^2 L_\tgt\right]^{-1} = \E\left[\triangledown^2\left[\frac{1}{N_\tgt}\sum_{i=1}^{N_\tgt} (Y_i - \bgb_a^\top \bX_i)^2\right]\right]^{-1} = \E\left[\frac{2}{N_\tgt}\sum_{i=1}^{n_i} (\bX_i^T \bX_i)\right]^{-1}$, \\
$\phi_\tgt(\bX_i,\bga^*) = \E\left[\bX_i^\top \pi_a(\bX_i,\bga^*) (1-\pi_a(\bX_i,\bga^*) \bX_i\right]^{-1} \{\bX_i^\top (I(A_i=a) - \pi_a(\bX_i,\bga^*) )\}$, \\
and $\triangledown f(\bX_i, Y_i, \bgb_a^*) = -2 \bX_i^\top (Y_i - \bgb_a^{*\top} \bX_i)$ for a linear outcome $Y_i$. When the outcome regression model is correctly specified, $D_1(\bga^*,\bgb_a^*) = 0$, and when the propensity score model is correctly specified, $D_2(\bga^*,\bgb_a^*) = 0$. \\

In the source sites, we must additionally consider estimation of the density ratio weights. Let $h(\bX_i,\bgg_k,\tau) = \bX_i \exp(\bgg_k^\top \bX_i) - \tau$, where $\bX = (1,X_1,...,X_p)^\top$ is a design matrix with $1$ in the first column and $\tau = (1, E_1(\bX))^\top$ is the covariate mean vector in the target population with $1$ in the first entry. By a Taylor series expansion, we can obtain that
\begin{align*}
    \sqrt{n_k}(\hat{\bgg}_k - \bgg_k^*) &= \frac{1}{\sqrt{N}\rho_k} \sum_{i=1}^N \Bigg\{H_1^{-1} h(\bX_i,\bgg_k^*,\tau^*) I(R_i=k) \\
    &\quad+ \sqrt{\frac{\rho_k}{\rho_\tgt}}H_1^{-1}(\bX_i-\tau) I(R_i=T) \Bigg\} + o_p(1),
\end{align*}
where $H_1 = \E\left[-\bX_i \exp(\bgg_k^{*\top} \bX_i) \bX_i^\top\right]$. \\

Then the source site influence function is
\begin{align*}
    &\sqrt{n_k}(\hat{\mu}^{(a)}_k - {\mu}^{(a)}_k) \\
    &= \frac{1}{\sqrt{N}\rho_k} \sum_{i=1}^{N} \Bigg\{ I(R_i=T)m_a(\bX_i,\bgb_a^*) - \mu^{(a)}_k \\
    &\quad+ \left[Y_i -m_a(\bX_i,\bgb_a^*)\right] \left[\frac{I(R_i=k, A_i=a)\omega_k(\bX_i,\bgg_k^*)}{\pi_{a,k}(\bX_i,\bga_k^*)} + 2D_2^k(\bga_k^*,\bgb_a^*) I_k \bX_i \right]  \\
    &\quad+ D_1^k(\bga_k^*) IF_i(\bX_i,\bga_k^*) + D_1^k(\bgg_k^*) IF_i(\bX_i,\bgg_k^*) \Bigg\} + o_p(1),
\end{align*}
where
$D_{2}^k(\bga_k^*,\bgb_a^*) = \E \left[\bX_i^T \left(\frac{I(R_i=k,A_i=a)\omega_k(\bX_i,\bgg_k^*)}{\pi_{a,k}(\bX_i,\bga_k^*)}-1 \right) \right]$, \\
$I_k = \left[\E \triangledown^2 L_k (\bgb_a^*) \right]^{-1} = \left[\E\left( \frac{2}{n_k} \sum_{i=1}^{n_k} \bX_i^T \bX_i \right))\right]^{-1}$, \\
$D_{1,\bga_k^*}^k = \E \left[\{Y_i-m_a(\bX_i,\bgb_a^*)\} I(R_i=k, A_i=a) \exp(\bgg_k^{*^\top} \bX_i) \bX_i \frac{\pi_{a,k}(\bX_i,\bga_k^*)-1}{\pi_{a,k}(\bX_i,\bga_k^*)} \right]$, \\
$IF_i(\bX_i,\bga_k^*) = \left[\E(\bX_i^\top \pi_{a,k}(\bX_i,\bga_k^*)(1-\pi_{a,k}(\bX_i,\bga_k^*)) \bX_i \right]^{-1} \bX_i \cdot (I(A_i=a)-\pi_{a,k}(\bX_i,\bga_k^*))$, \\
$D_{1,\bgg_k^*}^k = \E \left[\{Y_i-m_a(\bX_i,\bgb_a^*)\} I(R_i=k, A_i=a) \bX_i \exp(\bgg_k^{*^\top} \bX_i) \frac{\exp(\bga^{*^\top} \bX_i)}{1+\exp(\bga^{*^\top} \bX_i)} \right]$, \\
$IF_i(\bX_i,\bgg_k^*) = H_1^{-1} h(\bX_i,\bgg_k^*,\tau^*)I(R_i=k) + \sqrt{\frac{n_k}{N_\tgt}}H_1^{-1}I(R_i=T) (\bX_i-\tau)$.

\subsection*{Derivation of the Target Influence Function}
Suppose we have an estimator from the target site, $\hat{\mu}_{a,T}^T$ of the true $\mu_{a,T}$ for $a=0,1$. Suppose that the sample size in the target site is $N_\tgt$. In the propensity score model, suppose that $\hat{\bga} - \bga^* = o_p(1)$. In the outcome regression model, suppose also that $\hat{\bgb_a} - \bgb_a^* = o_p(1)$. Then the influence function for $\hat{\mu}_{a,T}$ can be decomposed as

\begin{align*}
    \sqrt{N_\tgt}\left(\hat{\mu}_{a,T}^T - \mu_{a,T}\right) &= \frac{1}{\sqrt{N_\tgt}}\sum_{i=1}^{N_\tgt} \left[\frac{I(A_i=a)Y_i}{\pi_a(X_i,\hat{\bga})} - \left(\frac{I(A_i=a)}{\pi_a(X_i,\hat{\bga})} - 1 \right) m_a(X_i,\hat{\bgb}_a) - \mu_{a,T} \right] \\
    &= \underbrace{\frac{1}{\sqrt{N_\tgt}}\sum_{i=1}^{N_\tgt}\left[\frac{I(A_i=a)Y_i}{\pi_a(X_i,\bga_k^*)} - \left(\frac{I(A_i=a)}{\pi_a(X_i,\bga_k^*)} - 1 \right) m_a(X_i,\bgb_a^*) - \mu_{a,T} \right]}_{T_1} \\
    &\quad+\underbrace{\frac{1}{\sqrt{N_\tgt}}\sum_{i=1}^{N_\tgt}\left[I(A_i=a)\left(\frac{1}{\pi_a(X_i,\hat{\bga})} - \frac{1}{\pi_a(X_i,\bga_k^*)}\right) \{Y_i - m_a(X_i,\bgb_a^*)\}   \right]}_{T_2} \\
    &\quad-\underbrace{\frac{1}{\sqrt{N_\tgt}}\sum_{i=1}^{N_\tgt}\left[ \left\{\frac{I(A_i=a)}{\pi_a(X_i,\bga_k^*)} - 1\right\}\{m_a(X_i,\hat{\bgb}_a)-m_a(X_i,\bgb_a^*)\} \right]}_{T_3} \\
     &\quad-\underbrace{\frac{1}{\sqrt{N_\tgt}}\sum_{i=1}^{N_\tgt}\left[I(A_i=a)\left(\frac{1}{\pi_a(X_i,\hat{\bga})}- \frac{1}{\pi_a(X_i,\bga_k^*)} \right) \{m_a(X_i,\hat{\bgb}_a)-m_a(X_i,\bgb_a^*)\} \right]}_{T_4}
\end{align*}

If either the outcome regression model or propensity score model is correctly specified, then $\E(T_1) = 0$. For $T_2$, when the outcome regression model is correctly specified but the propensity score model may be misspecified, $T_2 = o_p(1)$. To show this, note that since $\hat{\bga} \to \bga^*$, then $\frac{1}{\pi_a(X_i,\hat{\bga})}- \frac{1}{\pi_a(X_i,\bga^*)} = o_p(1)$. Let $D_i = I(A_i=a)\left(\frac{1}{\pi_a(X_i,\hat{\bga})} - \frac{1}{\pi_a(X_i,\bga^*)}\right) \{Y_i - m_a(X_i,\bgb_a^*)\}$. Given $X_i$, $i=1,...,N_\tgt$, $D_i \perp D_j$ for $i \neq j$ and $\E(D_i \mid X_i) = 0$. Since $Var(D_i \mid X_i) \leq \max(\pi_i(\bga^*) - \pi_i(\hat{\bga}))^2 \to o_p(1)$, then for any $\epsilon>0$, by Chebyshev's inequality, $$P_{D|X}\left( |N_\tgt^{-1/2} \sum_{i=1}^{N_\tgt} D_i | \geq \epsilon \right) \leq \frac{Var(D_i \mid X_i)}{\epsilon^2} \leq o_p(1),$$ so $T_2 = o_p(1)$. But when the outcome regression model is misspecified, then $\E(Y-m_a(X,\bgb_a^*) \neq 0$, and the uncertainty of $\hat{\bga}$ contributes to the uncertainty of $\hat{\mu}_{a,T}^T$, i.e., we need to consider the uncertainty contribution from the estimation of $\hat{\bga}$. By a similar argument, $T_3 = o_p(1)$ when the propensity score model is correctly specified. However, when it is misspecified, the influence function of $\hat{\bgb_a}$ must be considered as it will contribute to the influence function of $\hat{\mu}_{a,T}^T$. For $T_4$ to be $o_p(1)$, it suffices that the product $$\left\{{\pi_a(X_i,\hat{\bga})^{-1}}-{\pi_a(X_i,\bga^*)^{-1}} \right\} \cdot \{m_a(X_i,\hat{\bgb}_a)-m_a(X_i,\bgb_a^*)\}$$ be $O_p(n^{-d})$ where $d > 1/2$, which is easily satisfied if $\alpha$ and $\bgb_a$ are estimated by maximum likelihood, in which case the product converges at rate $O_p(n^{-1})$. In the general case where models may be misspecified, we need to plug-in the influence functions for $\hat{\bga}-\bga^*$ and $\hat{\bgb_a}-\bgb_a^*$.

Denote
$$g_0(X_i,\bga) := \frac{I(A_i=a)}{\pi_a(X_i,\bga)},$$ so that we can write $T_2$ as
\begin{align*}
    T_2 &= N_\tgt^{-1/2} \sum_{i=1}^{N_\tgt}\left[Y_i-m_a(X_i,\bgb_a^*)\right]\left[g_0(X_i,\hat{\bga})-g_0(X_i,\bga^*)\right] \\
    &= N_\tgt^{-1/2} \sum_{i=1}^{N_\tgt}\left[Y_i-m_a(X_i,\bgb_a^*)\right] \triangledown g_0(X_i,\bga^\prime)(\hat{\bga}-\bga^*) \\
    &= N_\tgt^{-1/2} \sum_{i=1}^{N_\tgt}\left[Y_i-m_a(X_i,\bgb_a^*)\right] \triangledown g_0(X_i,\bga^*)(\hat{\bga}-\bga^*) \\
    &\quad+ \underbrace{N_\tgt^{-1/2} \sum_{i=1}^{N_\tgt}\left[Y_i-m_a(X_i,\bgb_a^*)\right] \underbrace{\left[\triangledown g_0(X_i,\bga^\prime)- \triangledown g_0(X_i,\bga^*)\right]}_{o_p(\sqrt{N_\tgt})} (\hat{\bga}-\bga^*)}_{o_p(1)} \\
    &= N_\tgt^{-1} \sum_{i=1}^{N_\tgt}\left[Y_i-m_a(X_i,\bgb_a^*)\right] \triangledown g_0(X_i,\bga^*) \cdot \sqrt{N_\tgt}(\hat{\bga}-\bga^*) + o_p(1),
\end{align*}
where $||\bga^\prime-\bga^*|| < ||\hat{\bga}-\bga^*||$.

By the law of large numbers, $$N_\tgt^{-1} \sum_{i=1}^{N_\tgt}\left[Y_i-m_a(X_i,\bgb_a^*)\right] \triangledown g_0(X_i,\bga^*) \to E\left[\left[Y_i-m_a(X_i,\bgb_a^*)\right] \triangledown g_0(X_i,\bga^*)\right] = D_1(\bgb_a^*,\bga^*).$$ Further assume that $\sqrt{N_\tgt}(\hat{\bga}-\bga^*) = N_\tgt^{-1/2} \sum_{i=1}^{N_\tgt} \phi_\tgt(\bx_i,\bga) $ where $\phi_\tgt(\bx_i,\bga)$ is the influence function for $\hat{\bga}$. Then $$T_2 = N_\tgt^{-1/2}\sum_{i=1}^{N_\tgt}D_1(\bgb_a^*,\bga^*) \cdot \phi_\tgt(\bx_i,\bga) + o_p(1).$$
If the outcome regression model is correctly specified, then $$D_1 = E_{X,A}\{\triangledown g_0(X_i,\bga^*) E_{Y|X,A}\left[Y_i-m_a(X_i,\bgb_a^*)\right]\} = 0,$$ and $T_2 = o_p(1)$.

Now focusing on $T_3$, we can write
\begin{align*}
    T_3 &= \frac{1}{\sqrt{N_\tgt}}\sum_{i=1}^{N_\tgt}\left[ \left\{\frac{I(A_i=a)}{\pi_a(X_i,\bga^*)} - 1\right\}\{m_a(X_i,\hat{\bgb}_a)-m_a(X_i,\bgb_a^*)\} \right] \\
    &= \frac{1}{\sqrt{N_\tgt}}\sum_{i=1}^{N_\tgt}\left[ \left\{\frac{I(A_i=a)}{\pi_a(X_i,\bga^*)} - 1\right\} \triangledown m_a(X_i,\bgb_a^*) (\hat{\bgb_a}-\bgb_a^*) \right] + o_p(1) \\
    &= N_\tgt^{-1}\sum_{i=1}^{N_\tgt} \left[ \left\{\frac{I(A_i=a)}{\pi_a(X_i,\bga^*)} - 1\right\}\triangledown m_a(X_i,\bgb_a^*)\right] \cdot \sqrt{N_\tgt} (\hat{\bgb_a}-\bgb_a^*)  + o_p(1).
\end{align*}
By the law of large numbers, $$N_\tgt^{-1}\sum_{i=1}^{N_\tgt} \left[ \left\{\frac{I(A_i=a)}{\pi_a(X_i,\bga^*)} - 1\right\}\triangledown m_a(X_i,\bgb_a^*)  \right] \to E\left[\left\{\frac{I(A_i=a)}{\pi_a(X_i,\bga^*)} - 1\right\}\triangledown m_a(X_i,\bgb_a^*)\right] = D_2(\bga^*,\bgb_a^*).$$
How do we estimate ${\bgb}_a$? Suppose $\bgb_a$ is estimated within each site so that $$\hat{\bgb_a}_k = \argmin_{\bgb_a} L_k(\bgb_a) = \argmin_{\bgb_a} n_k^{-1} \sum_{i=1}^{n_k} f(X_i,Y_i,\bgb_a),$$
for some specified loss function $f$. For example, $f(x,y,\bgb_a) = (y-\bgb_a^\top x)^2$ for linear regression, and $f(x,y,\bgb_a) = \log\{1+\exp(\bgb_a^\top x)\} - y \bgb_a^\top x$ for logistic regression.
By a Taylor series expansion of the gradient, $$0 = \triangledown L_k(\hat{\bgb_a}_k) = \triangledown L_k(\bgb_a^*) + \triangledown^2 L_k(\bgb_a')(\hat{\bgb_a}_k - \bgb_a^*),$$ where $||\bgb_a'-\bgb_a^*|| \leq ||\hat{\bgb_a}_k - \bgb_a^*||$. This implies that $$\sqrt{n_k} E\triangledown^2L_k(\bgb_a^*) (\hat{\bgb_a}_k-\bgb_a^*) = \sqrt{n_k}\triangledown L_k(\bgb_a^*) + \underbrace{\sqrt{n_k}(\triangledown^2 L_k(\bgb_a')-E \triangledown^2 L_k(\bgb_a^*)) (\hat{\bgb_a}_k-\bgb_a^*)}_{o_p(1)},$$
since $\triangledown^2 L_k(\bgb_a')-E \triangledown^2 L_k(\bgb_a^*) = \underbrace{\triangledown^2 L_k(\bgb_a')-\triangledown^2 L_k(\bgb_a^*)}_{\text{by Lipschitz }C||\hat{\bgb_a}_k - \bgb_a^*|| \to O_p(n^{-1/2})} + \underbrace{\triangledown^2 L_k(\bgb_a^*) - E \triangledown^2 L_k(\bgb_a^*)}_{\text{by Concentration } O_p(n^{-1/2})}.$

Thus, $$ \sqrt{n_k} (\hat{\bgb_a}_k - \bgb_a^*) = \sqrt{n_k} \{E \triangledown^2 L_k(\bgb_a^*) \}^{-1} \triangledown L_k(\bgb_a^*) + o_p(1).$$
Denoting $\{E \triangledown^2 L_k(\bgb_a^*) \}^{-1}$ as $I_k$, we have
\begin{equation*}
    {\sqrt{n_k} (\hat{\bgb_a}_k - \bgb_a^*) = n_k^{-1/2}\sum_{i=1}^{n_k}I_k \cdot \triangledown f(x_i,y_i,\bgb_a) + o_p(1).}
\end{equation*}

If $\hat{\bgb}_a = \hat{\bgb}_{0a}$, using the outcome regression parameter fit in the target site, then $$\sqrt{N_\tgt} (\hat{\bgb}_{0a} - \bgb_a^*) = N_\tgt^{-1/2} \sum_{i=1}^{N_\tgt} I_\tgt \cdot \triangledown f(x_i,y_i,\bgb_a) + o_p(1).$$
Hence, we have $$T_3 = N_\tgt^{-1/2}\sum_{i=1}^{N_\tgt} D_2 \cdot I_\tgt \cdot \triangledown f(x_i,y_i,\bgb_a) + o_p(1).$$
If the propensity score model is correctly specified, then $$D_2 = E\left[ \left( \frac{I(A_i=a)}{\pi(X_i,\bga^*)} - 1 \right) \triangledown m_a(X_i,\bgb_a^*) \right] = 0,$$ and $T_3 = o_p(1).$ \\

Summarizing, since $T_4 = o_p(1)$, then combining $T_1$, $T_2$, and $T_3$, we obtain the influence function in the target site to be
\begin{align*}
    &\sqrt{N_\tgt}(\hat{\mu}_{a,T}^T - {\mu}_{a,T}) \\
    &= \frac{1}{\sqrt{N_\tgt}} \sum_{i=1}^{N_\tgt} \Bigg\{\frac{I(A_i=a)Y_i}{\pi_a(X_i,\bga^*)} - \left(\frac{I(A_i=a)}{\pi_a(X_i,\bga^*)}-1\right)m_a(X_i,\bgb_a^*) - \mu_{a,T} \\
    &\quad+ D_1 \phi_\tgt(x_i,\alpha) + D_2 I_\tgt \triangledown f(X_i,Y_i,\bgb_a^*) \Bigg\} + o_p(1)
\end{align*}

When both outcome regression and propensity score models are correct, $D_1=D_2=0$, so the influence function reduces to
\begin{align*}
    &\sqrt{N}(\hat{\mu}_{a,T}^T - {\mu}_{a,T}) \\
    &= {\frac{1}{\sqrt{N}\rho_\tgt} \sum_{i=1}^{N}I(R_i=T) \Bigg\{\frac{I(A_i=a)Y_i}{\pi_a(X_i,\bga^*)} - \left(\frac{I(A_i=a)}{\pi_a(X_i,\bga^*)}-1\right)m_a(X_i,\bgb_a^*) - \mu_{a,T} \Bigg\} + o_p(1).}
\end{align*}

When the outcome regression model is correct, but the propensity score model may be misspecified, $D_1=0$, so the influence function reduces to
\begin{align*}
    &\sqrt{N}(\hat{\mu}_{a,T}^T - {\mu}_{a,T}) \\
    &= {\frac{1}{\sqrt{N}\rho_\tgt} \sum_{i=1}^{N}I(R_i=T) \Bigg\{ \{Y_i-m_a(X_i,\bgb_a^*)\} \left[\frac{I(A_i=a)}{\pi_a(X_i,\bga^*)} + 2D_2 I_\tgt X_i \right] + m_a(X_i,\bgb_a^*) - \mu_{a,T} \Bigg\} + o_p(1)},
\end{align*}
since $\triangledown f(X_i,Y_i,\bgb_a^*) = 2X_i^\top(Y_i-m_a(X_i,\bgb_a^*))$.

When the propensity score model is correct, but the outcome regression model may be misspecified, $D_2=0$, so the influence function reduces to
\begin{align*}
    &\sqrt{N_\tgt}(\hat{\mu}_{a,T}^T - {\mu}_{a,T}) \\
    &= \frac{1}{\sqrt{N_\tgt}} \sum_{i=1}^{N_\tgt} \Bigg\{\frac{I(A_i=a)Y_i}{\pi_a(X_i,\bga^*)} - \left(\frac{I(A_i=a)}{\pi_a(X_i,\bga^*)}-1\right)m_a(X_i,\bgb_a^*) - \mu_{a,T} + D_1 \phi_\tgt(x_i,\alpha) \Bigg\} + o_p(1).
\end{align*}
It remains to obtain $\phi_\tgt(x_i,\alpha)$. We follow the general strategy to find $\hat{\bga}$ that solves the system of moment equations, $E\left[h(X_i,\alpha)\right] = 0$. Since the system is exactly identified, the variance can be estimated as $\widehat{Var}(\hat{\bga}) = n^{-1}B^{-1}(\hat{\bga})M(\hat{\bga})B^{-1}(\hat{\bga})^\top$, where $B(\hat{\bga}) = -n^{-1}\sum_{i=1}^n \frac{\partial h(X_i,\alpha)}{\partial \alpha^\top} |_{\alpha=\hat{\bga}}$ and $M(\hat{\bga}) = n^{-1}\sum_{i=1}^n h(X_i,\hat{\bga}) h(X_i,\hat{\bga})^\top$. Based on M-estimation theory and the theory of influence functions, $\widehat{Var}(\hat{\bga}) = n^{-1}(n^{-1}\sum_{i=1}^n IF_i(\hat{\bga}) IF_i(\hat{\bga})^\top)$. It is clear that $$IF_i(\alpha) = E\left[B(\alpha)\right]^{-1}h(X_i,\alpha).$$
For $\phi_\tgt(x_i,\alpha)$ where the propensity score is estimated by logistic regression, $P(A_i=1 \mid X_i) = \pi_a(X_i,\alpha) = \frac{\exp(\alpha^\top x)}{1+\exp(\alpha^\top x)}$. Then $h(X_i,\alpha)=X_i^\top (A_i-\pi_a(X_i,\alpha))$ and $\frac{\partial h(X_i,\alpha)}{\partial \alpha^\top} = -X_i^\top \left[\pi_a(X_i,\alpha)(1-\pi_a(X_i,\alpha))\right] X_i$. Hence $$\phi_\tgt(X_i,\alpha) = E\left[X_i^\top \pi_a(X_i,\alpha)(1-\pi_a(X_i,\alpha))X_i\right]^{-1} \cdot X_i^\top (I(A_i=a)-\pi_a(X_i,\alpha)).$$

The influence function $\sqrt{N}(\hat{\mu}_{a,T}^T - {\mu}_{a,T})$ reduces to
\begin{align*}
    \sqrt{N}(\hat{\mu}_{a,T}^T - {\mu}_{a,T})
    &= \frac{1}{\sqrt{N}\rho_\tgt} \sum_{i=1}^{N}I(R_i=T) \Bigg\{ \underbrace{\left(\frac{I(A_i=a)Y_i}{\pi_a(X_i,\bga^*)} - \mu_{a,T} \right)}_{\text{Mean } 0} \\
    &\quad- \underbrace{\left(\frac{I(A_i=a)}{\pi_a(X_i,\bga^*)}-1\right) \cdot \Big\left[m_a(X_i,\bgb_a^*) + G_1 G_2^{-1} \pi_a(X_i,\bga^*)X_i^\top \Big\right]  \Bigg\}}_{\text{Mean } 0 \text{ under correct PS model}} + o_p(1),
\end{align*}

where $$G_1 = E\left[\left[Y_i-m_a(X_i,\bgb_a^*)\right]I(A_i=a)\frac{X_i(\pi_a(X_i,\bga^*)-1)}{\pi_a(X_i,\bga^*)} \right]$$ and $$G_2 = E\left[X_i^\top\pi_a(X_i,\bga^*)(1-\pi_a(X_i,\bga^*))X_i\right].$$

\subsection*{Source Sites}
In the source sites, we must additionally consider estimation of the density ratio weights. Let $h(X_i,\bgg_k,\tau) = X_i \exp(\bgg_k^\top X_i) - \tau$, where $X = (1,X_1,...,X_p)^\top$ is a design matrix with $1$ in the first column and $\tau = (1, E_\tgt(X))^\top$ is the covariate mean vector in the target population with $1$ in the first entry.
\begin{align*}
    0 &= n_k^{-1} \sum_{i=1}^{n_k} h(X_i,\hat{\gamma}_k,\hat{\tau}) \\
    &= \frac{1}{\sqrt{n_k}}  \sum_{i=1}^{n_k} h(X_i,\bgg_k^*,\tau^*) + n_k^{-1} \sum_{i=1}^{n_k} \triangledown_{\bgg_k} h(X_i,\bgg_k^\prime, \tau^\prime) \cdot \sqrt{n_k} (\hat{\gamma}_k - \bgg_k^*) \\ &\quad+ n_k^{-1} \sum_{i=1}^{n_k} \triangledown_{\tau} h(X_i,\bgg_k^\prime,\tau^\prime) \cdot \sqrt{n_k}(\hat{\tau}-\tau^*) + o_p(1) \\
    &\to \frac{1}{\sqrt{n_k}}  \sum_{i=1}^{n_k} h(X_i,\bgg_k^*,\tau^*) + E\left[ \triangledown_{\bgg_k} h(X_i,\bgg_k^\prime, \tau^\prime) \right] \cdot \sqrt{n_k} (\hat{\gamma}_k - \bgg_k^*) \\ &\quad+ E\left[ \triangledown_{\tau} h(X_i,\bgg_k^\prime,\tau^\prime) \right] \cdot \sqrt{n_k}(\hat{\tau}-\tau^*) + o_p(1)
\end{align*}
Re-arranging for $\sqrt{n_k}(\hat{\gamma}_k - \bgg_k^*)$,
\begin{align*}
    \sqrt{n_k}(\hat{\gamma}_k - \bgg_k^*) &= \frac{1}{\sqrt{n_k}}  \sum_{i=1}^{n_k} H_1^{-1} h(X_i,\bgg_k^*,\tau^*) + H_1^{-1}\sqrt{n_k} (\hat{\tau}-\tau^*) \\
    &= \frac{1}{\sqrt{n_k}}  \sum_{i=1}^{n_k} H_1^{-1} h(X_i,\bgg_k^*,\tau^*) + \sqrt{\frac{\rho_k}{\rho_\tgt}} H_1^{-1}\sum_{i=1}^N \left[I(R_i \in T) (X_i-\tau)\right],
\end{align*}
where $H_1 = E\left[-\triangledown_{\bgg_k} h(X_i,\bgg_k^*,\tau^*)\right]$.

Then
\begin{align*}
    \sqrt{N} (\hat{\gamma}_k - \bgg_k^*) &= \frac{1}{\sqrt{N}\rho_k} \sum_{i=1}^N \Bigg\{H_1^{-1} h(X_i,\bgg_k^*,\tau^*) I(R_i=k) + \sqrt{\frac{\rho_k}{\rho_\tgt}}H_1^{-1}(X_i-\tau) I(R_i \in T) \Bigg\} + o_p(1),
\end{align*}
where $H_1 = E\left[-X \exp(\bgg_k^{*^\top} X) X^\top\right]$.

To summarize, when the outcome regression model, propensity score model, and density ratio model are all correctly specified, then
\begin{align*}
    \sqrt{N}(\hat{\mu}_{a,T}^r - {\mu}_{a,T}) &= \frac{1}{\sqrt{N}\rho_k} \sum_{i=1}^N  \Bigg\{ I(R_i=T) m_a(X_i,\bgb_a^*) - \mu_{a,T} \\ &\quad+ \frac{I(R_i=k, A_i=a) \omega_k(X_i,\bgg_k^*)}{\pi_a(X_i,\bga^*)} \left[Y_i - m_a(X_i,\bgb_a^*)\right]  \Bigg\}  + o_p(1)
\end{align*}

When only the outcome regression model is correctly specified, then

\begin{align*}
    &\sqrt{N}(\hat{\mu}_{a,T}^r - {\mu}_{a,T}) \\
    &= \frac{1}{\sqrt{N}\rho_k} \sum_{i=1}^{N} \Bigg\{ I(R_i=T)m_a(X_i,\bgb_a^*) - \mu_{a,T} \\ &\quad+ \{Y_i-m_a(X_i,\bgb_a^*)\} \left[\frac{I(R_i=k, A_i=a)\omega_k(X_i,\bgg_k^*)}{\pi_a(X_i,\bga^*)} + 2D_2^r I_k X_i \right] \Bigg\} + o_p(1),
\end{align*}
where $D_2^r = E\left[\left(\frac{I(A_i=a,T_i=k)\omega_k(X_i,\bgg_k^*)}{\pi_a(X_i,\bga^*)} - 1 \right)\triangledown m_a(X_i,\bgb_a^*) \right]$ and $I_k = \left[\E(\triangledown^2 L_k(\bgb_a^*))\right]^{-1}$.

When the propensity score model and density ratio model are correctly specified but the outcome regression model may be misspecified, then
\begin{align*}
    &\sqrt{N}(\hat{\mu}_{a,T}^r - {\mu}_{a,T}) \\
    &= \frac{1}{\sqrt{N}\rho_k} \sum_{i=1}^{N} \Bigg\{ I(R_i=T)m_a(X_i,\bgb_a^*) - \mu_{a,T} + \frac{I(R_i=k, A_i=a)\omega_k(X_i,\bgg_k^*)}{\pi_a(X_i,\bga^*)} \left[Y_i -m_a(X_i,\bgb_a^*)\right] \\
    &\quad+ \underbrace{E \left[\{Y_i-m_a(X_i,\bgb_a^*)\} I(R_i=k, A_i=a) \exp(\bgg_k^{*^\top}X_i)\frac{\pi_a(X_i,\bga^*)-1}{\pi_a(X_i,\bga^*)} \right]}_{D_{1,\bga^*}^r} \\
    &\quad \cdot \underbrace{\left[\E(X_i^\top \pi_a(X_i,\bga^*)(1-\pi_a(X_i,\bga^*))X_i \right]^{-1}X_i \cdot (I(A_i=a)-\pi_a(X_i,\bga^*))}_{IF_i(X_i,\bga^*)} \\
    &\quad+ \underbrace{E \left[\{Y_i-m_a(X_i,\bgb_a^*)\} I(R_i=k, A_i=a) X_i \exp(\bgg_k^{*^\top}X_i) \frac{\exp(\alpha^{*^\top}X_i)}{1+\exp(\alpha^{*^\top}X_i)} \right]}_{D_{1,\bgg_k^*}^r} \\
    &\quad \cdot \underbrace{H_1^{-1} h(X_i,\bgg_k^*,\tau^*)I(R_i=k) + \sqrt{\frac{n_k}{N_\tgt}}H_1^{-1}I(R_i= T) (X_i-\tau)}_{IF_i(X_i,\bgg_k^*)} \Bigg\} + o_p(1).
\end{align*}

\section*{Appendix II. Data-Adaptive Weights}
In this section, we prove that given a suitable choice for $\lambda$, i.e., $\lambda \asymp N^\nu$ with $\nu \in (0,1/2)$, then $\hat{\eta}_k = \argmin_{\eta_k} \hat{Q}_a(\boldsymbol{\eta})$ are adaptive weights such that  $\hat{\eta}_k - \eta_k^* = O_p(n_k^{-1/2})$ when $\bar{\delta}_k = 0$ and $P(\hat{\eta}_k = 0) \to 1$ when $\bar{\delta}_k \neq 0$. Recall that
\begin{equation} \label{eq:l1}
    \hat{Q}_a(\boldsymbol{\eta}) = \sum_{i=1}^N \left[ \hat{\xi}^{(a)}_{i,T} - \sum_{k \in T_c} \eta_k( \hat{\xi}^{(a)}_{i,T} -  \hat{\xi}_{i,k}^{(a)} - \hat{\delta}_k)\right]^2 + \lambda \sum_{k \in T_c} |\eta_k|\hat{\delta}_k^2,
\end{equation}

First consider the case when $\bar{\delta}_k = 0$. Then the $\hat{Q}_a(\boldsymbol{\eta})$ function reduces to the sum of the squared error term. By the Central Limit Theorem, the normalized and centered estimator converges in distribution to a mean $0$ normal distribution with asymptotic variance $\Sigma$ given by $$ \Sigma = E\left[ \hat{\xi}^{(a)}_{i,T} - \sum_{k \in T_c} \eta_k( \hat{\xi}^{(a)}_{i,T} -  \hat{\xi}_{i,k}^{(a)} - \hat{\delta}_k)\right]^2.$$
Solving for the minimizer of this asymptotic variance gives $\eta_k^*$, 
\begin{align*}
    0 &= \frac{\partial \Sigma}{\partial \eta_k} = 2 \eta_k E\left[\xi^{(a)}_{i,T} - \xi^{(a)}_{i,k}\right]^2 + 2E\left[\xi^{(a)}_{i,T}(\xi^{(a)}_{i,k}-\xi^{(a)}_{i,T})\right] \\
    & \implies \eta_k^* = \frac{E\left[\xi^{(a)2}_{i,T} - \xi^{(a)}_{i,T}\xi^{(a)}_{i,k}\right]}{E\left[\xi^{(a)}_{i,k}-\xi^{(a)}_{i,T}\right]^2} \\
    &\quad \quad \quad \quad= \frac{E\left[\xi^{(a)2}_{i,T}\right]}{E\left[\xi^{(a)}_{i,k}-\xi^{(a)}_{i,T}\right]^2}
\end{align*}
since $\xi^{(a)}_{i,T}$ and $\xi^{(a)}_{i,k}$ are independent and $E\left[\xi^{(a)}_{i,T}\right] = 0$.

Now consider the case when $\bar{\delta}_k \neq 0$. The asymptotic variance is given by $$E\left[\xi^{(a)}_{i,T} - \sum_{k=1}^K \eta_k(\xi^{(a)}_{i,T}- \xi^{(a)}_{i,k}) - \hat{\delta}_k)\right]^2 + \lambda \sum_{k=1}^K |\eta_k|\hat{\delta}_k^2.$$
For simplicity, consider the case where there are two sites in total, with one target site and one source site. Taking the derivative of the asymptotic variance with respect to $\eta_k$, we obtain 
\begin{align*}
    0 &= 2 \eta_k E\left[\xi^{(a)}_{i,T}-\xi^{(a)}_{i,k} - \hat{\delta}_k\right]^2 + 2E\left[\xi^{(a)}_{i,T}(\xi^{(a)}_{i,T}-\xi^{(a)}_{i,k} - \hat{\delta}_k)\right] \\
    &\implies \eta_k^* = 
   \frac{E\left[\xi^{(a)2}_{i,T}\right] + 1/2 \lambda \hat{\delta}_k^2 }{E\left[\xi^{(a)}_{i,T}-\xi^{(a)}_{i,k} - \hat{\delta}_k\right]^2}.
\end{align*}

Then this can be estimated by $$\hat{\eta}_k = \frac{\sum_{i=1}^n \hat{\xi}^{(a)2}_{i,T} + \lambda \hat{\delta}_k^2 /2n_k}{\sum_{i=1}^n \left[\hat{\xi}^{(a)}_{i,T} - \hat{\xi}^{(a)}_{i,k}\right]^2 + n \hat{\delta}_k^2},$$
where it can be seen that $Pr(\hat{\eta}_k = 0) \to 1$ as $n_k \to \infty$ if $\hat{\delta}_k \neq 0$.

\section*{Appendix III. Privacy-preserving Penalized Regression}
In this section, we show that in the federated setting, we can solve for the $\eta_k$ that minimizes the $Q_a(\boldsymbol{\eta})$ function without sharing patient-level information from the influence functions. Recall that $\xi^{(a)}_{i,T}$ is the influence function for the target site and $\xi^{(a)}_{i,k}$ is the influence function for source site $k$ such that
\begin{align*}
    \sqrt{N}(\hat{\mu}^{(a)}_{j} - \mu^{(a)}_{T}) &= \frac{1}{\sqrt{N}}\sum_{i = 1}^N \xi^{(a)}_{i,T} + o_p(1), \\
    \sqrt{N}(\hat{\mu}^{(a)}_{k} - \mu^{(a)}_{k}) &= \frac{1}{\sqrt{N}}\sum_{i=1}^N \xi^{(a)}_{i,k} + o_p(1),
\end{align*}
 
To estimate $\eta_k$, we minimized a weighted $\ell_1$ penalty function,
\begin{equation*} \label{eq:l1}
    \hat{Q}_a(\boldsymbol{\eta}) = \sum_{i=1}^N \left[ \hat{\xi}^{(a)}_{i,T} - \sum_{k \in T_c} \eta_k( \hat{\xi}^{(a)}_{i,T} -  \hat{\xi}_{i,k}^{(a)} - \hat{\delta}_k)\right]^2 + \lambda \sum_{k \in T_c} |\eta_k|\hat{\delta}_k^2,
\end{equation*}
where $\eta_k \geq 0$ and $\sum_{k=1}^K \eta_k = 1$, $\hat{\delta}_k = \hat{\mu}^{(a)}_{k} - \hat{\mu}^{(a)}_{T}$ is the estimated discrepancy from source site $k$, and $\lambda$ is a tuning parameter.

To see that patient-level influence function information is not required, denote $$\tilde{Y}_i = \hat{\xi}_{i,T}^{(a)}, \quad \tilde{\bX}_{i,k} = \hat{\xi}_{i,T}^{(a)}-\hat{\xi}_{i,k}^{(a)} - \hat{\delta}_k.$$
Then the first term of (\ref{eq:l1}) can be written as 
\begin{align*}
    (\tilde{Y} - \tilde{\bX} \eta_k)^\top (\tilde{Y} - \tilde{\bX} \eta_k) &= \tilde{Y}^\top \tilde{Y} + \eta_k^\top \tilde{\bX}^\top \tilde{\bX} \eta_k - 2 \eta_k^\top \tilde{\bX}^\top \tilde{Y} \\
    &= \underbrace{\sum_{k=1}^K \tilde{Y}_k^\top \tilde{Y}_k}_{S_{\tilde{Y}}} + \underbrace{\eta_k^\top \sum_{k=1}^K \tilde{\bX}_k^\top \tilde{\bX}_k \eta_k}_{S_{\tilde{\bX}}} - \underbrace{2 \eta_k^\top \sum_{k=1}^K \tilde{\bX}_k^\top \tilde{Y}_k}_{S_{\tilde{\bX Y}}},
\end{align*}
where $S_{\tilde{Y}}$ is a scalar, $S_{\tilde{\bX}}$ is a $p \times p$ matrix, and $S_{\tilde{\bX Y}}$ is a $1 \times p$-vector. Thus, it suffices to share only $S_{\tilde{Y}}$, $S_{\tilde{\bX}}$, and $S_{\tilde{\bX Y}}$ with the leading site.

\section*{Appendix IV. Additional Simulation Study Results}
Recall that in the simulation study, we considered five settings with $K \in \{10, 20, 50\}$ total sites, and five estimators: one estimator using data from the target site only (target), two estimators using all sites that weights each site proportionally to its sample size (ss (naive), ss), and two estimators that employ the federated algorithm with alternative solvers (DREAMFUL-$\ell_2$, DREAMFUL-$\ell_1$).

\subsection*{Distribution of observations}
In the sparse data setting, $\mathcal{D}_{\text{sparse}}$, for the source sites, if the sample size is greater than or equal to the $65 - K/10$ quantiles, or less than or equal to the $35 + K/10$ quantiles, we set $\mathrm{A}_{kp} = 2$ for $p = 1, 2$, and $\mathrm{A}_{kp} = 0$ for the remaining source sites and the target site. In the dense data setting, $\mathcal{D}_{\text{dense}}$, for the target site $k = 1$, for $p = 1, ..., P$, we let the location parameter $\Xi_{1p} = 0.15 + 0.05\frac{1-p}{P-1}$, the scale parameter $\Omega_{1p} = 1$, and the skewness parameter $\mathrm{A}_{1p} = 0$. For source sites $k = 2, ..., K$, if the sample size is greater than or equal to the third quartile, $n_k \geq \text{Q3}(\{n_k\}_{k = 1}^K)$, we set $\Xi_{kp} = 0.15 + 0.05\frac{1-p}{P-1}$, $\Omega_{kp} = 1$, $\mathrm{A}_{kp} = 3\times 2/P$. If the sample size is less than or equal to the first quartile, $n_k \leq \text{Q1}(\{n_k\}_{k = 1}^K)$, we set $\Xi_{kp} = 0.15 + 0.05\frac{1-p}{P-1}$, $\Omega_{kp} = 1$, $\mathrm{A}_{kp} = -1\times 2/P$. If the sample size is greater than the first quartile and less than the third quartile, $n_k \in (\text{Q1}(\{n_k\}_{k = 1}^K), \text{Q3}(\{n_k\}_{k = 1}^K))$, we set $\Xi_{kp} = 0.15 + 0.05\frac{1-p}{P-1}$, $\Omega_{kp} = 1$, $\mathrm{A}_{kp} = 0$. 
We plot the observations in target and source sites for the $\mathcal{D}_{\text{dense}}$ and $\mathcal{D}_{\text{sparse}}$ data generating mechanisms when $K = 10$ and $P = 2$.

\begin{figure}[H]

    \centering
\includegraphics[scale = 0.4]{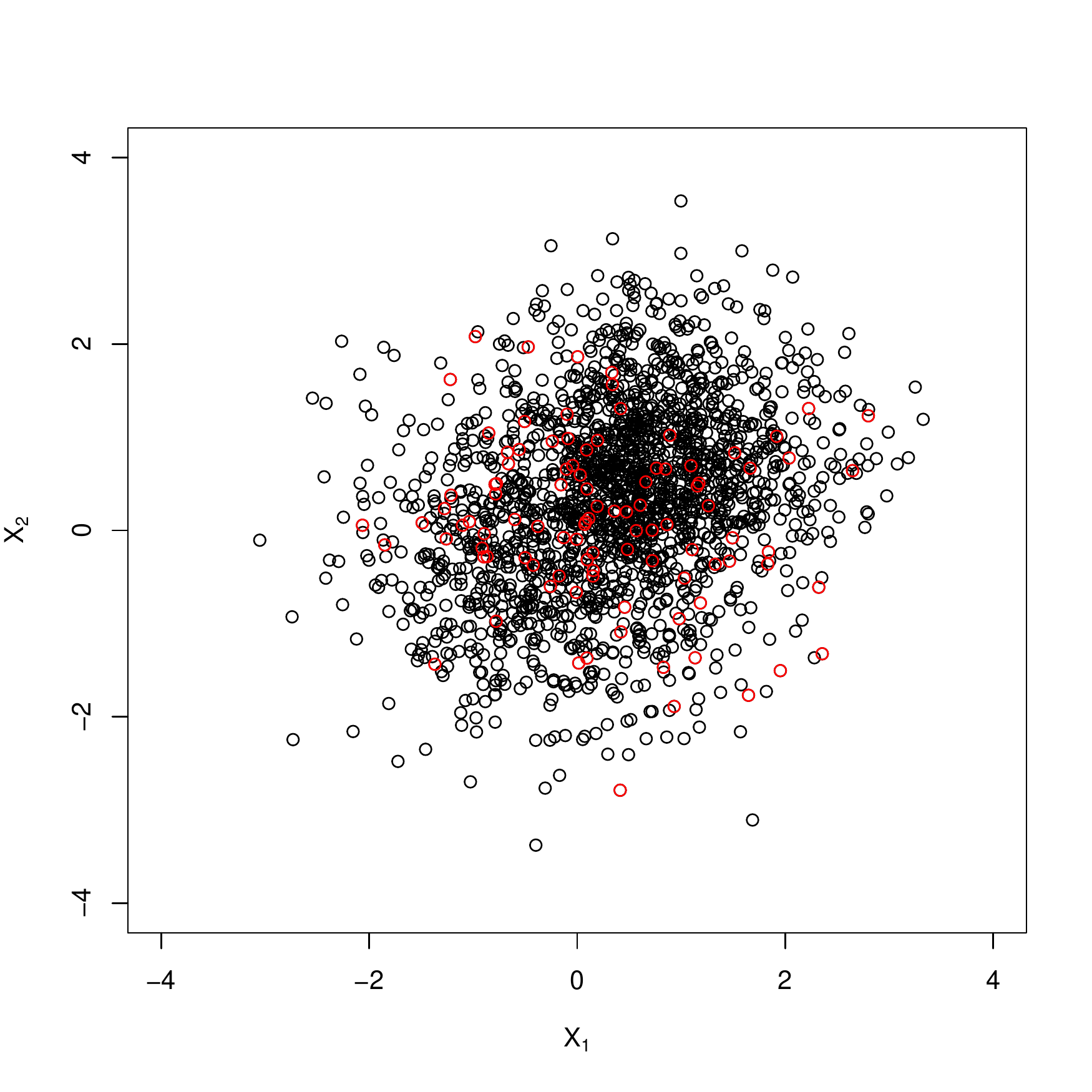} 
\includegraphics[scale = 0.4]{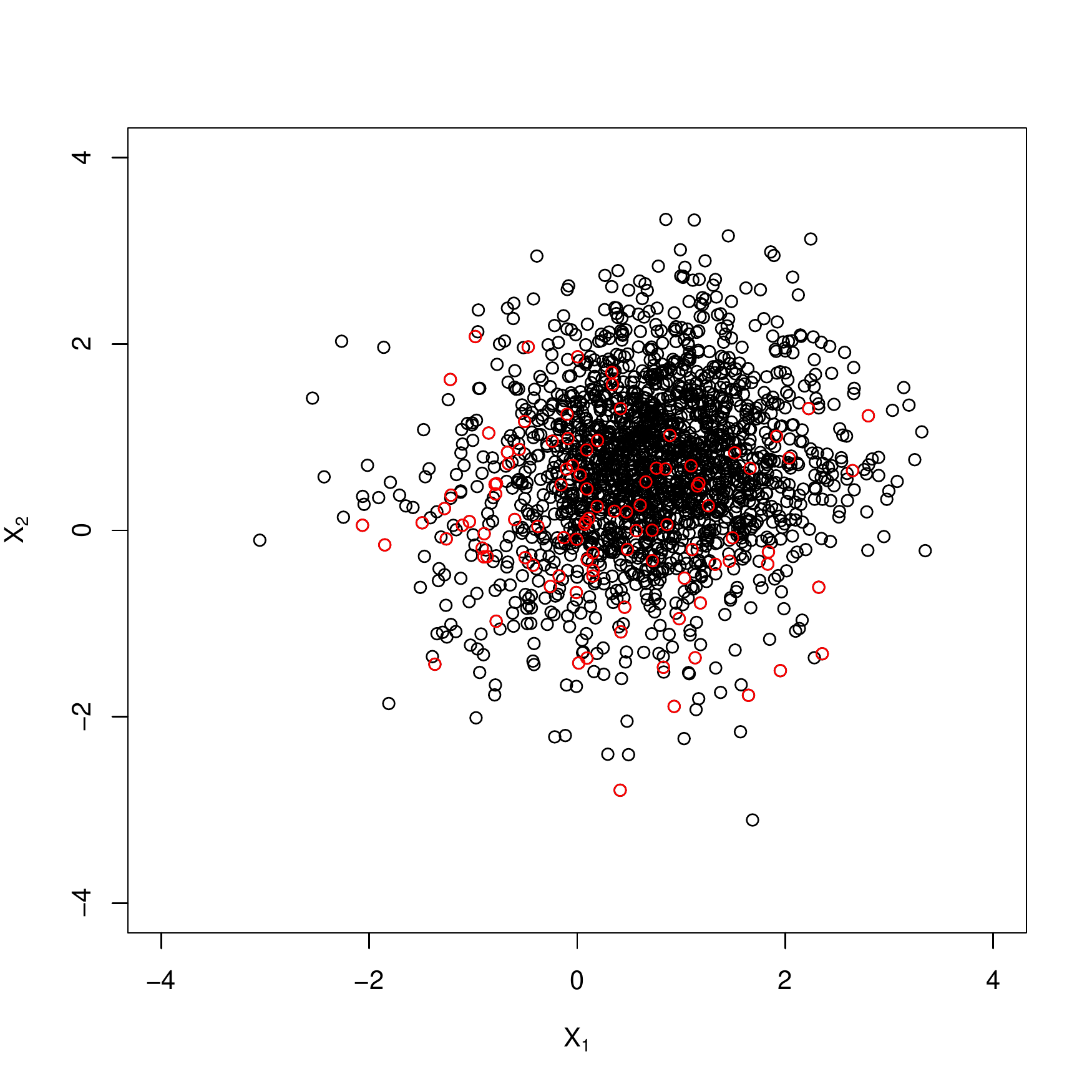}
\caption{Red (black) dots are target (source) site observations ($K = 10$, $P = 2$). (Left) Dense covariate setting. (Right) Sparse covariate setting. }
    \label{fig:Xdist}
\end{figure}

\subsection*{Alternative number of covariates and distributions}
In the main text, Table 1 reported simulation results for $\mathcal{D}_{\text{sparse}}$ in Settings I--V when $P = 2$ covariates across $1000$ simulations. Here, we present simulation results for $\mathcal{D}_{\text{sparse}}$ when $P = 10$ covariates and for $\mathcal{D}_{\text{dense}}$ when $P = \{2,10\}$.

    \begin{table}[H]
        \caption{See the main text.}
    \end{table}

\begin{center}
\begin{table}[H]
\caption{The absolute bias, root mean squared error, coverage and length of the 95\% confidence interval in 1000 simulation datasets for covariates distribution $\mathcal{D}_{\text{sparse}}$ when $P = 10$.}
\label{tab_the_practice_perspective}
\centering
\footnotesize
\setlength{\tabcolsep}{4pt}
\begin{tabular}{lrrrrrrrrrrrr}
  \hline
   & \multicolumn{12}{c}{Simulation scenarios} \\
     \cmidrule(lr){2-13} 
  &  \multicolumn{4}{c}{$K = 10$} &  \multicolumn{4}{c}{$K = 20$} &  \multicolumn{4}{c}{$K = 50$} \\
  \cmidrule(lr){2-5}  \cmidrule(lr){6-9} \cmidrule(lr){10-13} 
     & Bias & RMSE & Cov. & Len. & Bias & RMSE & Cov. & Len. & Bias & RMSE & Cov. & Len. \\
   \cmidrule(lr){1-1}  \cmidrule(lr){2-5}  \cmidrule(lr){6-9} \cmidrule(lr){10-13} 
Setting I &  &  &\\ 
\hspace{.15cm} target & 0.01 & 0.76 & 96.00 & 3.22 & 0.00 & 0.71 & 97.60 & 3.24 & 0.00 & 0.76 & 97.30 & 3.25\\
\hspace{.15cm} ss (naive) & 0.87 & 0.88 & 27.20 & 1.46 & 0.90 & 0.91 & 2.40 & 0.87 & 0.92 & 0.92 & 0.00 & 0.46\\
\hspace{.15cm} ss     & 0.10 & 0.42 & 99.30 & 2.61 & 0.09 & 0.30 & 99.00 & 1.69 & 0.08 & 0.21 & 96.60 & 0.97\\
\hspace{.15cm} $\text{DREAMFUL}-\ell_2$   & 0.36 & 0.51 & 96.40 & 2.26 & 0.41 & 0.52 & 87.00 & 1.56 & 0.44 & 0.50 & 54.00 & 0.90\\
\hspace{.15cm} $\text{DREAMFUL}-\ell_1$ & 0.16 & 0.55 & 96.10 & 2.17 & 0.18 & 0.51 & 93.20 & 1.78 & 0.14 & 0.55 & 81.90 & 1.50\\
   \cmidrule(lr){1-1}  \cmidrule(lr){2-5}  \cmidrule(lr){6-9} \cmidrule(lr){10-13} 
Setting II &  &  &\\ 
\hspace{.15cm} target & 0.01 & 0.82 & 95.90 & 3.32 & 0.01 & 0.74 & 97.40 & 3.31 & 0.00 & 0.79 & 97.20 & 3.33 \\
\hspace{.15cm} ss (naive) & 0.87 & 0.88 & 28.60 & 1.52 & 0.90 & 0.91 & 2.60 & 0.89 & 0.92 & 0.92 & 0.00 & 0.48\\
\hspace{.15cm} ss     & 0.10 & 0.42 & 99.30 & 2.68 & 0.09 & 0.31 & 99.30 & 1.71 & 0.08 & 0.21 & 97.00 & 0.99\\
\hspace{.15cm} $\text{DREAMFUL}-\ell_2$   & 0.37 & 0.52 & 96.60 & 2.36 & 0.43 & 0.53 & 87.60 & 1.61 & 0.46 & 0.51 & 54.00 & 0.92\\
\hspace{.15cm} $\text{DREAMFUL}-\ell_1$ & 0.16 & 0.57 & 95.00 & 2.25 & 0.19 & 0.52 & 93.50 & 1.80 & 0.13 & 0.59 & 80.60 & 1.56\\
   \cmidrule(lr){1-1}  \cmidrule(lr){2-5}  \cmidrule(lr){6-9} \cmidrule(lr){10-13} 
Setting III &  &  &\\ 
\hspace{.15cm} target & 0.04 & 0.76 & 95.60 & 3.24 & 0.06 & 0.74 & 96.40 & 3.24 & 0.06 & 0.79 & 96.70 & 3.29\\
\hspace{.15cm} ss  (naive)   & 0.84 & 0.86 & 30.80 & 1.52 & 0.87 & 0.88 & 3.70 & 0.92 & 0.89 & 0.89 & 0.00 & 0.49\\
\hspace{.15cm} ss     & 0.07 & 0.44 & 99.20 & 2.81 & 0.07 & 0.32 & 99.70 & 1.84 & 0.06 & 0.22 & 97.40 & 1.06
\\
\hspace{.15cm} $\text{DREAMFUL}-\ell_2$   & 0.31 & 0.48 & 97.00 & 2.27 & 0.37 & 0.48 & 91.00 & 1.60 & 0.40 & 0.47 & 62.80 & 0.92\\
\hspace{.15cm} $\text{DREAMFUL}-\ell_1$ & 0.10 & 0.55 & 96.10 & 2.22 & 0.13 & 0.49 & 94.40 & 1.81 & 0.09 & 0.56 & 82.90 & 1.52\\
   \cmidrule(lr){1-1}  \cmidrule(lr){2-5}  \cmidrule(lr){6-9} \cmidrule(lr){10-13} 
Setting IV &  &  &\\ 
\hspace{.15cm} target & 0.10 & 0.83 & 95.30 & 3.33 & 0.12 & 0.78 & 96.10 & 3.30 & 0.12 & 0.84 & 95.90 & 3.38 \\ 
\hspace{.15cm} ss  (naive)   & 0.81 & 0.83 & 38.80 & 1.63 & 0.85 & 0.86 & 5.60 & 0.96 & 0.86 & 0.87 & 0.20 & 0.51 \\ 
\hspace{.15cm} ss     & 0.05 & 0.44 & 99.70 & 2.89 & 0.04 & 0.32 & 99.90 & 1.87 & 0.03 & 0.21 & 98.30 & 1.06 \\ 
\hspace{.15cm} $\text{DREAMFUL}-\ell_2$   & 0.29 & 0.48 & 97.70 & 2.42 & 0.36 & 0.48 & 92.30 & 1.68 & 0.39 & 0.46 & 64.50 & 0.96 \\
\hspace{.15cm} $\text{DREAMFUL}-\ell_1$ & 0.07 & 0.56 & 95.70 & 2.31 & 0.09 & 0.50 & 93.60 & 1.85 & 0.06 & 0.58 & 82.00 & 1.54 \\
   \cmidrule(lr){1-1}  \cmidrule(lr){2-5}  \cmidrule(lr){6-9} \cmidrule(lr){10-13} 
Setting V &  &  &\\ 
\hspace{.15cm} target & 0.01 & 0.82 & 95.90 & 3.32 & 0.01 & 0.74 & 97.40 & 3.31 & 0.00 & 0.79 & 97.10 & 3.33\\
\hspace{.15cm} ss (naive) & 0.84 & 0.86 & 30.60 & 1.53 & 0.87 & 0.88 & 3.40 & 0.90 & 0.89 & 0.89 & 0.00 & 0.49\\
\hspace{.15cm} ss     & 0.08 & 0.44 & 99.50 & 2.88 & 0.06 & 0.32 & 99.60 & 1.85 & 0.06 & 0.22 & 97.70 & 1.05\\
\hspace{.15cm} $\text{DREAMFUL}-\ell_2$   & 0.34 & 0.50 & 96.70 & 2.33 & 0.40 & 0.50 & 89.10 & 1.60 & 0.43 & 0.49 & 57.70 & 0.92\\
\hspace{.15cm} $\text{DREAMFUL}-\ell_1$ & 0.14 & 0.56 & 95.40 & 2.27 & 0.16 & 0.53 & 93.10 & 1.81 & 0.14 & 0.59 & 81.10 & 1.52\\
\hline
\end{tabular}
\vspace{.25cm}
\footnotesize{
\begin{flushleft}
\end{flushleft}
}
\end{table}
\end{center}

\begin{center}
\begin{table}[H]
\caption{The absolute bias, root mean squared error, coverage and length of the 95\% confidence interval in 1000 simulation datasets for covariates distribution $\mathcal{D}_{\text{dense}}$ when $P = 2$.}
\label{tab_the_practice_perspective}
\centering
\footnotesize
\setlength{\tabcolsep}{4pt}
\begin{tabular}{lrrrrrrrrrrrr}
  \hline
   & \multicolumn{12}{c}{Simulation scenarios} \\
     \cmidrule(lr){2-13} 
  &  \multicolumn{4}{c}{$K = 10$} &  \multicolumn{4}{c}{$K = 20$} &  \multicolumn{4}{c}{$K = 50$} \\
  \cmidrule(lr){2-5}  \cmidrule(lr){6-9} \cmidrule(lr){10-13} 
     & Bias & RMSE & Cov. & Len. & Bias & RMSE & Cov. & Len. & Bias & RMSE & Cov. & Len. \\
   \cmidrule(lr){1-1}  \cmidrule(lr){2-5}  \cmidrule(lr){6-9} \cmidrule(lr){10-13} 
Setting I &  &  &\\ 
\hspace{.15cm} target & 0.00 & 0.69 & 98.20 & 3.10 & 0.00 & 0.69 & 98.20 & 3.10 & 0.00 & 0.69 & 98.10 & 3.10 \\
\hspace{.15cm} ss (naive) & 0.35 & 0.39 & 91.50 & 1.31 & 0.28 & 0.31 & 75.60 & 0.77 & 0.27 & 0.28 & 22.90 & 0.42 \\
\hspace{.15cm} ss     & 0.00 & 0.44 & 99.80 & 2.60 & 0.00 & 0.31 & 98.90 & 1.53 & 0.01 & 0.21 & 95.50 & 0.88 \\
\hspace{.15cm} $\text{DREAMFUL}-\ell_2$   & 0.08 & 0.26 & 98.60 & 1.42 & 0.06 & 0.20 & 95.10 & 0.86 & 0.07 & 0.17 & 87.30 & 0.50 \\
\hspace{.15cm} $\text{DREAMFUL}-\ell_1$ & 0.04 & 0.41 & 97.30 & 1.68 & 0.03 & 0.38 & 94.80 & 1.30 & 0.03 & 0.44 & 82.00 & 1.15\\
   \cmidrule(lr){1-1}  \cmidrule(lr){2-5}  \cmidrule(lr){6-9} \cmidrule(lr){10-13} 
Setting II &  &  &\\ 
\hspace{.15cm} target & 0.01 & 1.10 & 96.60 & 4.35 & 0.01 & 1.10 & 96.60 & 4.35 & 0.00 & 1.10 & 96.60 & 4.35 \\
\hspace{.15cm} ss (naive) & 0.36 & 0.42 & 96.40 & 1.98 & 0.29 & 0.33 & 91.20 & 1.18 & 0.27 & 0.29 & 58.60 & 0.63 \\
\hspace{.15cm} ss     & 0.00 & 0.53 & 99.90 & 3.79 & 0.00 & 0.37 & 99.60 & 2.21 & 0.01 & 0.25 & 97.70 & 1.23 \\
\hspace{.15cm} $\text{DREAMFUL}-\ell_2$   & 0.08 & 0.34 & 99.40 & 2.31 & 0.07 & 0.24 & 98.80 & 1.36 & 0.07 & 0.18 & 95.20 & 0.76 \\
\hspace{.15cm} $\text{DREAMFUL}-\ell_1$ & 0.04 & 0.54 & 98.50 & 2.29 & 0.04 & 0.48 & 94.30 & 1.65 & 0.03 & 0.60 & 72.40 & 1.34\\
   \cmidrule(lr){1-1}  \cmidrule(lr){2-5}  \cmidrule(lr){6-9} \cmidrule(lr){10-13} 
Setting III &  &  &\\ 
\hspace{.15cm} target & 0.04 & 0.70 & 96.60 & 3.09 & 0.04 & 0.70 & 96.60 & 3.09 & 0.04 & 0.71 & 96.50 & 3.09 \\
\hspace{.15cm} ss (naive) & 0.32 & 0.36 & 94.80 & 1.34 & 0.25 & 0.27 & 85.00 & 0.79 & 0.23 & 0.24 & 40.10 & 0.43 \\
\hspace{.15cm} ss     & 0.03 & 0.46 & 99.80 & 2.68 & 0.03 & 0.31 & 99.10 & 1.57 & 0.03 & 0.21 & 96.10 & 0.90 \\
\hspace{.15cm} $\text{DREAMFUL}-\ell_2$  & 0.04 & 0.25 & 99.10 & 1.44 & 0.02 & 0.20 & 95.80 & 0.87 & 0.02 & 0.15 & 90.00 & 0.51 \\
\hspace{.15cm} $\text{DREAMFUL}-\ell_1$ & 0.00 & 0.41 & 97.20 & 1.67 & 0.01 & 0.39 & 92.70 & 1.30 & 0.02 & 0.46 & 78.60 & 1.14\\
   \cmidrule(lr){1-1}  \cmidrule(lr){2-5}  \cmidrule(lr){6-9} \cmidrule(lr){10-13} 
Setting IV &  &  &\\ 
\hspace{.15cm} target & 0.32 & 1.17 & 95.30 & 4.38 & 0.32 & 1.17 & 95.30 & 4.38 & 0.32 & 1.18 & 95.30 & 4.38 \\
\hspace{.15cm} ss (naive)& 0.14 & 0.26 & 99.90 & 2.19 & 0.05 & 0.18 & 99.40 & 1.29 & 0.05 & 0.11 & 99.10 & 0.68 \\
\hspace{.15cm} ss     & 0.18 & 0.58 & 100.00 & 3.94 & 0.20 & 0.43 & 99.50 & 2.29 & 0.18 & 0.32 & 96.30 & 1.27 \\
\hspace{.15cm} $\text{DREAMFUL}-\ell_2$   & 0.15 & 0.37 & 98.70 & 2.43 & 0.16 & 0.29 & 97.30 & 1.41 & 0.16 & 0.23 & 90.30 & 0.78  \\
\hspace{.15cm} $\text{DREAMFUL}-\ell_1$ & 0.21 & 0.59 & 95.50 & 2.35 & 0.22 & 0.53 & 89.00 & 1.64 & 0.25 & 0.64 & 66.40 & 1.25\\
   \cmidrule(lr){1-1}  \cmidrule(lr){2-5}  \cmidrule(lr){6-9} \cmidrule(lr){10-13} \cmidrule(lr){14-17}
Setting V &  &  &\\ 
\hspace{.15cm} target & 0.01 & 1.10 & 96.60 & 4.35 & 0.01 & 1.10 & 96.60 & 4.35 & 0.00 & 1.10 & 96.60 & 4.35 \\
\hspace{.15cm} ss (naive) & 0.30 & 0.37 & 98.20 & 1.96 & 0.24 & 0.29 & 95.40 & 1.18 & 0.21 & 0.24 & 79.00 & 0.63 \\
\hspace{.15cm} ss     & 0.01 & 0.54 & 99.90 & 3.84 & 0.01 & 0.37 & 99.70 & 2.24 & 0.01 & 0.25 & 98.10 & 1.25 \\
\hspace{.15cm} $\text{DREAMFUL}-\ell_2$   & 0.05 & 0.34 & 99.40 & 2.29 & 0.05 & 0.24 & 99.00 & 1.35 & 0.04 & 0.17 & 96.00 & 0.75 \\
\hspace{.15cm} $\text{DREAMFUL}-\ell_1$ & 0.03 & 0.53 & 98.10 & 2.27 & 0.03 & 0.47 & 93.60 & 1.62 & 0.02 & 0.61 & 71.40 & 1.31\\
\hline
\end{tabular}
\vspace{.25cm}
\footnotesize{
\begin{flushleft}
\end{flushleft}
}
\end{table}
\end{center}

\begin{center}
\begin{table}[H]
\caption{The absolute bias, root mean squared error, coverage and length of the 95\% confidence interval in 1000 simulation datasets for covariates distribution $\mathcal{D}_{\text{dense}}$ when $P = 10$.}
\label{tab_the_practice_perspective}
\centering
\footnotesize
\setlength{\tabcolsep}{4pt}
\begin{tabular}{lrrrrrrrrrrrr}
  \hline
   & \multicolumn{12}{c}{Simulation scenarios} \\
     \cmidrule(lr){2-13} 
  &  \multicolumn{4}{c}{$K = 10$} &  \multicolumn{4}{c}{$K = 20$} &  \multicolumn{4}{c}{$K = 50$} \\
  \cmidrule(lr){2-5}  \cmidrule(lr){6-9} \cmidrule(lr){10-13} 
     & Bias & RMSE & Cov. & Len. & Bias & RMSE & Cov. & Len. & Bias & RMSE & Cov. & Len. \\
   \cmidrule(lr){1-1}  \cmidrule(lr){2-5}  \cmidrule(lr){6-9} \cmidrule(lr){10-13} 
Setting I &  &  &\\ 
\hspace{.15cm} target & 0.13 & 2.63 & 95.30 & 7.46 & 0.13 & 2.63 & 95.30 & 7.46 & 0.13 & 2.63 & 95.30 & 7.46\\
\hspace{.15cm} ss (naive) & 0.22 & 0.46 & 99.50 & 5.09 & 0.19 & 0.33 & 99.00 & 2.89 & 0.18 & 0.25 & 94.90 & 1.40\\
\hspace{.15cm} ss     & 0.00 & 0.70 & 100.00 & 7.65 & 0.01 & 0.47 & 99.80 & 4.27 & 0.01 & 0.31 & 99.70 & 2.19\\ 
\hspace{.15cm} $\text{DREAMFUL}-\ell_2$   & 0.06 & 0.53 & 99.80 & 6.31 & 0.07 & 0.38 & 99.40 & 3.45 & 0.07 & 0.25 & 98.40 & 1.71\\
\hspace{.15cm} $\text{DREAMFUL}-\ell_1$ & 0.01 & 0.90 & 97.80 & 5.79 & 0.02 & 0.80 & 92.50 & 3.56 & 0.02 & 0.96 & 65.10 & 2.35\\
   \cmidrule(lr){1-1}  \cmidrule(lr){2-5}  \cmidrule(lr){6-9} \cmidrule(lr){10-13} 
Setting II &  &  &\\ 
\hspace{.15cm} target & 0.02 & 3.38 & 93.90 & 10.21 & 0.02 & 3.38 & 93.90 & 10.21 & 0.02 & 3.39 & 93.90 & 10.21 \\
\hspace{.15cm} ss (naive) & 0.22 & 0.61 & 99.90 & 8.24 & 0.19 & 0.45 & 99.50 & 4.65 & 0.19 & 0.31 & 98.60 & 2.22\\
\hspace{.15cm} ss     & 0.00 & 1.03 & 99.90 & 12.80 & 0.01 & 0.69 & 100.00 & 6.97 & 0.01 & 0.42 & 99.90 & 3.38 \\
\hspace{.15cm} $\text{DREAMFUL}-\ell_2$   & 0.11 & 0.75 & 100.00 & 10.32 & 0.11 & 0.53 & 99.70 & 5.52 & 0.12 & 0.35 & 99.30 & 2.67\\
\hspace{.15cm} $\text{DREAMFUL}-\ell_1$ & 0.10 & 1.24 & 97.90 & 8.71 & 0.10 & 1.12 & 94.00 & 5.32 & 0.09 & 1.39 & 65.80 & 3.33 \\
   \cmidrule(lr){1-1}  \cmidrule(lr){2-5}  \cmidrule(lr){6-9} \cmidrule(lr){10-13} 
Setting III &  &  &\\ 
\hspace{.15cm} target & 0.12 & 2.06 & 93.20 & 7.08 & 0.12 & 2.06 & 93.20 & 7.08 & 0.12 & 2.06 & 93.20 & 7.08 \\
\hspace{.15cm} ss  (naive)   & 0.19 & 0.46 & 99.70 & 4.12 & 0.17 & 0.33 & 99.00 & 2.41 & 0.16 & 0.25 & 95.30 & 1.23 \\
\hspace{.15cm} ss     & 0.03 & 0.87 & 100.00 & 8.11 & 0.01 & 0.59 & 100.00 & 4.63 & 0.02 & 0.38 & 99.90 & 2.42
\\
\hspace{.15cm} $\text{DREAMFUL}-\ell_2$   & 0.02 & 0.53 & 99.40 & 5.23 & 0.04 & 0.37 & 99.60 & 2.92 & 0.04 & 0.24 & 98.80 & 1.51 \\
\hspace{.15cm} $\text{DREAMFUL}-\ell_1$ & 0.02 & 0.90 & 97.10 & 4.54 & 0.01 & 0.83 & 93.20 & 2.97& 0.04 & 0.99 & 65.90 & 2.12\\
   \cmidrule(lr){1-1}  \cmidrule(lr){2-5}  \cmidrule(lr){6-9} \cmidrule(lr){10-13} 
Setting IV &  &  &\\ 
\hspace{.15cm} target & 0.59 & 2.92 & 92.30 & 9.90 & 0.59 & 2.92 & 92.30 & 9.90 & 0.58 & 2.92 & 92.30 & 9.91\\
\hspace{.15cm} ss (naive)  & 0.34 & 0.70 & 99.50 & 7.18 & 0.38 & 0.57 & 98.70 & 4.19 & 0.37 & 0.47 & 94.50 & 2.10\\
\hspace{.15cm} ss     & 0.57 & 1.36 & 99.80 & 14.04 & 0.57 & 1.03 & 99.70 & 7.93 & 0.54 & 0.77 & 98.20 & 3.90\\
\hspace{.15cm} $\text{DREAMFUL}-\ell_2$  & 0.49 & 0.91 & 99.20 & 9.19 & 0.47 & 0.71 & 98.20 & 4.92 & 0.45 & 0.57 & 94.00 & 2.44\\
\hspace{.15cm} $\text{DREAMFUL}-\ell_1$ & 0.50 & 1.42 & 95.10 & 7.18 & 0.52 & 1.32 & 88.20 & 4.62 & 0.53 & 1.56 & 64.60 & 3.56 \\
   \cmidrule(lr){1-1}  \cmidrule(lr){2-5}  \cmidrule(lr){6-9} \cmidrule(lr){10-13} 
Setting V &  &  &\\ 
\hspace{.15cm} target & 0.02 & 3.38 & 93.90 & 10.21 & 0.02 & 3.38 & 93.90 & 10.21  & 0.02 & 3.39 & 93.90 & 10.21\\
\hspace{.15cm} ss (naive) & 0.04 & 0.59 & 99.80 & 8.33 & 0.05 & 0.42 & 99.90 & 4.74 & 0.05 & 0.42 & 99.90 & 4.74 \\
\hspace{.15cm} ss     & 0.29 & 1.28 & 99.90 & 15.48 & 0.25 & 0.89 & 100.00 & 8.57 & 0.25 & 0.59 & 99.80 & 4.12 \\
\hspace{.15cm} $\text{DREAMFUL}-\ell_2$   & 0.11 & 0.77 & 100.00 & 10.37 & 0.08 & 0.53 & 99.70 & 5.52 & 0.10 & 0.35 & 99.10 & 2.68 \\
\hspace{.15cm} $\text{DREAMFUL}-\ell_1$ & 0.02 & 1.30 & 97.00 & 8.68 & 0.04 & 1.15 & 94.00 & 5.33 & 0.01 & 1.39 & 66.60 & 3.33\\
\hline
\end{tabular}
\vspace{.25cm}
\footnotesize{
\begin{flushleft}
\end{flushleft}
}
\end{table}
\end{center}

\subsection*{Weights for the dense covariate setting}
\begin{figure}[H]
    \centering
\includegraphics[scale = 0.52]{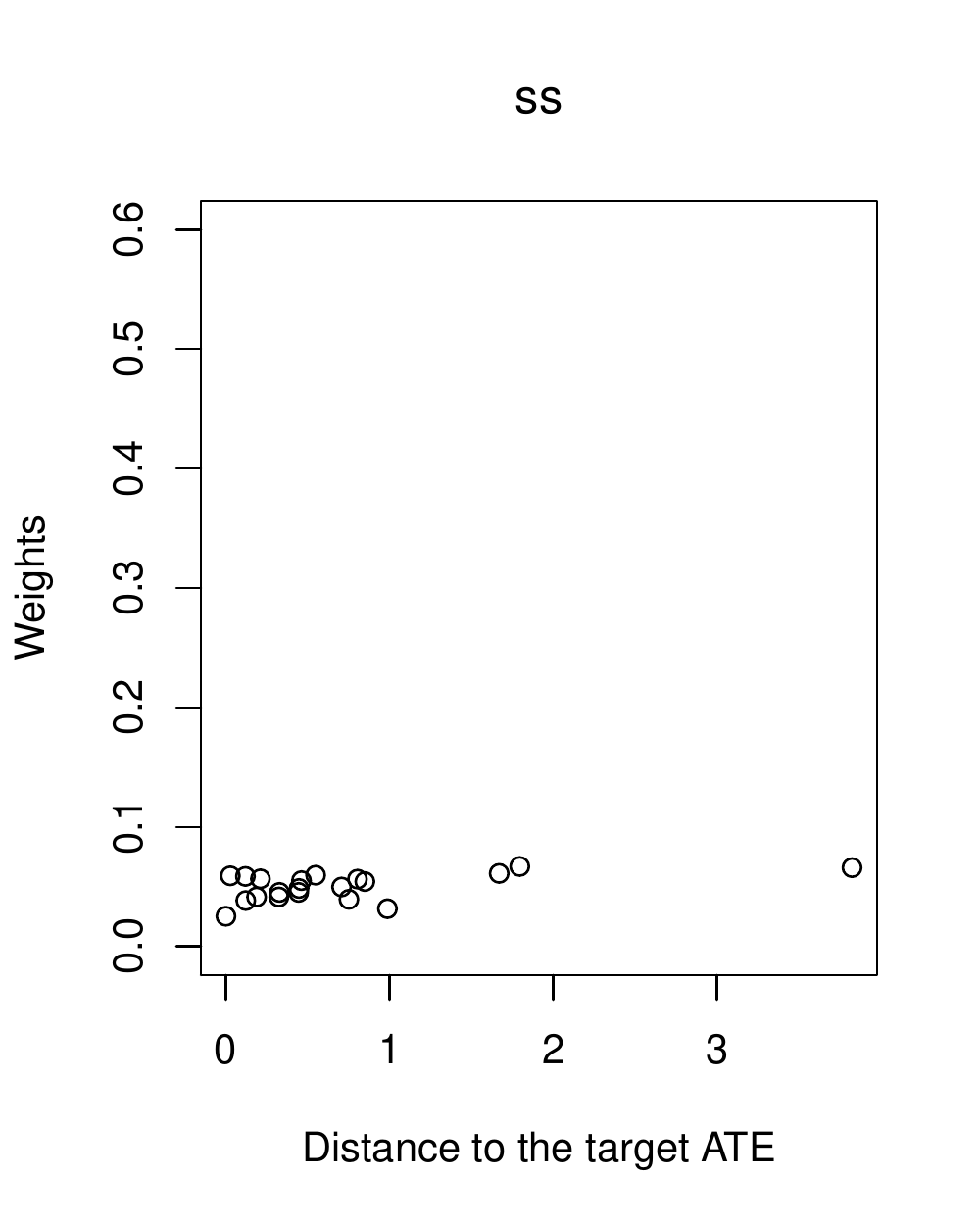} 
\includegraphics[scale = 0.52]{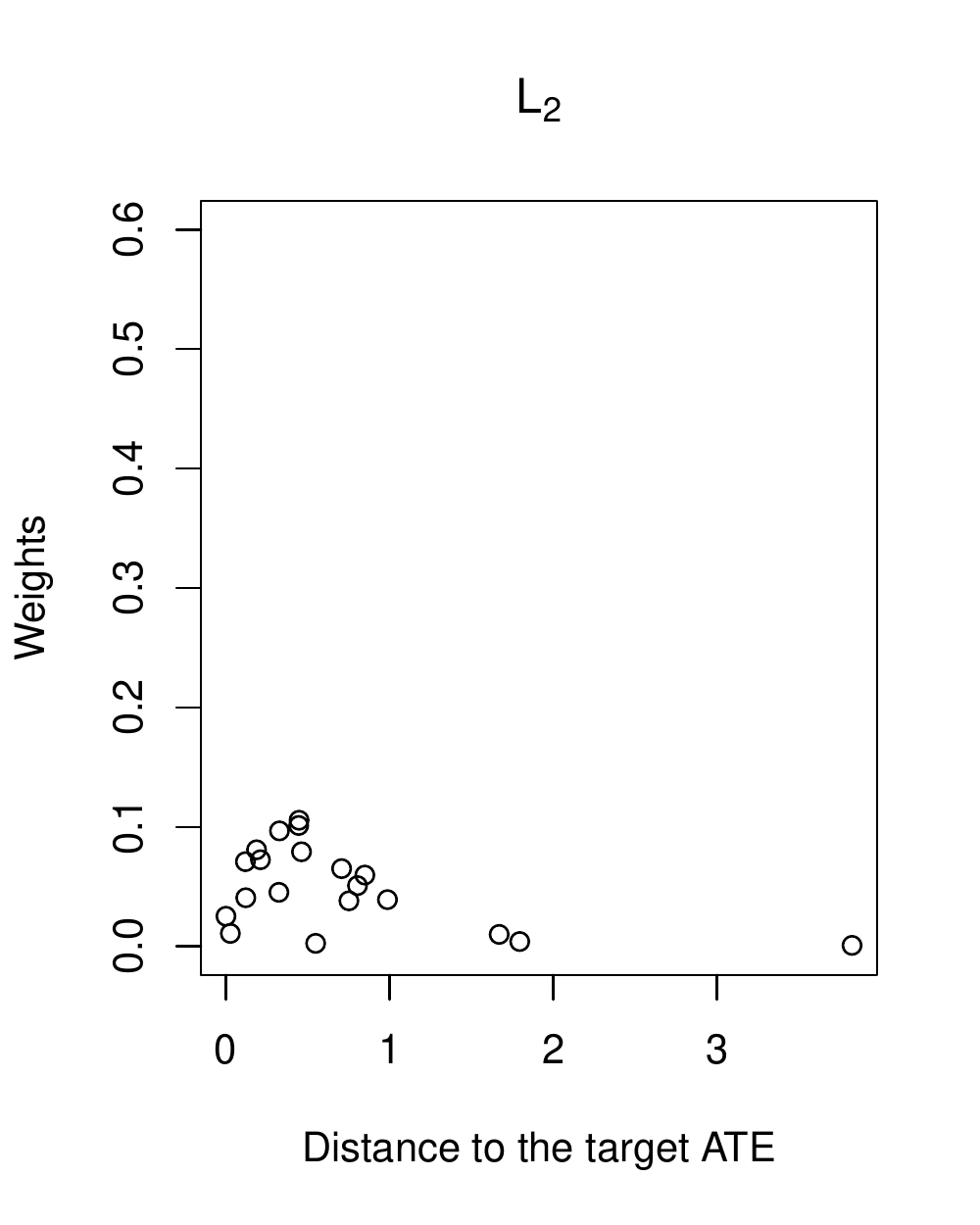}
\includegraphics[scale = 0.52]{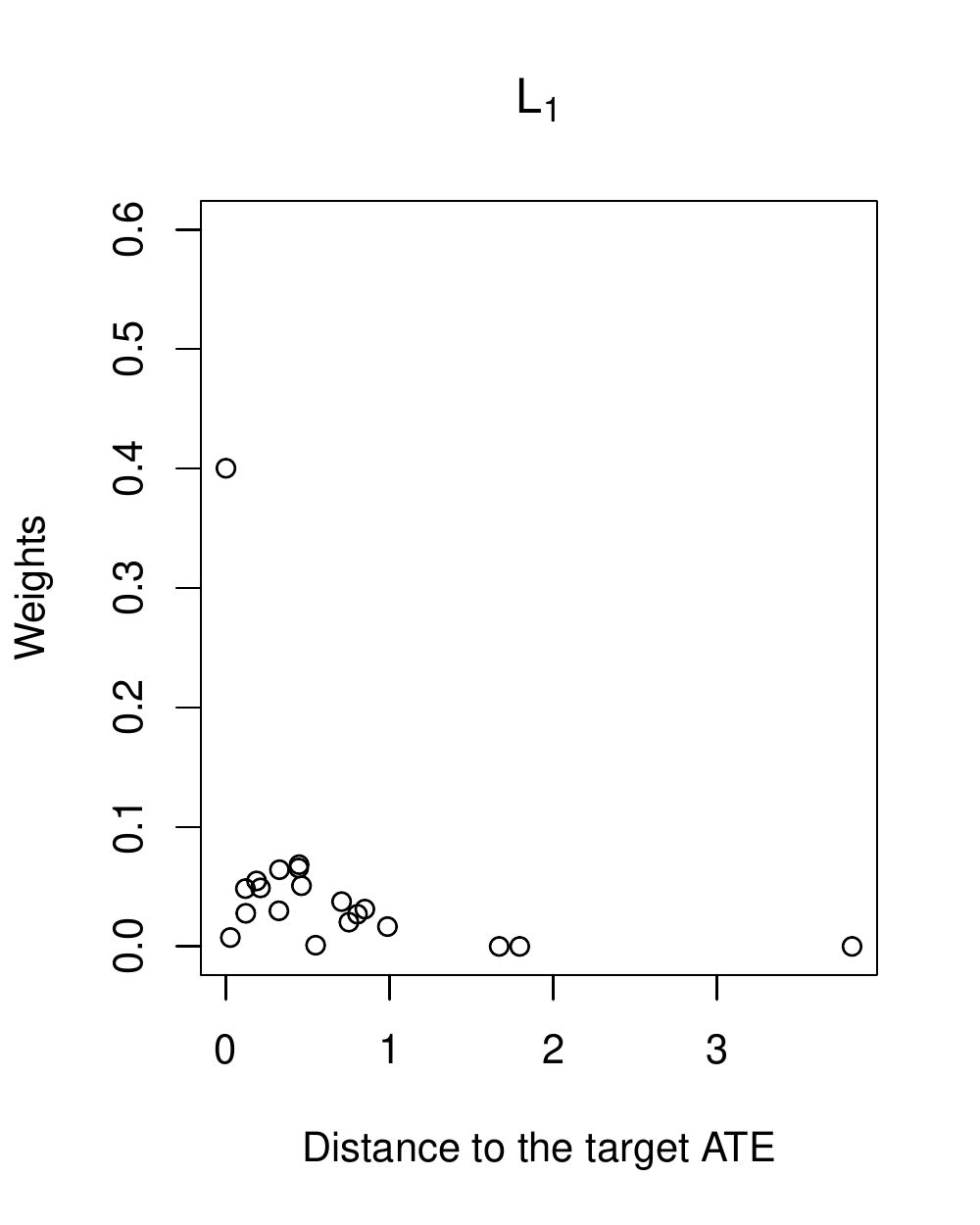} \\
\vspace{-0.5cm}
\includegraphics[scale = 0.60]{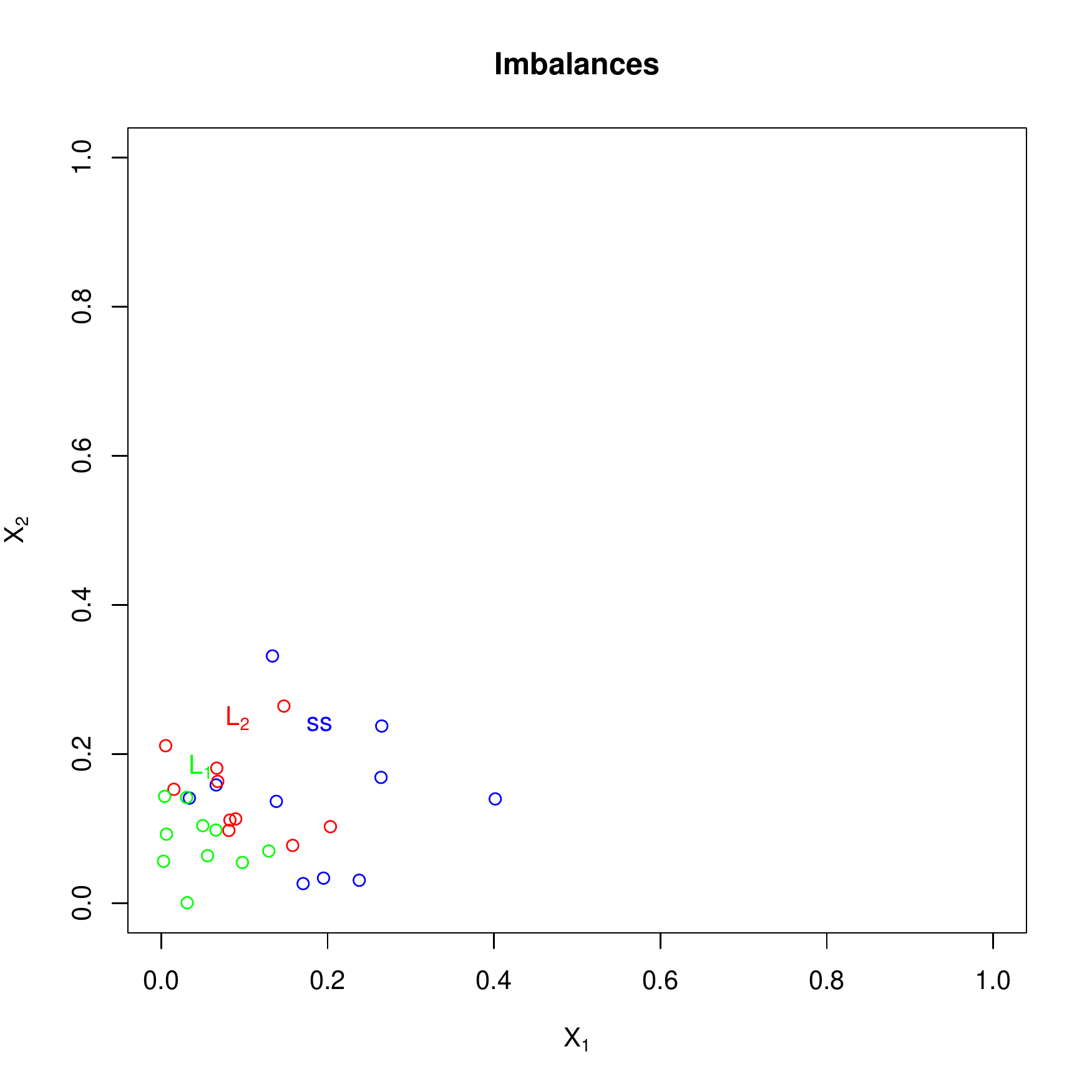} 
\vspace{-0.5cm}
\caption{The $\mathcal{D}_{\text{dense}}$ setting where the $\text{DREAMFUL}-\ell_2$ estimator has smaller RMSE, the $\text{DREAMFUL}-\ell_1$ estimator and $\text{DREAMFUL}-\ell_2$ estimator have comparable biases.}
    \label{fig:osqp}
\end{figure}

Recall that in the main text, we displayed the $\eta_k$ weights for $k=1,...,K$ when $K=20$ sites in the $\mathcal{\text{sparse}}$ setting. Here, we illustrate the $\eta_k$ weights, again for $K = 20$ sites, but for the $\mathcal{\text{dense}}$ setting. Figure \ref{fig:osqp} summarizes the results. The $\text{DREAMFUL}-\ell_1$ estimator places about $40\%$ of the weight on the target site and drops some sites entirely that have large discrepancy compared to the target site TATE. The ss estimator has the same weights as in the $\mathcal{D}_\text{sparse}$ setting. As in the $\mathcal{D}_\text{sparse}$ setting, the $\text{DREAMFUL}-\ell_2$ estimator produces weights between the $\text{DREAMFUL}-\ell_1$ estimator and the ss estimator. In terms of covariate imbalance, there is less difference between the $\text{DREAMFUL}-\ell_1$ and $\text{DREAMFUL}-\ell_2$ estimators, so the $\text{DREAMFUL}-\ell_2$ estimator may be preferable as it has a larger effective sample size and thus smaller variance since it drops fewer sites. In the $\mathcal{D}_\text{dense}$ setting, the differences in covariate imbalances using the $\text{DREAMFUL}-\ell_1$, $\text{DREAMFUL}-\ell_2$, and ss estimators are less pronounced. In this setting, the $\text{DREAMFUL}-\ell_2$ estimator is preferred to the $\text{DREAMFUL}-\ell_1$ estimator due to its smaller RMSE.

\section*{Appendix V. Example of Covariate Imbalance in the Real Data Analysis}
To illustrate the covariate imbalances using the ss, DREAMFUL-$\ell_1$, and DREAMFUL-$\ell_2$ estimators, we plot the imbalances of two covariates (proportion of female vs. proportion of ANT-MI-1), where each dot represents one of the $51$ hospitals being the target of interest. The results show that the DREAMFUL-$\ell_1$ estimator has the least amount of covariate imbalance, followed by the DREAMFUL-$\ell_2$ estimator, with the ss estimator displaying the largest covariate imbalance.

\begin{figure}[H]
\begin{center}
\includegraphics[scale = 0.7]{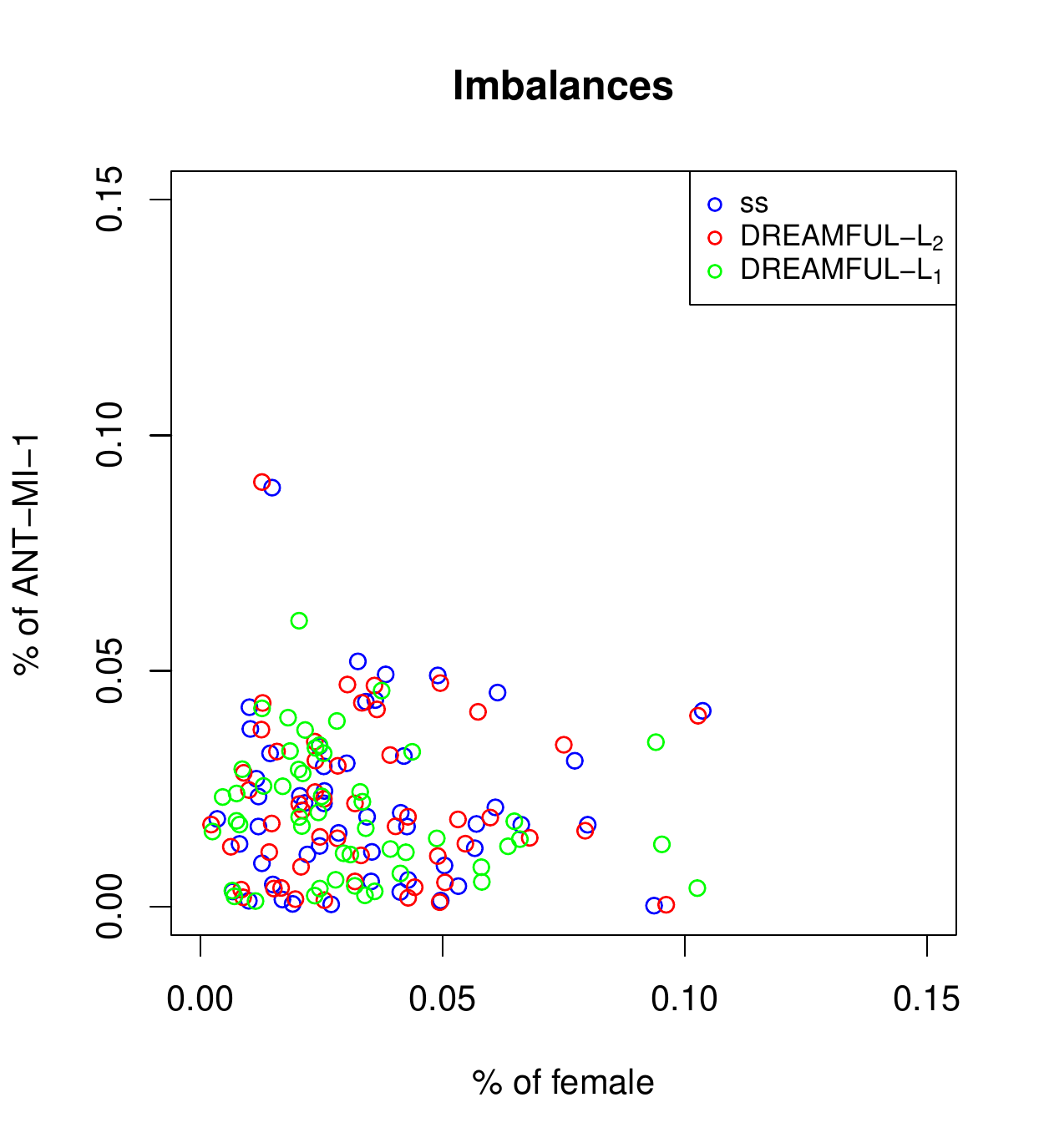} 
\end{center}
\vspace{-0.5cm}
    \caption{Imbalances of two covariates (proportion of female vs proportion of ANI-MI-1) for all 51 hospitals. Each dot represent one of the hospitals being a target.}
    \label{fig:weights}
\end{figure}